\documentclass[longauth]{aa}  
\usepackage{graphicx}
\usepackage{txfonts}
\usepackage{color}
\usepackage{xcolor}
\usepackage{soul}
\sethlcolor{orange}
\definecolor{byzantine}{rgb}{0.74, 0.2, 0.64}

\usepackage{hyperref}
\usepackage{longtable}
\hypersetup{
    colorlinks=true,
    citecolor=blue,
    linkcolor=blue,
    urlcolor=blue,
}
\usepackage{comment}

\defcitealias{Nascimbeni2022}{N22}
\defcitealias{Montalto2021}{M21}

\begin{document} 

   \title{The PLATO field selection process}

   \subtitle{II. Characterization of LOPS2, the first long-pointing field}

   \author{
   V.~Nascimbeni\thanks{E-mail: valerio.nascimbeni@inaf.it} \inst{\ref{inst1}}$^{\href{https://orcid.org/0000-0001-9770-1214}{\includegraphics[scale=0.5]{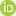}}}$ \and
   G.~Piotto\inst{\ref{inst2},\ref{inst1}}$^{\href{https://orcid.org/0000-0002-9937-6387}{\includegraphics[scale=0.5]{orcid.jpg}}}$ \and   
   J.~Cabrera \inst{\ref{inst5}} \and
   M.~Montalto  \inst{\ref{inst13}}\and
   S.~Marinoni\inst{\ref{inst10},\ref{inst11}}  \and
   P.~M.~Marrese \inst{\ref{inst10},\ref{inst11}} \and
   C.~Aerts \inst{\ref{inst3}} \and
   G.~Altavilla\inst{\ref{inst10},\ref{inst11}} \and
   S.~Benatti \inst{\ref{inst19}}\and
   A.~B\"orner \inst{\ref{inst4}} \and
   M.~Deleuil \inst{\ref{inst6}}\and
   S.~Desidera \inst{\ref{inst1}} \and 
   L.~Gizon  \inst{\ref{inst7}} \and
   M.~J.~Goupil \inst{\ref{inst8}} \and
   V.~Granata\inst{\ref{inst1},\ref{inst2}} \and
   A.~M.~Heras \inst{\ref{inst9}}\and
   D.~Magrin \inst{\ref{inst1}} \and
   L.~Malavolta \inst{\ref{inst2},\ref{inst1}} \and
   J.~M.~Mas-Hesse \inst{\ref{inst12}}\and
   H.~P.~Osborn \inst{\ref{inst14}} \and
   I.~Pagano \inst{\ref{inst13}}\and
   C.~Paproth \inst{\ref{inst4}}\and
   D.~Pollacco \inst{\ref{inst15}}\and
   L.~Prisinzano \inst{\ref{inst19}}\and
   R.~Ragazzoni \inst{\ref{inst2},\ref{inst1}}  \and
   G.~Ramsay \inst{\ref{inst16}}\and
   H.~Rauer \inst{\ref{inst5},\ref{inst18}}\and
   A.~Tkachenko \inst{\ref{inst3}} \and
   S.~Udry \inst{\ref{inst17} 
   }
          }

   \institute{
   INAF -- Osservatorio Astronomico di Padova, vicolo dell'Osservatorio 5, 35122 Padova, Italy \label{inst1} \and
   Dipartimento di Fisica e Astronomia ``Galileo Galilei'', Universit\`a degli Studi di Padova, Vicolo dell'Osservatorio 3, 35122 Padova, Italy \label{inst2} \and
   Deutsches Zentrum f\"ur Luft- und Raumfahrt (DLR), Institut f\"ur Planetenforschung, Rutherfordstra{\ss}e 2, 12489 Berlin-Adlershof, Germany \label{inst5} \and   
   INAF -- Osservatorio Astrofisico di Catania, Via S. Sofia 78, 95123, Catania, Italy \label{inst13} \and
   INAF -- Osservatorio Astronomico di Roma, Via Frascati, 33, 00078 Monte Porzio Catone (RM), Italy \label{inst10} \and
   SSDC-ASI, Via del Politecnico, snc, 00133 Roma, Italy \label{inst11} \and
   Institute of Astronomy, KU Leuven, Celestijnenlaan 200D, 3001, Leuven, Belgium \label{inst3}\and
   INAF–Osservatorio Astronomico di Palermo, Piazza del Parlamento, 1, I-90129, Palermo, Italy \label{inst19} \and
   Deutsches  Zentrum  f\"ur  Luft-  und  Raumfahrt  (DLR),  Institut  f\"ur  Optische  Sensorsysteme, Rutherfordstra{\ss}e  2, 12489 Berlin-Adlershof, Germany \label{inst4} \and   
   Aix-Marseille Universit\'e, CNRS, CNES, Laboratoire d’Astrophysique de Marseille, Technop\^{o}le de Marseille-Etoile, 38, rue Fr\'ed\'eric Joliot-Curie, 13388 Marseille cedex 13, France \label{inst6} \and
   Max-Planck-Institut f\"ur Sonnensystemforschung, Justus-von-Liebig-Weg~3, 37077~G\"ottingen, Germany \label{inst7}\and
   LESIA, CNRS UMR 8109, Universit\'e Pierre et Marie Curie, Universit\'e Denis Diderot, Observatoire de Paris, 92195 Meudon, France\label{inst8}\and
   European Space Agency (ESA), European Space Research and Technology Centre (ESTEC), Keplerlaan 1, 2201 AZ Noordwijk, The Netherlands \label{inst9} \and
   Centro de Astrobiolog\'{\i}a (CSIC--INTA), Depto. de Astrof\'{\i}sica, 28692 Villanueva de la Ca\~nada, Madrid, Spain \label{inst12}\and
   Center for Space and Habitability, University of Bern, Bern, Switzerland. \label{inst14} \and
   Department of Physics, University of Warwick, Gibbet Hill Road, Coventry CV4 7AL, UK \label{inst15} \and
   Armagh Observatory \& Planetarium, College Hill, Armagh, BT61 9DG, UK \label{inst16} \and
   Institute of Geological Sciences, Freie Universit\"at Berlin, Malteserstra{\ss}e 74-100, 12249 Berlin, Germany \label{inst18} \and
   Institute of Astronomy, KU Leuven, Celestijnenlaan 200D, 3001 Leuven, Belgium \label{inst20} \and
   Observatoire de Gen\`eve, Universit\'e de Gen\`eve, Chemin Pegasi 51, 1290 Sauverny, Switzerland \label{inst17} 
   }

   \date{Submitted 20 September 2024 / Accepted 11 January 2025}

  \abstract{PLAnetary Transits and Oscillations of stars (PLATO) is an ESA M-class mission to be launched by the end of 2026 to discover and characterize transiting planets around bright and nearby stars, and in particular habitable rocky planets hosted by solar-like stars. Over the mission lifetime, an average of 8\% of the science data rate will be allocated to Guest Observer programs selected by ESA through public calls. Hence, it is essential for the community to know in advance where the observing fields will be located. 
  In a previous paper, we identified two preliminary long-pointing fields (LOPN1 and LOPS1) for PLATO, respectively in the northern and southern hemispheres. Here we present LOPS2, a slightly adjusted version of the southern field that has recently been selected by the PLATO Science Working Team as the first field to be observed by PLATO for at least two continuous years, following the scientific requirements. In this paper, we describe the astrophysical content of LOPS2 in detail, including known planetary systems, bright stars, variables, binary stars, star clusters, and synergies with other current and future facilities.}


   \maketitle

\nolinenumbers

\section{Introduction}

PLAnetary Transits and Oscillations of stars (PLATO; \citealt{Rauer2024}) is an ESA M-class mission primarily designed to detect a large number of planetary systems hosted by nearby and bright stars through the transit technique. Among many other aims, the goal of this mission is the discovery and confirmation of Earth-like planets, that is, habitable rocky planets hosted by solar-type dwarfs \citep{Heller2022,Matuszewski2023}. Very accurate stellar parameters, including stellar age at 10\% accuracy,\footnote{For solar-like stars.} will be extracted from the light curves themselves through asteroseismological analysis \citep{Cunha2021,Betrisey2023}. This will enable the discovered planetary systems to be placed in a consistent evolutionary context. Furthermore, follow-up and confirmation of candidate planets discovered by PLATO is part of the mission through the Ground-based Observation Program (GOP) of the PLATO Consortium. We refer the reader to \citet{Rauer2024} for a recent, detailed review of the mission.

\begin{figure}
    \centering
    \includegraphics[width=0.95\columnwidth]{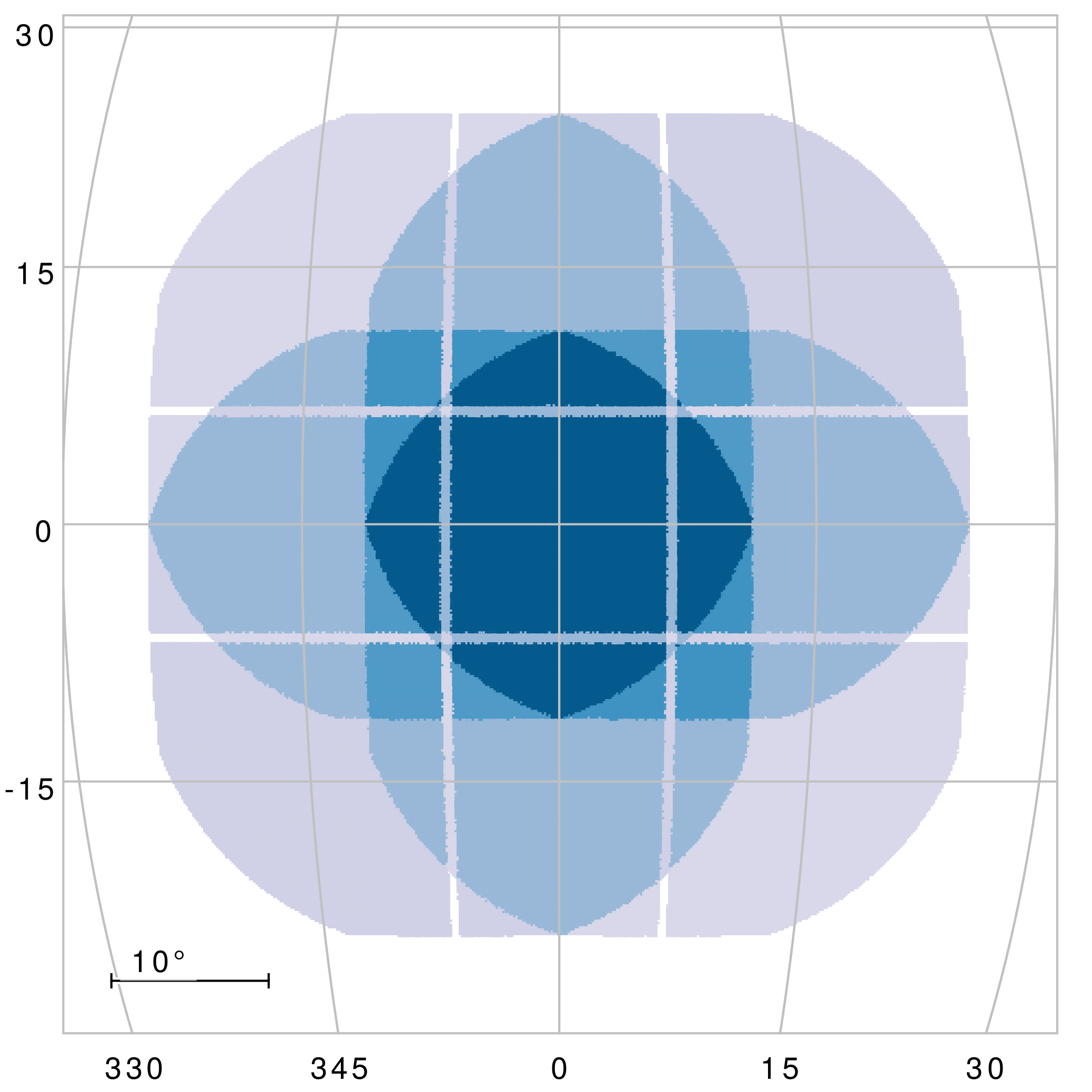}
    \caption{Geometry of the PLATO field FOV. In this figure, the field is centered at the origin (0,0) of a generic spherical reference frame (units are degrees; projection is orthographic). The number of ``normal'' cameras covering a given line of sight is color coded. The four blue shades, from dark to light, map the regions covering 325/153/847/824~deg$^2$ observed respectively with 24, 18, 12, six cameras (and corresponding to four, three, two, and one group of six co-pointing telescopes each).}
    \label{fov}
\end{figure}

During its nominal four-year mission, PLATO will continuously monitor one or two pre-selected fields for at least two years each during the so-called Long-duration Observation Phase (LOP). The mission has been designed to have the capability to observe additional fields for shorter intervals, at least two months each (Step-and-stare Observation Phase; SOP).
The 24 ``normal'' cameras (NCAMs) of PLATO are not all co-aligned, but rather split into four groups with an angular offset (9.2$^\circ$) with respect to the satellite bore-sight \citep{Pertenais2021}. As a consequence, the overall field of view (FOV) of about $2\,149$~$\textrm{deg}^2$ is covered by a variable number of NCAMs, from six to 24, depending on the specific line of sight considered, as illustrated by Fig.~\ref{fov}. In terms of sky area, approximately 325~deg$^2$ will be covered by 24 NCAMs, 153~deg$^2$ by 18 NCAMs, 847~deg$^2$ by 12 NCAMs, and 824~deg$^2$ by six NCAMs (assuming a corrected FOV of $19.2^\circ$ and CCD gaps of 1.3~mm; \citealt{Pertenais2021}), although the exact numbers could slightly change according to the actual optical performances once in flight.

Due to telemetry constraints, PLATO will not be able to download full-frame images at high cadence.\footnote{Full-frame images will be acquired during the commissioning phase, at least once every 90 days.} Instead, it will perform a significant part of the photometric analysis onboard. Only light curves, centroid data, and a number of stamp-like ``imagettes'' will be transmitted back to Earth, meaning that targets have to be selected in advance. PLATO targets are divided into four main samples, summarized in Table~\ref{tab:samples}, with different requirements on magnitude, spectral type, and noise-to-signal ratio (NSR); three of such samples are composed by main-sequence or subgiant stars with a spectral type from F5 to K7 (\texttt{P1}, \texttt{P2}, \texttt{P5}), while a fourth one contains only M dwarfs (\texttt{P4}). The PLATO Scientific Requirement Document\footnote{ESA reference: PTO-EST-SCI-RS-0150.} (SciRD) identifies \texttt{P1} ($V<11$, $\textrm{NSR} < 50$~ppm in one hour) as the highest-priority sample, and sets a requirement of at least $15\,000$ \texttt{P1} targets to be observed during the LOPs. We refer to the SciRD, \citet{Nascimbeni2022}, and \citet[][hereafter quoted as \citetalias{Nascimbeni2022} and \citetalias{Montalto2021}]{Montalto2021} for a complete summary of the definitions and requirements on the PLATO samples.

PLATO is currently on schedule to be launched at the end of 2026 and to start its routine scientific operations in mid-2027. Most raw and calibrated data products from PLATO will be made publicly available as soon as possible.\footnote{With the exception of GO programs and reserved targets allocated to the Plato Mission Consortium, within their proprietary time.} The astronomical community has the opportunity to be deeply involved at different stages: preparation, data analysis, and follow-up. PLATO data products will be released quarterly. Each set of data from a three-month observation period will be made available approximately six months later. For around $20\,000$ bright targets prioritized for ground-based follow-up observations, the data products will be delivered roughly 1.2~years after the corresponding three-month observation period. On top of that, over the mission lifetime, an average of 8\% of the PLATO science data rate will be allocated to Guest Observer (GO) programs selected by ESA through open calls, so it is essential to select and communicate each field that is going to be pointed at with a proper time margin.

In a previous work (Paper I; \citetalias{Nascimbeni2022}), we described all the steps of the process that led to the identification of two provisional fields (labeled LOPN1 and LOPS1) as candidates for the LOP. Back then, it was still unknown which field would be selected as the first one. Also, it was anticipated that the final LOP fields could have had small changes in position and/or rotation angle. In this paper, we present LOPS2, the long-pointing field finally chosen by the ESA PLATO Science Working Team (PSWT) as the first field to be observed by PLATO. After summarizing the fine-tuning process and the general properties of LOPS2 in Section~\ref{sec:finetune}, we review and discuss its astrophysical content in Section~\ref{sec:content}. Finally, the conclusions and some prospects for the future work to be done are given in Section~\ref{sec:conclusions}. A glossary of the most common acronyms used throughout this work is compiled in Table~\ref{table:glossary} in the Appendix.

\section{The choice of the first long-pointing field}\label{sec:finetune}

Paper I gave a very detailed description of the process that led to the selection of LOPS1 and LOPN1 as candidate long-duration fields for PLATO. In order to fine-tune their position/rotation angle and to help the SWT to select the first LOP, three main criteria were considered:
\begin{itemize}
    \item \emph{Duty cycle:} The PLATO attitude is defined by the pointing direction and the rotation around the mean line of sight. Because of the geometry of the overlapping cameras, different rotation angles result in different sets of stars observed with a given photometric precision (with a different number of cameras). For pointing directions with ecliptic latitudes between $63^\circ<|\beta|<70^\circ$, the choice of the rotation angle around the mean line of sight is constrained by the geometry of the spacecraft. There is always a possible rotation angle that meets the duty cycle requirements (93\% of time on-target), but some angles result in incompatible attitudes. For pointing directions with $|\beta|>70^\circ$ all choices of the rotation angle are compatible with the duty cycle requirements.
    \item \emph{Rotation angle}: the choice of a different rotation angle should be investigated for any technical or scientific advantage.
    \item \emph{Synergies:} the field with the most interesting synergies with other facilities should be given higher priority, in particular with regard to the radial-velocity follow-up process, considering the time frame of the mission.
\end{itemize}

In what follows, we discuss the impact of these three key ingredients on the final decision. The final decision is presented and discussed in Section~\ref{sec:lops2first}.

\subsection{Duty cycle}

As anticipated in Paper I, Section 2.4, the need of keeping the solar panels of PLATO at a nominal level of solar irradiation while constantly staring at the same LOP field for $\geq 2$~years implies that the roll angle of the spacecraft must be moved at regular intervals throughout the year. Since the PLATO field, by design, shows a $90^\circ$ symmetry, it is natural to perform $90^\circ$ rotations each quarter, also known as quarterly rolls. A similar observing strategy was already devised for the original \emph{Kepler} mission \citep{Borucki2010}. The ecliptic latitude $\beta$ is the crucial parameter to be investigated, as it determines the incidence angle of the solar rays on the panels. There is a critical value of $|\beta|$ above which an observing quarter can start at any moment, without any need of interruptions before the first roll. 

Following a detailed assessment of the problem with the PLATO mission team, we concluded that the threshold must be set at $|\beta|>69^\circ.671$ for the geometrical center of the field, that is, slightly more stringent than the formal requirement of $|\beta|>63^\circ$ previously applied in Paper I. While this is already fulfilled by LOPN1 ($\beta \simeq 75^\circ.85$; Table~2 of Paper I), LOPS1 is slightly below that value, being at $\beta \simeq -66^\circ.30$. As a working hypothesis, we examined, within the ``compliant'' region of our HEALPix level-4 grid (Fig.~5 of Paper I), the grid point at $\beta < -70^\circ$ and closest to LOPS1 (\#2189 in our original HEALPix numbering scheme; $\beta\simeq -71^\circ.1$). We renamed this field LOPS2, and verified that:
\begin{itemize}
    \item LOPS2 is centered at the same galactic latitude as LOPS1 ($b\simeq -24^\circ.6$), but shifted by $\sim 5^\circ.1$ westward along galactic longitude, leaving the average stellar crowding (and hence the expected false positive ratio; \citealt{Bray2023}) essentially unchanged;
    \item LOPS2 is centered at a slightly larger declination in modulus ($\delta\simeq -47^\circ.9$ vs.~$-42^\circ.9$), with no significant impact on the follow-up strategy;
    \item the prioritization metric as defined in Paper I is 1\% smaller (0.980 vs.~0.990), that is, the difference is negligible;
    \item the number of P1 targets in LOPS2, defined as in Table~\ref{tab:samples} and evaluated with the most updated version of the PLATO Input Catalog (PIC) (v.2.0.0), is 0.4\% lower, but still well within the margins of SciRD requirements.
\end{itemize}
In other words, LOPS2 appears as an optimal alternative, having properties very similar to LOPS1, but allowing us to be much more efficient during mission operations. For example, the in-orbit commissioning will take place before the start of the nominal science operations. In order to avoid constraints on the choice of the rotation angle for commissioning (and hence in the duty cycle), the pointing direction for the commissioning field must have $|\beta|>70^\circ$, which was an incompatible constraint with LOPS1. We therefore consider LOPN1 (as defined in Paper I) and LOPS2 (as defined above; detailed coordinates in Table~\ref{tab:lops2}) as candidate LOP fields hereafter.

\begin{table}\centering
\caption{Definitions of the main PLATO stellar samples.}
\begin{tabular}{c|cp{1.55cm}cc}
\hline\hline
sample & SpT & lum.~class & mag. limit & noise \\ \hline 
P1 & F5-K7 & dwarfs and subgiants & $V<11$ & $<50$~ppm \\ 
P2 & F5-K7 & dwarfs and subgiants & $V<8.5$ & $<50$~ppm \\ 
P4 & M & dwarfs & $V<16$ & --- \\ 
P5 & F5-K7 & dwarfs and subgiants & $V<13$ & --- \\ \hline
\end{tabular}\label{tab:samples}
\tablefoot{The rows give: the name of the PLATO stellar sample, the spectral type, the luminosity class, the limiting magnitude in the $V$ band, and the noise limit in ppm in one hour (see also Section~\ref{sec:main_targets}).}
\end{table}

Although it has a negligible impact on the duty cycle, it is worth reminding that the the dates of the quarterly rotations along the orbit are uniquely determined by the choice of the attitude. This is a consequence of the duty cycle requirement and the geometry of the spacecraft. For LOPS2 and LOPN1, the quarterly rotations will take place at the end of January, April, July, and October (the exact dates differ by few days in both hemispheres). For a scheduled launch end 2026 and considering that the commissioning shall take less than 90 days, it could very well be that the first quarter is significantly shorter than the nominal duration of approximately 90 days, similar to the situation experienced by \emph{CoRoT} and \emph{Kepler}.

\subsection{Rotation angle}\label{sec:rotation}

In paper I, both fields were presented at zero rotation angle, which, in our convention, implies that one side of the field is parallel to the galactic plane. We investigated the impact of the rotation angle $\varphi$ by generating nine simulated catalogs for both LOPN1 and LOPS2 by varying the angle by $10^\circ$ intervals (from $\varphi=0^\circ$ to $80^\circ$). For the resulting 18 catalogs, extracted from the PIC 2.0.0, we added a NSR column (calculated with PINE; \citealt{Borner2024}), in order to estimate the number of P1-P2-P4-P5 counts and to assess the visibility of our main targets of interest. 

As expected, given the large size of the PLATO field and its fair degree of radial symmetry, $\varphi$ has a mostly negligible impact on the number of targets. For both LOPN1 and LOPS2 the P1 counts are constant within $\sim1\%$, with the highest number corresponding to the $\varphi =0^\circ$ case (Fig.~\ref{fig:rotation_counts}, left panel). On the other hand, P4 and P5 counts are virtually unaffected, showing just $<0.5\%$ variations, that is  at the level of the predicted Poissonian noise (Fig.~\ref{fig:rotation_counts}, right panel). We can conclude that $\varphi$ has no practical effects on the planet yield of PLATO.

It is worth considering whether there is any other reason to rotate the field, for instance due to specific astrophysical objects we would like to include, or avoid, in the LOP field. As for the former, we verified that 1) the number of known planetary systems is not significantly affected by $\varphi$ for LOPN1, and is actually maximized at $\varphi=0^\circ$-$10^\circ$ for LOPS2 due to an overdensity of TESS-discovered planets lying in the galactic southwest corner of the field (as we show in Section~\ref{sec:known}); and 2) no open cluster among the ones of potential interest for PLATO (i.~e., nearby, sparse, and mature) slips in or out of the field for any rotation angle choice, as it is shown in Section~\ref{sec:content} for LOPS2. 
More in general, the range of galactic latitudes probed is always $0^\circ<|b|\lesssim 50^\circ$ regardless of $\varphi$, and therefore any property correlated with $b$, such as the amount of stellar crowding and the diversity of stellar populations probed (in terms of age, kinematics, metallicity, etc.) is approximately the same. 

Finally, it makes sense to investigate whether there is any way to rotate the field in order to move any extremely bright star ($V<1$) off the silicon. Paper I identified Deneb and Vega as the brightest sources on LOPN1, and Sirius and Canopus for LOPS1. Moving from LOPS1 to LOPS2 also implies that Sirius is now left outside the focal plane for any choice of $\varphi$, while there is no consequence on the other three stars, that are unavoidable and always monitored with the same number of NCAMs (18-24 for Vega, 6 for Deneb, 24 for Canopus) for almost any possible rotation angle.\footnote{It could be possible in principle to fit Deneb within a $\sim30 '$ CCD gap of the northern field at $\varphi\simeq 11^\circ$, but the technical feasibility of such a choice and its possible unintended consequences (see for instance \citealt{Porterfield2016}) need to be assessed. This will be done at a further stage, should LOPN1 be scheduled for observation.} Fig.~\ref{fig:rotation_fields} (but also Section~\ref{sec:bright} below) investigates this point a bit further, looking at how the number of bright stars (up to $V<3$) in LOPS2 changes as a function of $\varphi$, at $10^\circ$ intervals from $0^\circ$ to $80^\circ$ (the symmetry of the PLATO field makes it invariant after $90^\circ$ rotations). The nominal $\varphi=0^\circ$ case is the only one for which all the five $1<V<2$ stars ($\delta$~CMa, $\epsilon$~CMa, $\gamma_2$~Vel, $\delta$~Vel, $\epsilon$~Car) always land on 6 NCAMs regions, while for instance four of them fall on 12 NCAMs in the $\varphi=40$-$50^\circ$ case. Finally, also the distribution of $2<V<3$ stars does not change significantly as a function of $\varphi$.

\begin{figure*}
    \centering
    \includegraphics[width=1.9\columnwidth]{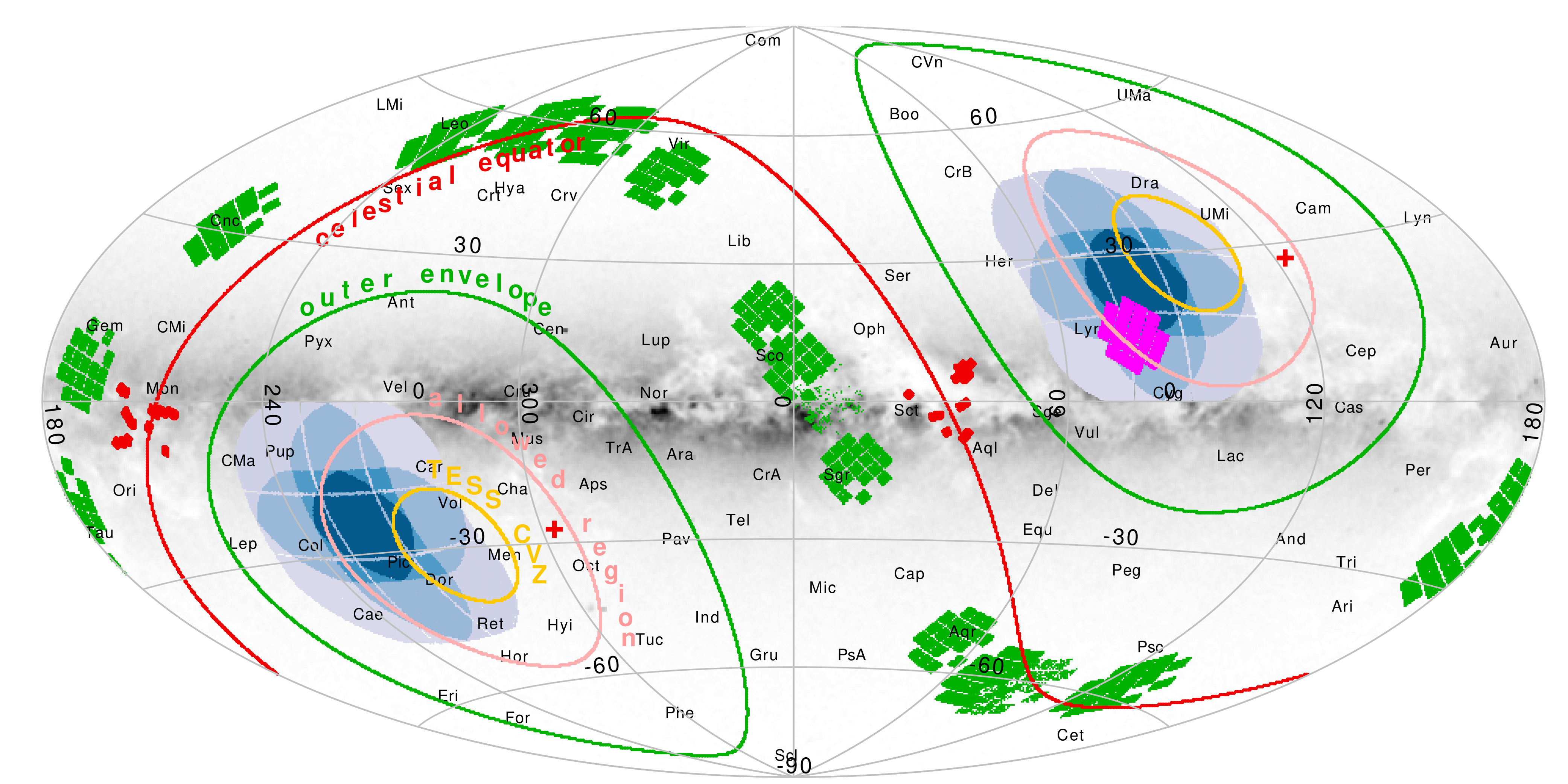}
    \caption{All-sky Aitoff projection in galactic coordinates of LOPS2 and LOPN1, showing the formal constraints for the selection of the PLATO LOP fields and the synergies with other missions. The two pink circles represent the $|\beta|>63^\circ$ technical requirement for the center of the LOP fields (``allowed region''), implying that the overall envelopes of every allowed field choice are two ecliptic caps at $|\beta|\gtrsim38^\circ$ (green circles). LOPS2 (lower left) and LOPN1 (upper right) are plotted with blue shades according to the number of co-pointing cameras, as in Fig.~\ref{fov}. The footprints of CoRoT (red), \emph{Kepler} (magenta), and K2 (green) are over-plotted together with the TESS continuous viewing zone at $|\beta|\gtrsim78^\circ$ (yellow circle). The background gray layer is color coded according to the areal density of $G<13.5$ stars from \emph{Gaia} DR3. The celestial equator and poles are marked with a red line and crosses, respectively. This sky chart is also plotted in equatorial and Ecliptic coordinates in Fig.~\ref{fig:full_sky_equ}-\ref{fig:full_sky_ecl}.}
    \label{fig:full_sky}
\end{figure*}

In summary, we conclude that there is no reason to change the rotation angle of both LOPN1 and LOPS2 from its initial value of $\varphi=0^\circ$. We implicitly assume zero rotation (i.e., one side of the field almost tangent to the galactic plane) hereafter, as plotted in the all-sky projection in Fig.~\ref{fig:full_sky}.


\subsection{Synergies and follow-up}
\label{sec:synergies}

The last possible criterion remaining to guide the choice between LOPN1 and LOPS2 are the synergies with other past, present and forthcoming observing facilities, including (and with a particular attention to) those to be involved with the PLATO follow-up process during and after the nominal scientific observations of the first field, approx.~from mid-2027 with a minimum duration of two years.

As for the previous space-based transit surveys, we first remind that no overlap between a LOP field and K2 or CoRoT \citep{Howell2014,Auvergne2009} is possible at all, since the latter fields are all located close to the ecliptic by design, inaccessible during the LOP phase (Fig.~\ref{fig:full_sky}). The TESS continuous viewing zone (CVZ) at $|\beta|>78^\circ$, on the other hand, is monitored at 100\% in LOPN1 and $\gtrsim 90\%$ in LOPS2, mostly in six- and 12-camera regions. Finally, the \emph{Kepler} field is fully included in the LOPN1, almost entirely with at least 12 NCAMs and up to 24.
While the scientific case of following-up the \emph{Kepler} TTV systems with PLATO is very solid \citep{JontofHutter2021b,JontofHutter2021a}, we emphasize that it should not be seen as a main driver of the mission, but rather as a potential source of interesting additional science.\footnote{It is worth mentioning that the follow-up of the \emph{Kepler} field is part of the science goals of the proposed Earth-2.0 mission \citep{Ge2022}.} Indeed, the last word is on the follow-up availability and timeliness, since the confirmation of the PLATO candidates and eventual mass determination is part of the mission products (and goals) through the Ground-based Observation Program (GOP) of the PLATO Consortium.

We already noted in Paper I that the range of declination spanned by any compliant LOP field (including LOPN1 and LOPS2) hinders the follow-up of a large fraction of the northern field with most southern facilities and vice versa, with the exception of a strip in the low-$|\delta|$ region of the fields, whose size depends on the maximum acceptable airmass $X$ and the geographical latitude $\psi$ of the observing site. Such a strip can be typically monitored from the opposite hemisphere only for 1-2 months per year at reasonably low airmass values ($1.5 \lesssim X< 2.0$; see Paper I), making ultra-high precision ($<1$~m/s) RV measurements unfeasible or exceedingly difficult for the vast majority of P1 targets. A detailed analysis on how the visibility of LOPS2 and its sub-regions covered by six, 12, 18, 24 NCAMs is dependent on $\psi$ and $X$ is shown in Fig.~\ref{fig:reachable}. As a reference value, we note that at the Observatorio Roque de Los Muchachos (ORM) only 16\% of LOPS2 is reachable at $X<2$ (and only in the 6-12 NCAMs region), while at Mauna Kea 36\% of LOPS2 can be observed with the same requirement (but only 8\% of the 24~NCAMs region can be reached).

A survey of the high-precision RV spectrographs available to the European community reveals that most of them are located in Chile: they include the already operational HARPS \citep{Mayor2003} and ESPRESSO \citep{Pepe2021}; and the planned ANDES, whose first light at the ESO Extremely Large Telescope (ELT) is currently foreseen in 2031 \citep{Marconi2022}. More in general, large-scale facilities such as ELT will also host other instruments with a huge potential in the follow-up process, such as coronagraphs and infrared spectrographs. Conversely, no large telescope of the 30-m class is currently at the construction stage in any northern observatory.

It is worth mentioning two other important survey projects in the south that could have interesting implications for the PLATO preparation and/or follow-up: LSST \citep{Ivezic2019} at the Vera C.~Rubin Observatory, which will map the whole southern hemisphere (including LOPS2) beginning in 2025, and 4MOST at the VISTA telescope \citep{deJong2012}, starting its consortium surveys also in 2024. The southern hemisphere will also host the Giant Magellan Telescope (GMT; \citealt{Johns2012}), which is scheduled to be operational in the early 2030s.

\subsection{LOPS2 as first LOP field}
\label{sec:lops2first}

\begin{figure*}
    \centering
    \includegraphics[width=0.9\columnwidth]{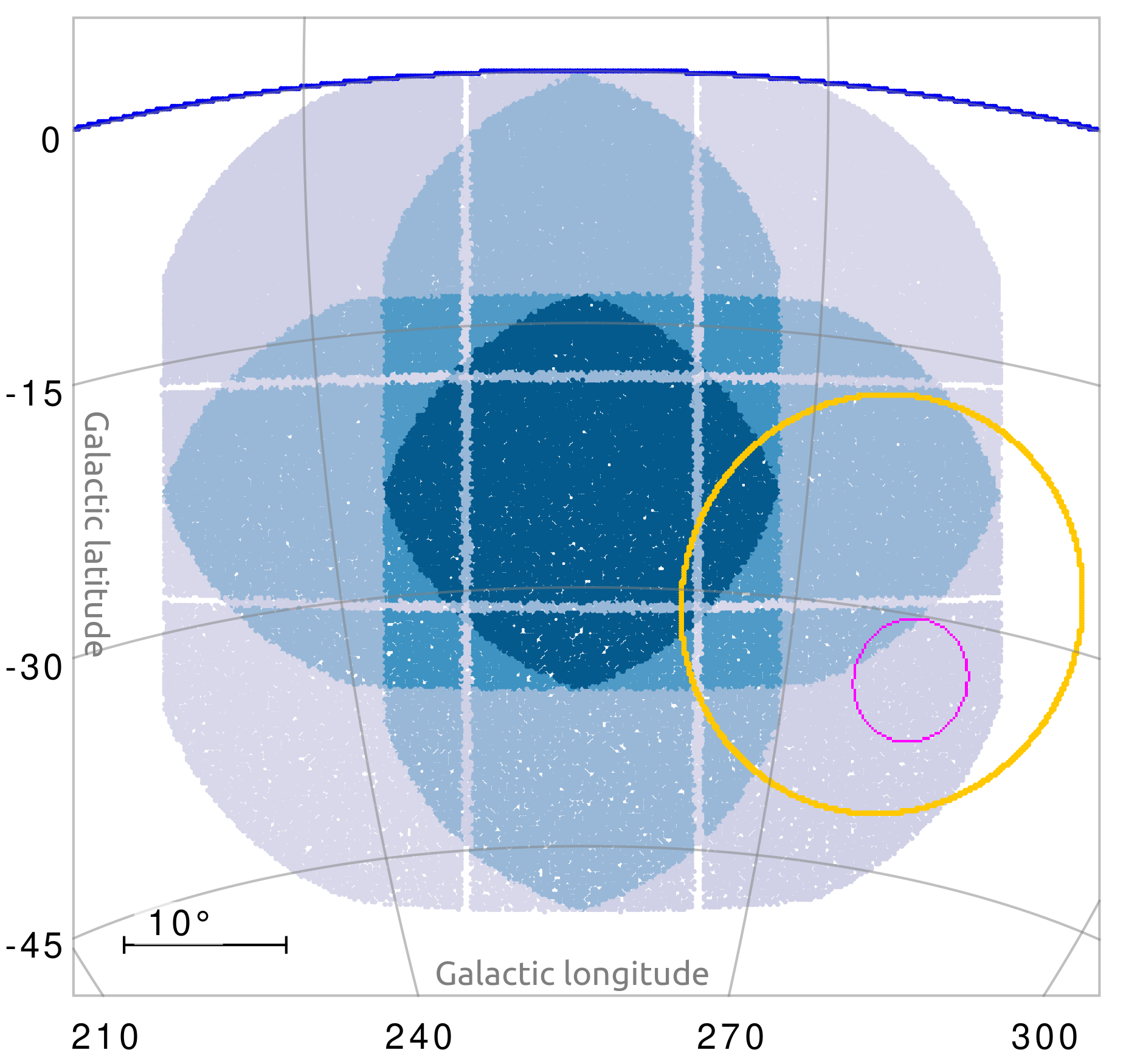}\hspace{5mm}
    \includegraphics[width=0.9\columnwidth]{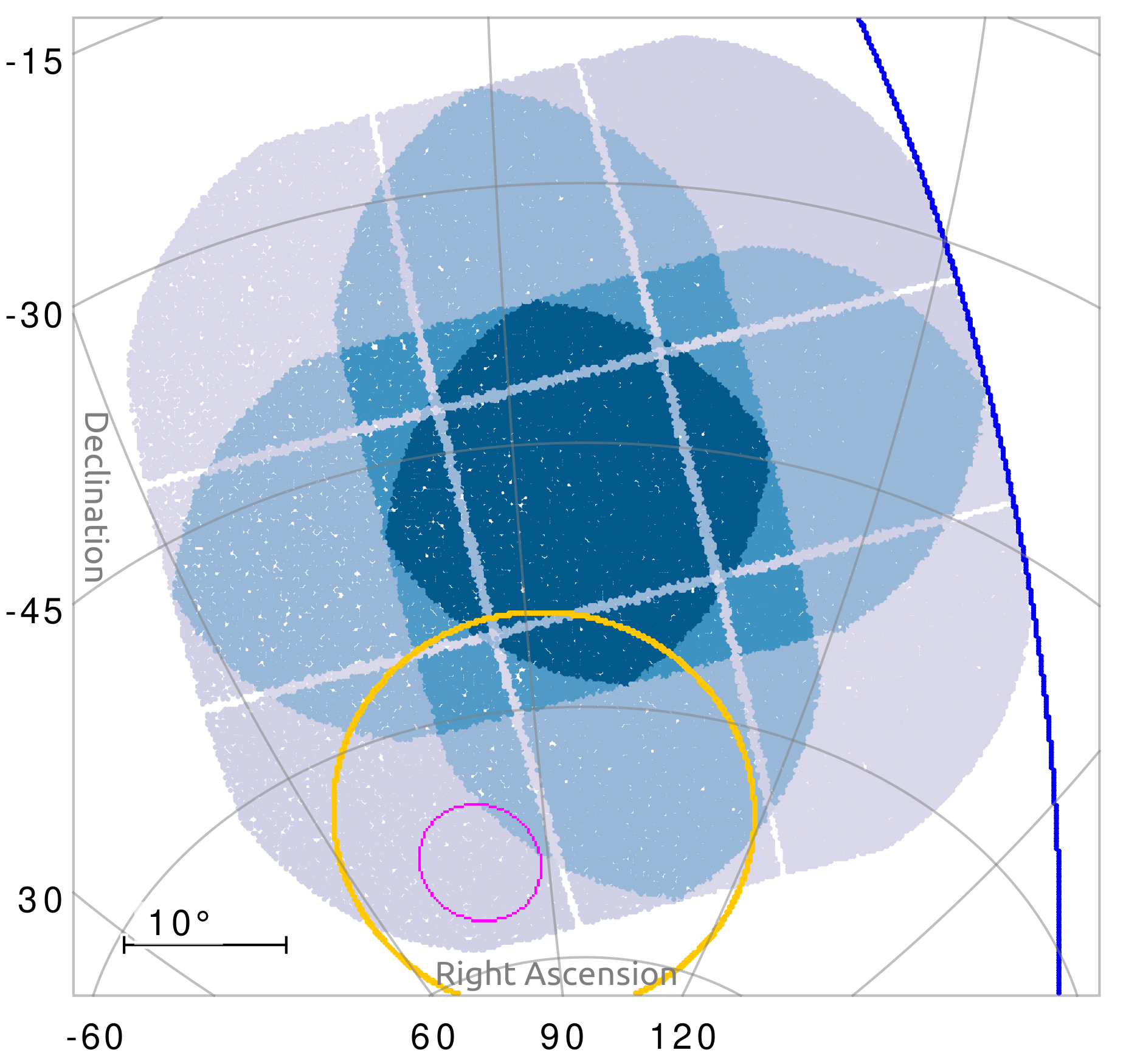}
    \caption{LOPS2 footprint. \emph{Left panel:} Orthographic projection in galactic coordinates centered on LOPS2.  The field is color coded according to the number of NCAMs as in Fig.~\ref{fov}. The yellow circle at $\beta=-78^\circ$ approximates the southern TESS continuous viewing zone (CVZ), while the magenta circle is centered on the Large Magellanic cloud. \emph{Right panel:} Same as in left panel but in equatorial coordinates.}
    \label{fig:lops2}
\end{figure*}

On the basis of what has been discussed in the previous Section, in June 2023 the SWT has formally approved LOPS2 (at $\varphi=0^\circ$) as the first LOP field. A full-sky map in galactic coordinates including LOPS2 and LOPN1 is presented in Fig.~\ref{fig:full_sky}; the same full-sky map is also plotted in equatorial and Ecliptic coordinates in Fig.~\ref{fig:full_sky_equ}-\ref{fig:full_sky_ecl}; a zoomed-in chart centered on LOPS2 is plotted in Fig.~\ref{fig:lops2} in both galactic and equatorial coordinates. All the relevant quantities for LOPS2 are listed in Table~\ref{tab:lops2}, including the number of targets for each PLATO sample. The latter are calculated by adopting PIC~2.0.0 as input catalog and conservatively assuming an end-of-life scenario with 22 surviving NCAMs (``EOL 22''). We refer the reader to Section~\ref{sec:favorite} in the Appendix on how to check if a given object falls within the LOPS2 boundaries.

\begin{table}\centering
\caption{Some properties of the LOPS2 field.}
\begin{tabular}{crl}
\hline\hline
parameter & value & notes \\ \hline 
 $\alpha$ [deg]& 95.31043 & ICRS \rule{0pt}{16pt}\\ 
 $\alpha$ [hms]& 06:21:14.5 & ICRS\\ 
 $\delta$ [deg]& $-$47.88693 & ICRS\\
 $\delta$ [dms]& $-$47:53:13 & ICRS\\
 $l$ [deg] & 255.9375 & IAU 1958\rule{0pt}{16pt}\\
 $b$ [deg] & $-$24.62432 & IAU 1958\\
 $\lambda$ [deg] & 101.05940 & Ecliptic \rule{0pt}{16pt}\\
 $\beta$ [deg] & $-$71.12242 & Ecliptic \\ 
P1 targets & 8\,235 & SciRD req.~7\,500 \rule{0pt}{16pt} \\ 
P2 targets & 699 & SciRD req.~500\\ 
P4 targets & 12\,415 & SciRD req.~2\,500\\
P5 targets & 167\,149 & SciRD req.~122\,500 \\ \hline
\end{tabular}\label{tab:lops2}
\tablefoot{The rows give the coordinates for the field center in equatorial, galactic, and ecliptic reference frames and the number of targets available in the P1-P2-P4-P5 samples. The latter are calculated from PIC~2.0.0 assuming the ``EOL 22'' scenario (see Section~\ref{sec:lops2first}).}
\end{table}

We mention that LOPS2 includes most of the TESS southern CVZ (as we further discuss in Section~\ref{sec:tess}), the Large Magellanic Cloud (LMC; but not the SMC), the southern Ecliptic pole (but not the celestial one), and one of its sides is a $49^\circ$-long strip parallel to the galactic plane and reaching up to $0^\circ.25$ from it; the opposite side goes as far as $b\simeq -49^\circ$. The range of declination spanned by LOPS2 is $-73^\circ.8<\delta< -20^\circ.8$, with 90\% of its area lying within the $-64^\circ.0<\delta< -28^\circ.6$ range. Six constellations are covered entirely, or almost entirely by LOPS2: Pic, Col, Dor, Cae, Ret, Pup; smaller parts of CMa, Car, Vol, Men, Hor, Eri, Lep, Vel, Hyi are also overlapped.

\begin{figure*}
    \centering
    \includegraphics[width=0.95\columnwidth]{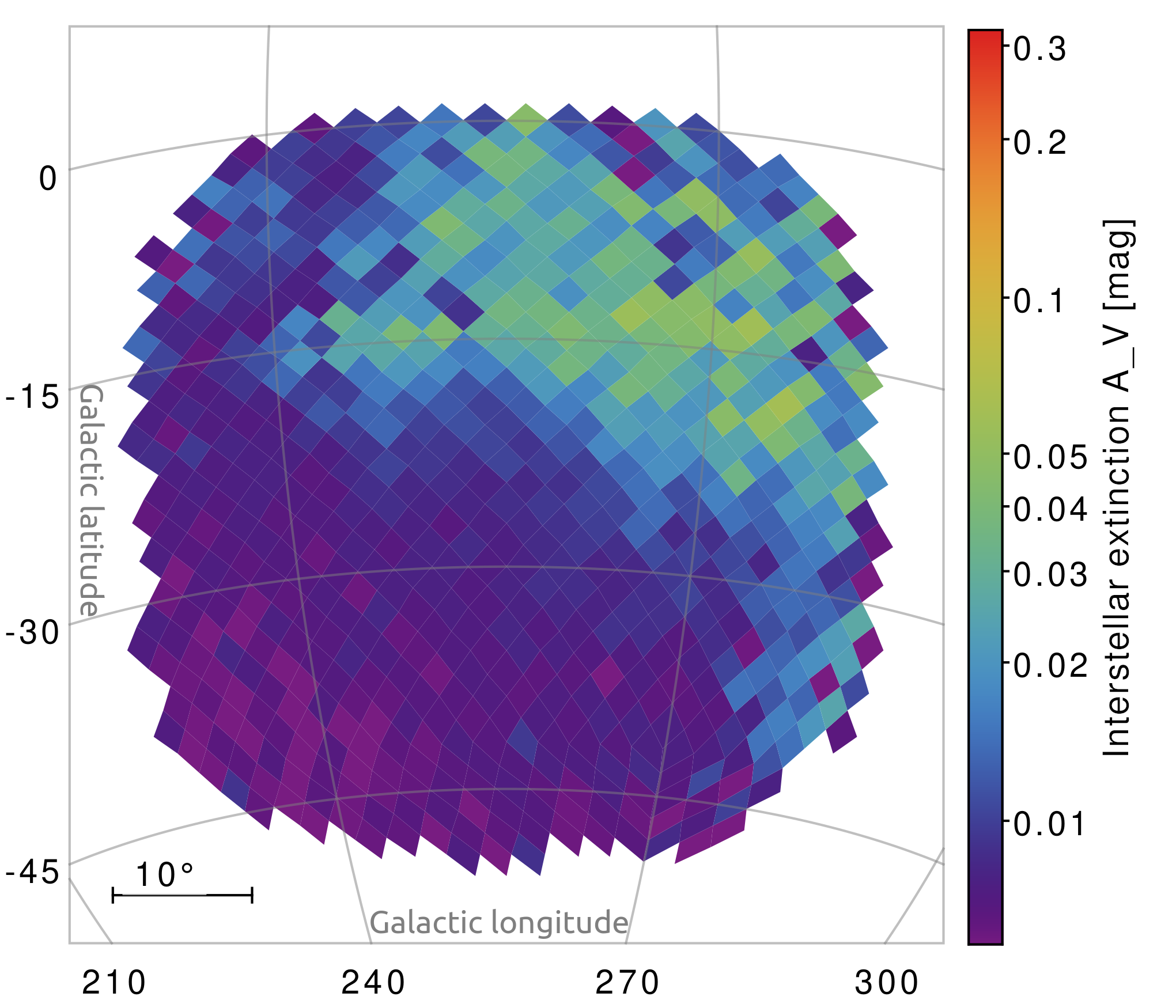}\hspace{5mm}
    \includegraphics[width=0.95\columnwidth]{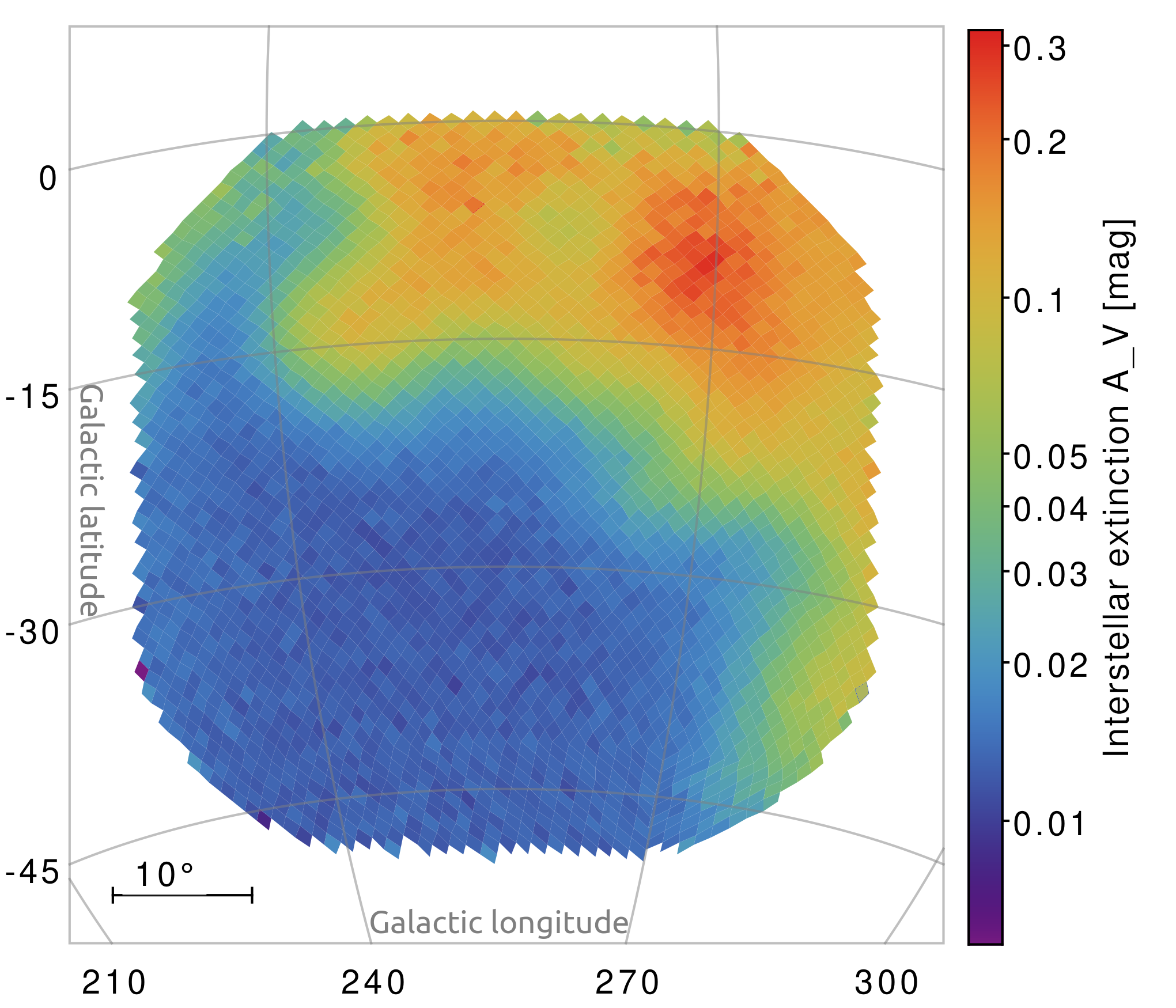}
    \caption{Interstellar extinction properties of LOPS2. \emph{Left panel:} Average $V$-band extinction $A_V$ of the P1+P2 targets from PIC~2.0.0, calculated over HEALPix level-5 cells and represented in a color scale. \emph{Right panel:} Same but for the P5 sample and over a finer level-6 grid. The high-extinction bump at $A_V\simeq 0.3$ is due to the closest parts of the Vela-Puppis region \citep{CantatGaudin2019}.}
    \label{fig:av}
\end{figure*}

The average interstellar extinction coefficient $A_V$ and its spatial dependence is shown in Fig.~\ref{fig:av} for both the P1+P2 and P5 samples of LOPS2. In the former case we always got $A_V<0.07$; in the latter sample, a clear structure with $A_V\simeq 0.3$ appears, corresponding to some parts of the Vela-Puppis region \citep{CantatGaudin2019}. Clearly, the main reason for the larger extinction of the P5 sample is the much larger average distance of its stars, due to a much fainter magnitude limit ($V<13$ for the P5 sample, $V<11$ for P1). No significant excess of extinction can be seen  on these maps in the LMC region, because even our faintest and intrinsically more luminous targets lie always within 3~kpc from the Sun: the LMC is just a background object for the PLATO planet-hunting survey. See also Section~\ref{sec:main_targets} for some details on the distance distribution.

\section{Astrophysical content of LOPS2}\label{sec:content}

\subsection{Main PLATO targets}\label{sec:main_targets}

Adopting PIC v.~2.0.0 as input catalog, and calculating the NSR through the PINE code \citep{Borner2024}, we can calculate the counts for the four main samples of PLATO (defined according to the formal requirements of the SciRD):
\begin{itemize}
    \item $8\,235$ P1 targets, i.~e., FGK dwarfs and subgiants with $\textrm{NSR}\leq 50$~ppm in one hour and $V\leq 11$;
    \item $699$ P2 targets, i.~e., FGK dwarfs and subgiants with $\textrm{NSR}\leq 50$~ppm in one hour and $V\leq 8.5$;
    \item $12\,415$ P4 targets,  i.~e., M dwarfs with $V\leq 16$;
    \item $167\,149$ P5 targets, i.~e., FGK dwarfs and subgiants with $V \leq 13$.
\end{itemize}
The definitions listed above are also summarized in Table~\ref{tab:samples}.
When the samples are computed as disjoint sets (that is, P2 is excluded from P1, and P1+P2 is excluded from P5), in LOPS2 we have 699 P2 targets, $7\,536$ P1 targets and $158\,914$ P5 targets. The total sample of available targets in LOPS2 is therefore $179\,564$ FGKM stars. We note that, with some exceptions and unless otherwise stated, targets in LOPS2 that are not included in the PLATO samples will not be observed by default and should be proposed in the GO program. The detailed policy will be communicated in advance of the first GO call.
According to the ESA Science Management Plan (available on the ESA website, but also summarized in \citealt{Rauer2024}, Section 10) the first public release of the PIC, which also includes flags for P1-P2-P4-P5, will be released to the community nine months before the launch, concurrently with the call for the GO proposals. According to the current mission schedule, this is expected to happen in spring 2026.

\begin{figure*}
    \centering
    \includegraphics[width=0.95\columnwidth]{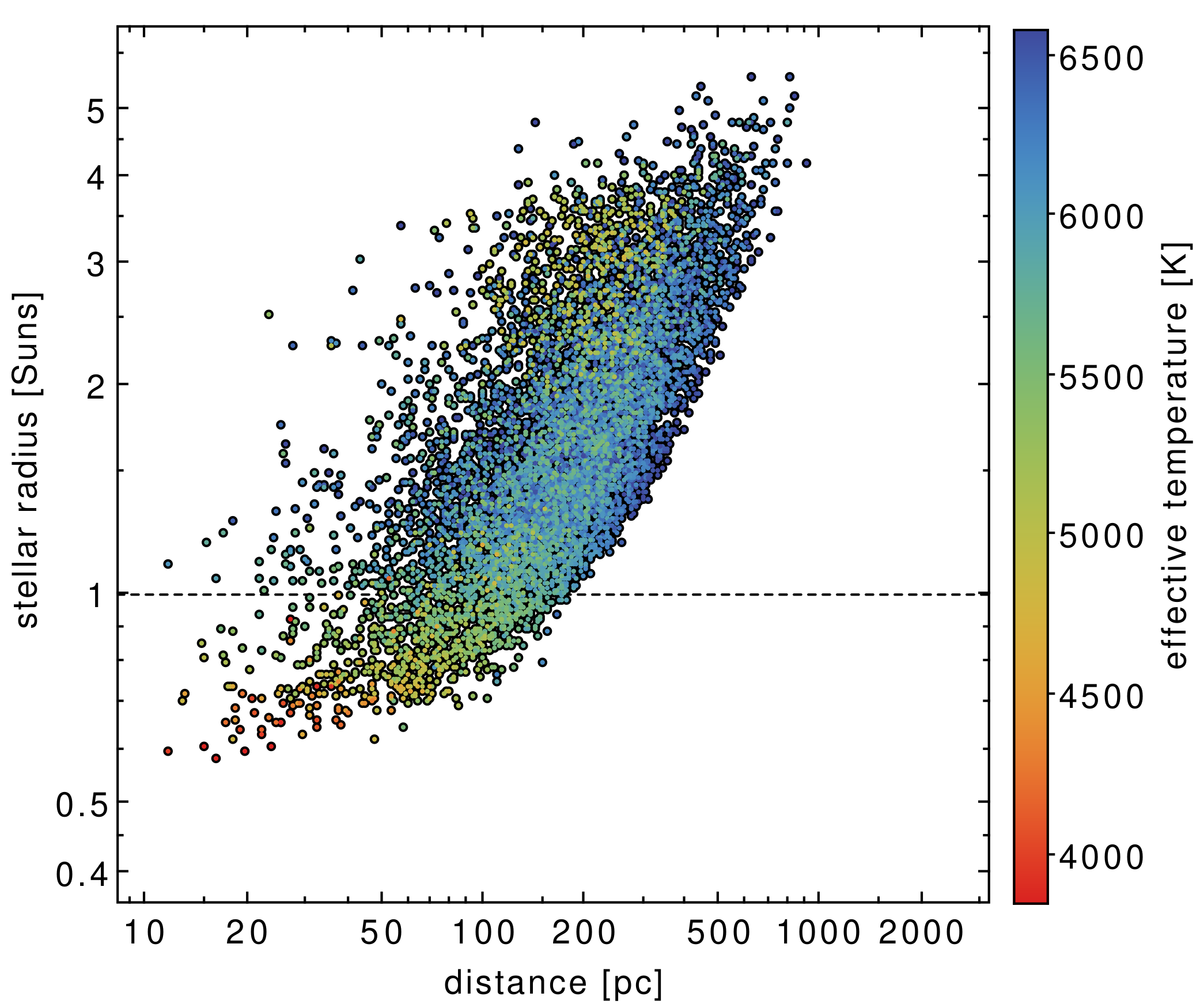}\hspace{5mm}
    \includegraphics[width=0.95\columnwidth]{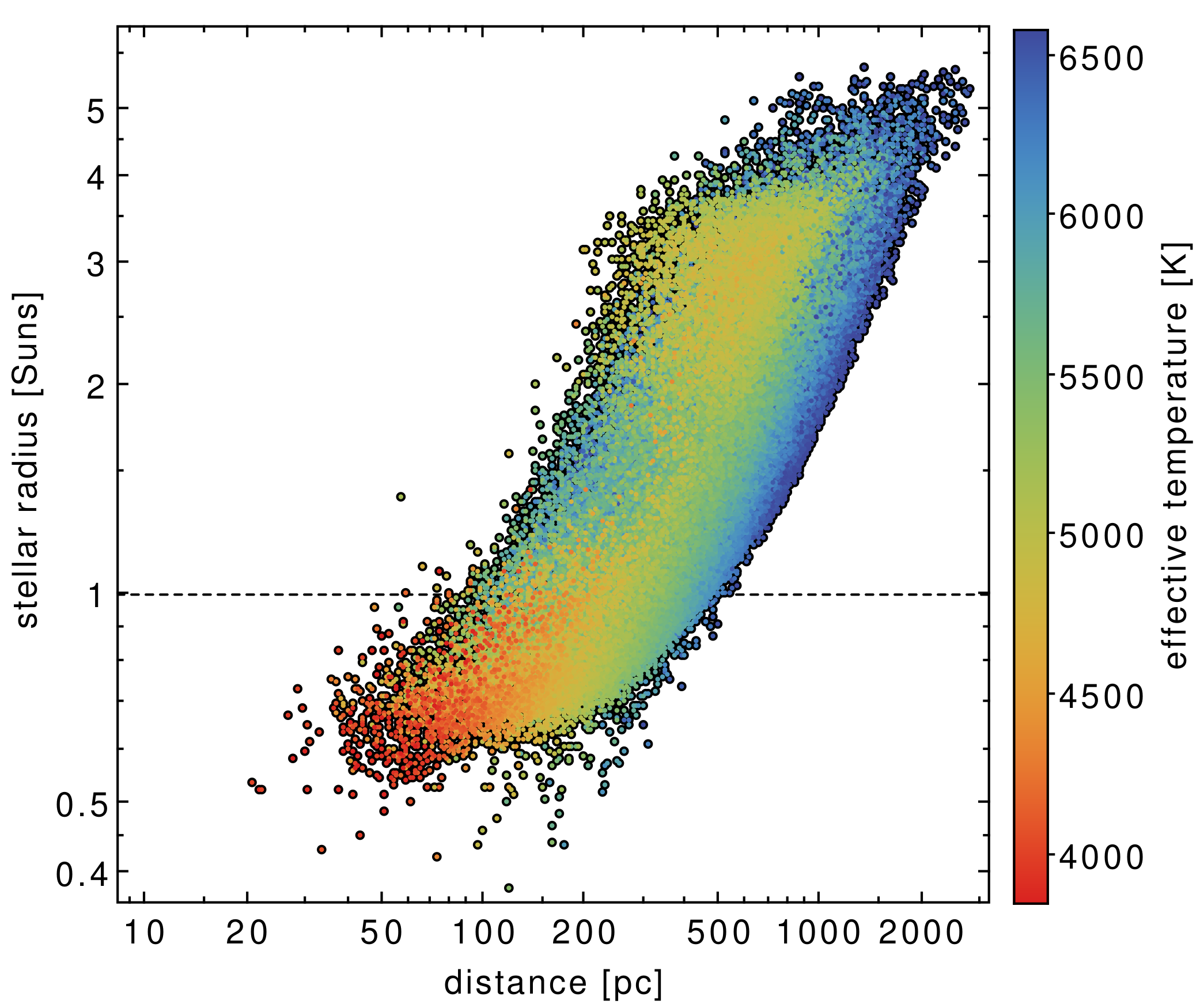}
    \caption{Distance of the P1/P2 and P5 samples in LOPS2. \emph{Left panel:} Distribution of the $8\,235$ P1+P2 stars in LOPS2 in the ($R_\star$, distance) plane. The stellar effective temperature from the PIC v2.0.0 is color coded. \emph{Right panel:} Same but for the P5 sample ($167\,149$ stars; P1/P2 not included). The distribution moves to larger distances due to the fainter magnitude limit of P5 ($V<13$ vs.~$V<11$).}
    \label{fig:distance_prad}
\end{figure*}

The distances of individual solar-type LOPS2 targets (i.e., samples P1+P2 and P5) are plotted in Fig.~\ref{fig:distance_prad}, while the statistical distribution in terms of distance, interstellar extinction $A_V$ and magnitude is shown in the histograms of Fig.~\ref{fig:distance_ext}-\ref{fig:magnitudes}, for both the whole PIC and for a few subsamples of interest. Unsurprisingly for a magnitude-limited sample, and as also shown in Paper I, late-type, main-sequence targets such as G/K dwarfs are on average much closer to us with respect to early-type stars and (sub-)giants, due to their fainter absolute magnitude. As a consequence, the former have also lower interstellar extinction values with respect to the latter (right panel of Fig.~\ref{fig:distance_ext}).

The histograms of some other astrophysical properties of the P1+P2 and P5 samples are plotted in Fig.~\ref{fig:histograms}-\ref{fig:histograms_hz}. The spatial distributions of the P1, P2, P4, P5 samples are plotted in the four panels of Fig.~\ref{fig:densitymaps}: P2 and P4 stars are almost perfectly homogeneously distributed, because they both lie relatively close to the Sun, and are purely magnitude-limited samples.\footnote{Formally, the P2 definition also includes a $\textrm{NSR}\leq50$~ppm requirement but with the current design it is always satisfied even at its $V=8.5$ faint limit.} P5 is also magnitude-limited, but being on average much fainter is also farther from the Sun, and the consequent line-of-sight effects due to the galactic disk are revealed by the significant density gradient along galactic latitude. Finally, P1 is concentrated toward the inner region of LOPS2, observed with 18/24 NCAMs, where the noise requirement is easier to be met at equal magnitude.

\subsection{Bright stars}\label{sec:bright}

\begin{figure*}
    \centering
    \includegraphics[width=0.9\columnwidth]{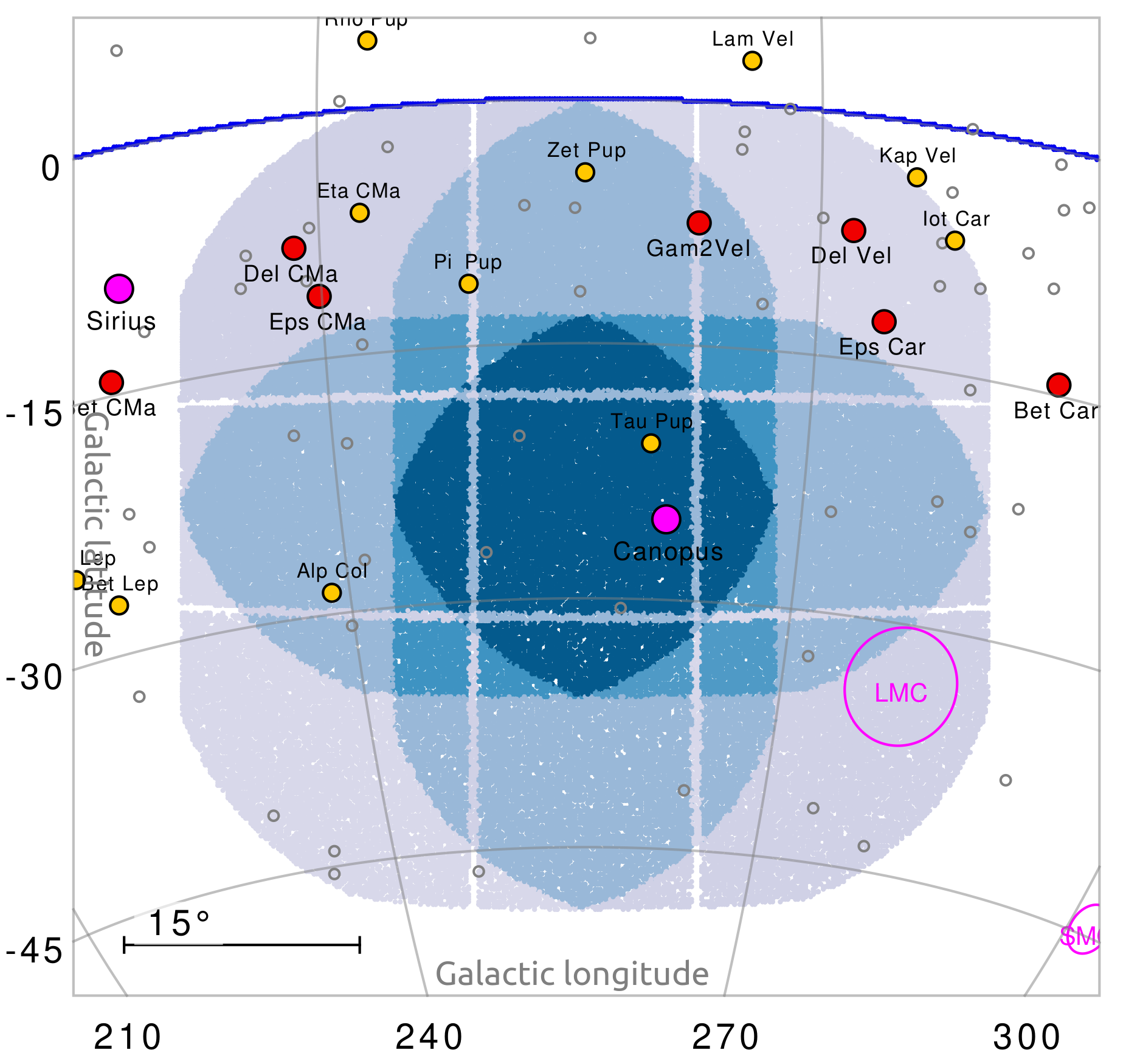}\hspace{5mm}
    \includegraphics[width=0.9\columnwidth]{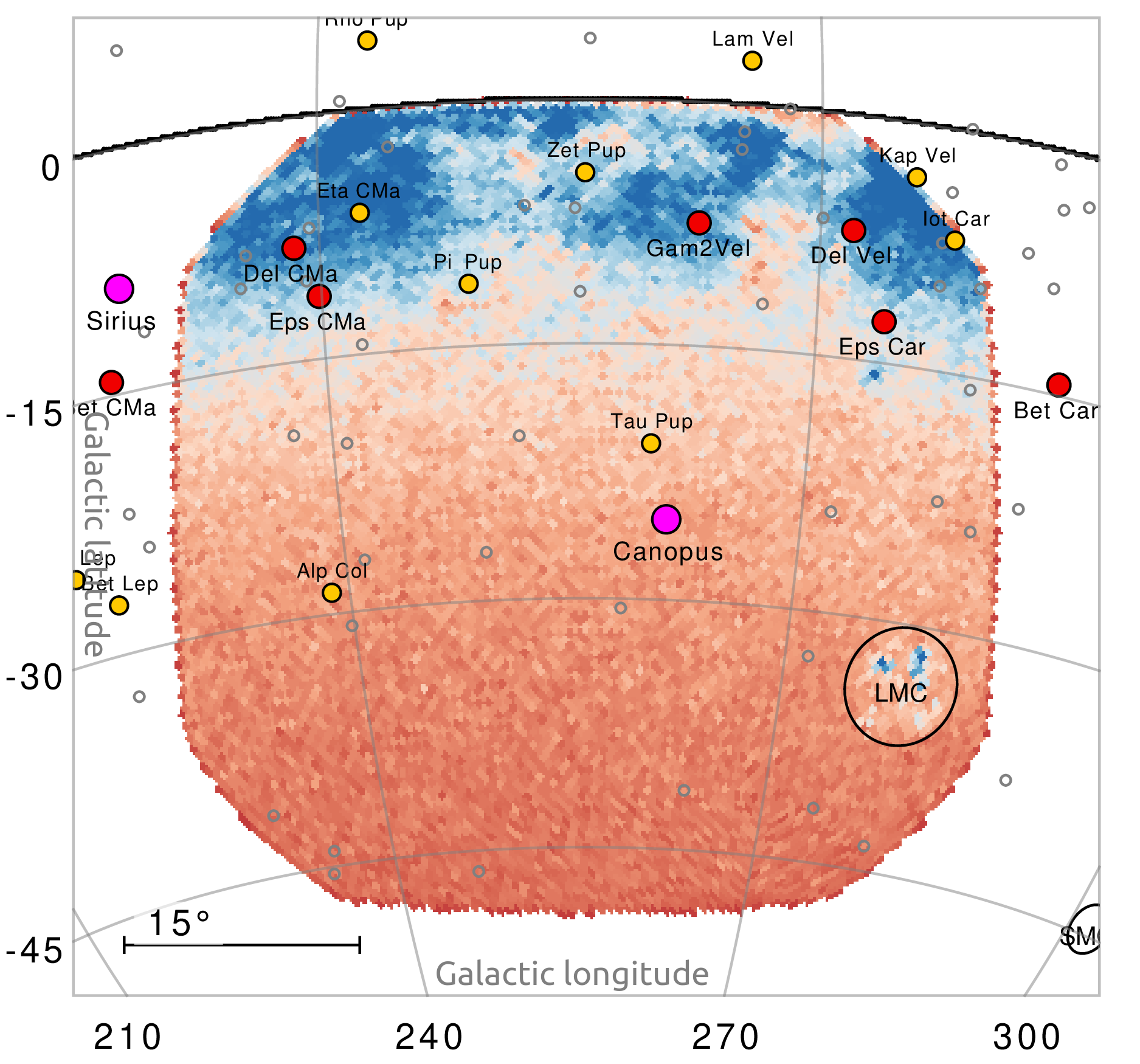}
    \caption{Bright stars and stellar crowding in LOPS2. \emph{Left panel:} Orthographic projection in galactic coordinates centered on LOPS2. The field is color coded according to the number of NCAMs as in Fig.~\ref{fov}. The brightest stars from the Yale Bright Star Catalog are plotted as circles of different colors: magenta ($V<1$), red ($1<V<2$), yellow ($2<V<3$), gray ($3<V<4$). \emph{Right panel:} Same but the footprint of LOPS2 is color coded according to the areal density of $G<13.5$ stars from \emph{Gaia} DR3. The color scale increases proportionally to density from orange to blue with a transition at $\sim 400$ targets per $\textrm{deg}^2$, at $b\simeq -15^\circ$.}
    \label{fig:bright}
\end{figure*}

As part of the field selection process, and as anticipated in Section~\ref{sec:rotation}, we investigated the presence of very bright stars on LOPS2. We used the Yale Bright Star Catalogue version 5 revised (YBSC; \citealt{Hoffleit1995}) as input, being a reasonably complete and accurate census of all the stars brighter than $V\simeq 6.5$. 
The strategy of PLATO for moderately saturated stars imaged on the NCAMs (i.e., between magnitudes 4 and 8 in the PLATO Vegamag system, and approximately\footnote{The exact saturation limit is largely determined by factors other than the color term $P-V$, including the position on the FOV, the PSF shape, the actual in-orbit CCD gain and optical transmission, etc.} also in the  $4\lesssim V\lesssim 8$ range) is to use larger imagettes (saturated star windows) that extend up to $256\times 6$ pixels in size to extract the photometry, if the additional telemetry cost needed to download more data is justified. For stars brighter than magnitude 4, a different strategy is being discussed. For example, smearing photometry is technically possible \citep{White2017,Pope2019}, but alternatives are being considered.

In any case, knowing the position of such bright sources could be of interest since their bright halos, bleeding columns and smearing trails can potentially contaminate any other fainter target in their neighborhood. A discussion on the technical issues related to CCD saturation, blooming, non-linearity as well as optical ghosting is discussed in \citet{Verhoeve2016,Pertenais2022,Jannsen2024} and will be better explored in the forthcoming papers of this series.

Among the 712 YBSC stars lying in the LOPS2, only 47 are brighter than $V=4$ (Fig.~\ref{fig:bright}, left plot). With the only exception of Canopus = $\alpha$~Car (also by far the brightest one; $V=-0.72$), Phact = $\alpha$~Col, and $\tau$~Pup, all the 12 extremely bright stars at $V<3$ are located close to the galactic plane at $|b|<15^\circ$, where also the average stellar density increases very rapidly as the line of sight crosses the galactic disk (Fig.~\ref{fig:bright}, right plot). Not surprisingly, all of them are early-type and/or evolved giants, so none would fit the spectral type requirement to be included in the PLATO P2 sample. 

Scrolling down the list, $\zeta$ Dor A = HD~33262 at $V=4.68$ is the brightest star in LOPS2 to be included in the PIC: it is the F7V component of a very wide visual binary. It is also known as a young and active star surrounded by a debris disk \citep{DobsonRobinson2011}. B~Car = HD~68456 = HR~3220 comes as a close second at $V\simeq 4.73$; it is a single-lined spectroscopic binary whose F7V primary was also identified as a field blue straggler by \citet{Fuhrmann2011}, transferring mass to its low-mass white dwarf companion on a $P=899$~d orbit. 212~Pup = HD~64379, at $V=5.05$, is the brightest binary system made of main-sequence stars (F7V+K5V) in our sample, with a narrow separation ($3"$). Finally, the brightest single solar twin in the P2 sample is HD~59967 ($V=6.66$), a young G2V star \citep{LorenzoOliveira2018} with a debris disk \citep{Pearce2022}, showing evident high levels of stellar activity and with no detected planetary companions from FEROS \citep{Zakhozhay2022} and HARPS \citep{Grandjean2023}.

\subsection{Nearby stars}

Among the 562 known stars within 10~pc from the Sun (\citealt{Reyle2021}, 2023-02-06 version), 14 are located within LOPS2. Only seven of them are bright enough ($V\lesssim 16$) to be reliably monitored by PLATO.\footnote{The excluded targets are all at $G\gtrsim 20$, that is, way too faint for PLATO to deliver reliable photometry.} Six of them are M dwarfs or sub-dwarfs, spanning spectral types from sdM1V to M4.5V; sorted by increasing $G$ magnitude and spectral type: GJ~191, CD-44~3045 B and A, AP~Columbae, L~230-188, SCR~J0740-4257. The seventh one is LAWD~26 = GJ~293, a bright hydrogen-rich white dwarf (WD; DA; $G\sim 13.7$, $V\sim 14.1$), spectroscopically confirmed by \citet{Bell1964} and recently identified by \citet{Sanderson2022} as one of the most promising white dwarf targets for the discovery of astrometric planets with \emph{Gaia}.

Other nearby and reasonably bright WDs can be identified from the catalog by \citet{GentileFusillo2021}: 86 targets with $G<15$ lie within the LOPS2 footprint.\footnote{We emphasize that WDs will not be included in the main PIC subsamples (P1-P2-P4-P5) due to the spectral class requirement (Section~\ref{sec:main_targets}), unless they are flagged as planetary hosts or candidate hosts (See Section~\ref{sec:known}). Within the technical limits, though, they can be included as ``scientific calibration and validation'' targets (scvPIC; \citealt{Zwinze2024}) or as GO targets.} Among these, and besides the already mentioned LAWD~26, four additional stars show a parallax $\pi> 50$~mas, that is, they are closer than 20~pc: two of them were also spectroscopically confirmed by \citet{Obrien2023}, both as DA spectral types.

It is worth mentioning that Kapteyn's star = GJ~191 ($V\sim 8.8$, $K\sim 5.0$) is an old M1 sub-dwarf at only 3.9~pc from the Sun, and also the nearest known halo star \citep{Kotoneva2005}. It will be observed by PLATO through 12 cameras, and with an average NSR estimated around 18.4~ppm in 1~h according to the current noise budget calculations. Two planetary candidates were claimed to be discovered through RVs around GJ~191 by \citet{Anglada2014}, but were later disproved as due to stellar activity by \citet{Robertson2015} and \citet{Bortle2021}, who also noted that the star is photometrically very stable. As is discussed in Section~\ref{sec:known}, a coordinated RV follow-up of this target could take advantage of the simultaneous PLATO coverage to disentangle the contributions of stellar activity and the Keplerian signal from planets, if any.

CD-44 3045 A and B, also known as GJ 257, represent an equal-mass visual binary (M3V + M3V) with an angular separation of $2.4"$ at 2016.0 according to \emph{Gaia} DR3. Hence, they will fall on the same $\sim 15"$ pixel of PLATO, and observed as a single object in combined light within the P4 sample.

\subsection{Known planetary systems}\label{sec:known}

It is of primary interest to identify all the already known planets and planetary candidates within LOPS2, since 1) it makes the vetting and follow-up process much more efficient, 2) there are many science cases for which a long-term and/or extreme-precision follow-up of already known planets can yield compelling scientific results (see for instance \citealt{JontofHutter2021a}), and 3) the unprecedented photometric accuracy of PLATO may allow us to discover additional transiting planets around planet-hosting stars and/or confirm low-SNR candidates and mono-transits \citep{Magliano2024}. 

For the reasons mentioned above, it has been decided that the final PIC will include all confirmed and candidate exoplanets known at the compilation date, so all the planetary systems mentioned in the following sections, if technically feasible with PLATO, will be forced into the catalog regardless whether they meet the P1-P2-P4-P5 SciRD requirement or not. We mention that a preliminary review of the known planetary systems located in LOPS2 was also presented by \citet{Eschen2024}, although limited to transiting planets only and with a particular focus on the PLATO vs.~TESS synergy. We note that their overall numbers are roughly consistent with ours, but a direct comparison is not possible because a combination of different catalogs and/or unreported catalog versions were employed in their match.

\subsubsection{Transiting planets}
\label{sec:transiting_planets}

We retrieved the list of known planets from Exo-MerCat (\citealt{Alei2020}, Alei et al.~in prep.), a meta-catalog built by comparing and merging with an automated process scanning three input catalogs: Exoplanet Encyclopedia,\footnote{\url{https://exoplanet.eu/home/}} the NASA Exoplanet Archive\footnote{\url{https://exoplanetarchive.ipac.caltech.edu/index.html}} and the Open Exoplanet Catalog.\footnote{\url{https://www.openexoplanetcatalogue.com/}} In this Section we focus on the transiting sub-sample only, which is the most important one given the scientific opportunity they represent for PLATO.

\begin{figure*}
    \centering
    \includegraphics[width=0.9\columnwidth]{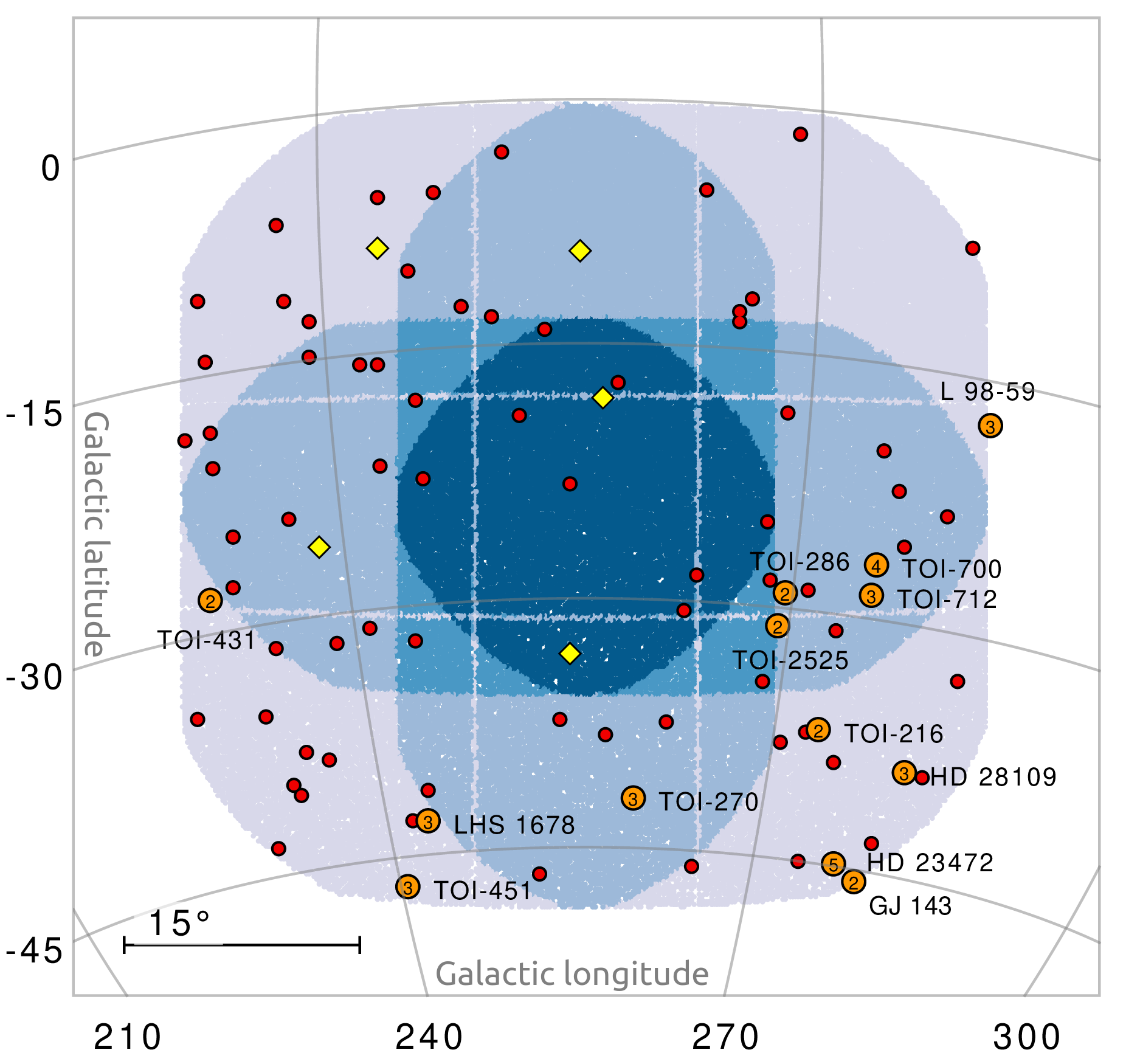}\hspace{5mm}
    \includegraphics[width=0.9\columnwidth]{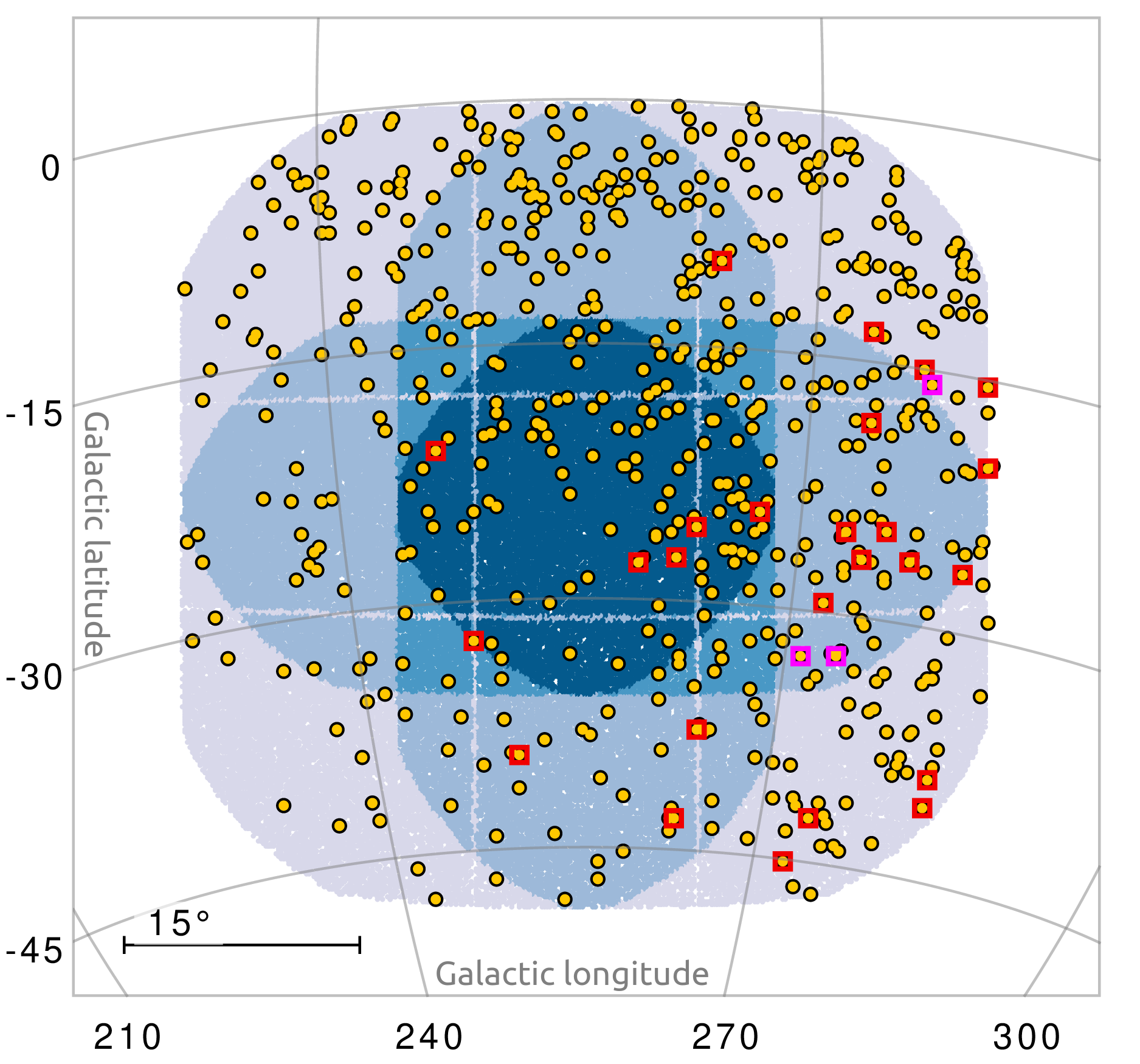}
    \caption{Known planetary systems with at least one transiting planet in LOPS2 (orthographic projection in galactic coordinates). \emph{Left panel:} Entries from the Exo-MerCat database (Section~\ref{sec:known}): single transiting planets (red circles), systems with multiple transiting planets (orange circles with labels; the number inside the circles is the multiplicity), transiting brown dwarfs (yellow diamonds). \emph{Right panel:} Same projection but TESS candidates from the TOI database are plotted with yellow circles. Systems with multiple transiting candidates are marked with red squares (doublets) or magenta squares (triplets).}
    \label{fig:lops2_trans}
\end{figure*}

Overall, 108 confirmed transiting planets are located in LOPS2 (Fig.~\ref{fig:lops2_trans}, left panel and Table~\ref{table:transiting_planets}); they belong to 84 distinct planetary systems. Among the 13 systems with multiple transiting planets there are five doublets (GJ~143, TOI-216, TOI-286, TOI-431, TOI-2525), six triplets (TOI-270, TOI-451, TOI-712, LHS~1678, L98-59\footnote{L98-59 and GJ~143 lie very close to a LOPS2 external border (Fig.~\ref{fig:lops2_trans}, left plot), therefore their inclusion in the target list (and the delivered photometric precision) will depend on the actual pointing accuracy and system performance.} and HD~28109), and two high-multiplicity systems that look as perfect showcases for the photometric performances of PLATO:
\begin{itemize}
    \item TOI-700, a four-planet system hosted by a relatively bright ($V\simeq 13$) M2V star, including two super-Earths in or close to the habitable zone \citep{Gilbert2020,Gilbert2023};
    \item HD~23472 = TOI-174, a packed five-planet system \citep{Trifonov2019,Teske2021} orbiting around a $V\simeq 9.7$ K dwarf, made of three confirmed super-Earths (with at least a suspected 5:3 mean-motion resonance) and two candidate high-density ``super-Mercuries'', among the least massive planets to ever have been detected through RVs \citep{Barros2022}.
\end{itemize} 
Among the mentioned  multiple transiting systems we also identify three systems hosted by relatively young stars ($<10^9$~years): TOI-451, belonging to the Pisces Eridanus stream (120~Myr; \citealt{Newton2021}), and the field stars TOI-201 (870~Myr; \citealt{Hobson2021}) and TOI-712 (830~Myr; \citealt{Vach2022}).

There are also five additional entries in Exo-MerCat whose estimated masses fall within the brown dwarf (BD) regime (HIP~33609b, TOI-569b, TOI-811b, TOI-1496b) or across the planet/BD transition (HATS-70b; \citealt{Zhou2019}). They are plotted as yellow diamonds in Fig.~\ref{fig:lops2_trans}.

\begin{figure*}
    \centering
    \includegraphics[width=1.0\columnwidth]{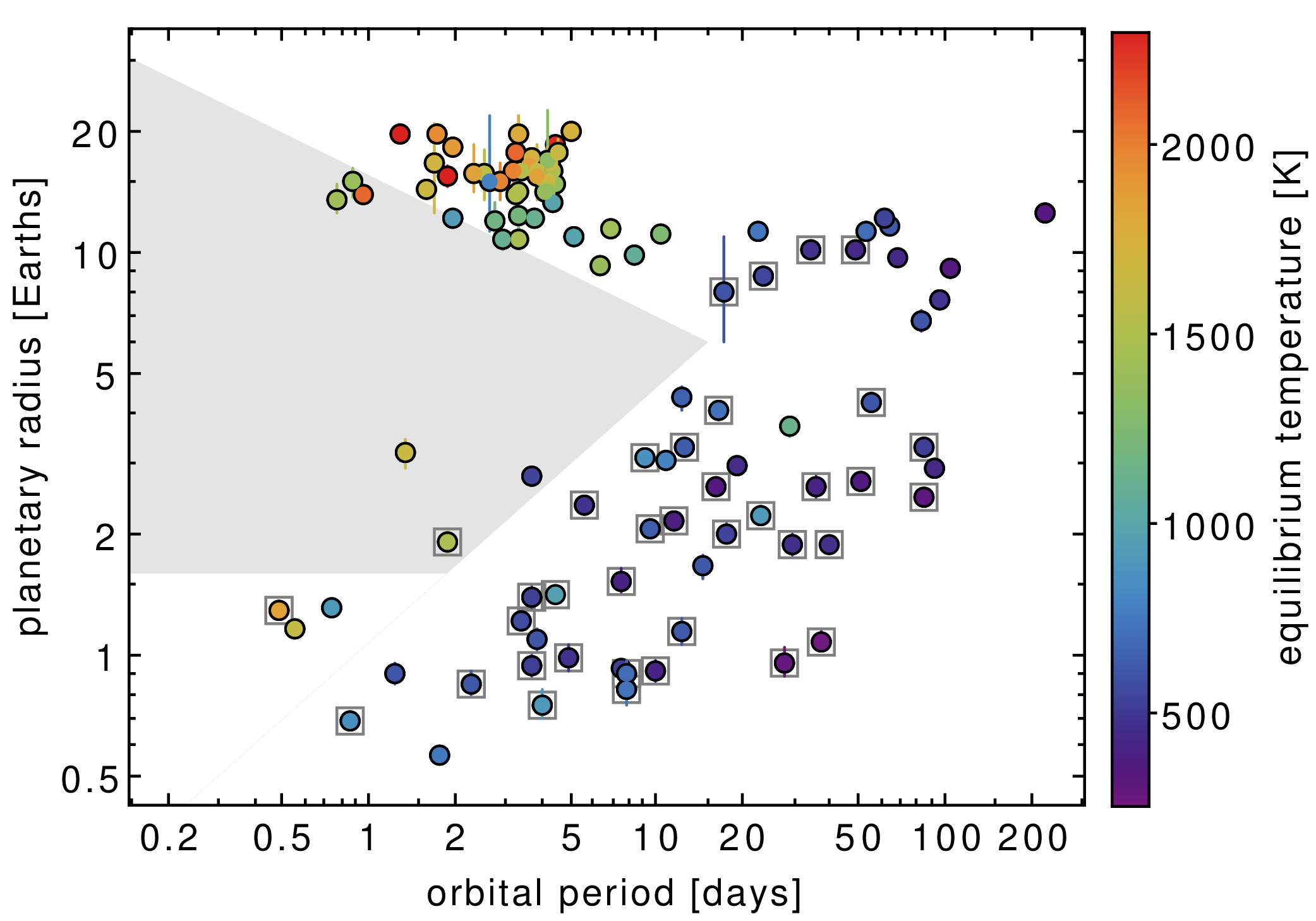}\hspace{3mm}
    \includegraphics[width=1.0\columnwidth]{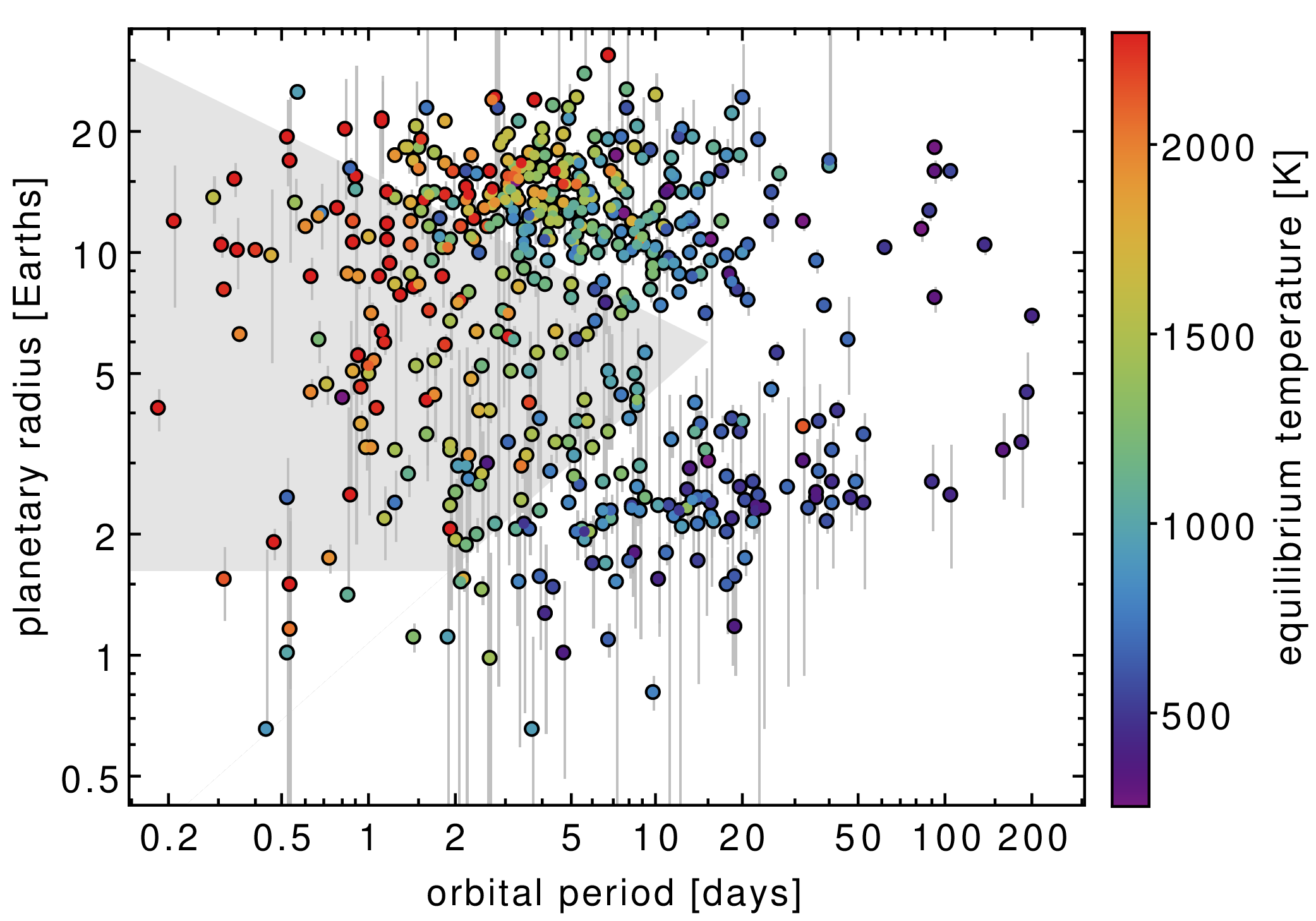}
    \caption{Known planetary systems with at least one transiting planet in LOPS2. \emph{Left panel:} Period-radius diagram with the 108 transiting planets identified from Exo-MerCat and plotted in the left panel of Fig.~\ref{fig:lops2_trans}. The ``Neptunian desert'' region as identified by \citet{Mazeh2016} and with the modified lower boundary by \citet{Deeg2023} is plotted in light gray. Planets belonging to multiple systems are marked with a gray square. \emph{Right panel:} Period-radius diagram with the 612 candidate transiting planets from the TESS TOI database and yet unpublished, plotted in the right panel of Fig.~\ref{fig:lops2_trans}}
    \label{fig:lops2_trans_prad}
\end{figure*}

A closer look to the planetary parameters of the 108 transiting planets (Fig.~\ref{fig:lops2_trans_prad}, left panel) reveals a vast diversity of planetary properties. Adopting some common definitions from the literature, we can split our sample into four main classes:
\begin{itemize}
    \item 47 hot Jupiters (HJ; $R_\mathrm{p}>6$~$R_\oplus$, $P<10$~d). This sub-sample is unlikely to increase significantly in the future, given the almost perfect sensitivity of TESS and the existing ground-based surveys in this region of the parameter space. Notably, there are three ultra-hot Jupiters (UHJs; $T_\mathrm{eq}>2200$~K): WASP-121b \citep{Delrez2016}, KELT-25b \citep{Rodriguez2020}, and the already mentioned HATS-70b. WASP-121b is of particular interest since it spawned an impressively vast literature on its atmospheric characterization, including the study of its optical phase curve \citep{Bourrier2020,Daylan2021} and claims of temporal variability as well \citep{Wilson2021};
    \item 14 warm or cool Jupiters (WJ/CJ); $R_\mathrm{p}>6$~$R_\oplus$, $P>10$~d), eight of them on orbital periods longer than 50~d. The longest period planet of our whole LOPS2 sample belongs to this sample: TOI-4562b, a temperate Jupiter analog on a $P\simeq 225$~d orbit \citep{Heitzmann2023} recently claimed to show TTVs by \citet{Fermiano2024};
    \item 24 Neptunians and mini-Neptunes ($1.7$~$R_\oplus<R_\mathrm{p}<6$~$ R_\oplus$), including at least one planet undoubtedly within the ``hot Neptune'' desert (\citealt{Mazeh2016}; gray area in Fig.~\ref{fig:lops2_trans_prad}): NGTS-4b \citep{West2019} plus two other more marginally so (TOI-451b, TOI-269b);
    \item 23 rocky planets ($R_\mathrm{p}<1.7 R_\oplus$), including four ultra short-period ones (USP; $P\lesssim 1$~d): TOI-206b ($P=0.74$~d; \citealt{Giacalone2022}), LHS~1678b ($P=0.86$~d; \citealt{Silverstein2024}), TOI-431b ($P=0.49$~d; \citealt{Osborn2021}) and TOI-500b ($P=0.55$~d; \citealt{Serrano2022}). The latter will be observed close to the center of LOPS2 through 24 NCAMs, so PLATO will be able to gather at least $\approx 1300$ full phase curves at $\textrm{NSR}\simeq33$~ppm in one hour.
\end{itemize}
We calculated for each entry of the Exo-MerCat sample the so called \emph{ephemeris drift} $\sigma (T_0)$ at epoch 2027.0, that is, the expected 1-$\sigma$ error on the transit prediction (according to the most accurate ephemeris available) at the approximated epoch when PLATO is supposed to start its scientific operations. The median is $\sigma (T_0)\sim 7$~min.  Only ten planets show a drift larger than two hours; not surprisingly, they are mostly small and/or long-period planets observed by TESS on a limited number of sectors and not accessible by ground-based facilities. At $\sigma (T_0)\simeq 1.8$~d, an extreme case is TOI-1338b = BEBOP-1b (the first circumbinary planet discovered by TESS, one of the only two known multiple planetary system of this kind; \citealt{Kostov2020,Standing2023}), virtually ``lost'' for the time being. For each of all these loose-ephemeris targets, the first two or three months of PLATO photometry will deliver a new, extremely accurate ephemeris, crucial for the community to setup any further follow-up observations.

\subsubsection{Transiting candidate planets}
\label{sec:tois}

Currently, TESS \citep{Ricker2015} is by far the largest provider of candidate planets within LOPS2. It makes sense to quantify how many. In any case, they will be included into the PIC, not only because some fraction of them\footnote{Quantifying the false positive ratio (FPR) of the TOI database is non-trivial because the vetting process is partially done in a manual fashion \citep{Kunimoto2023}. We note, however, that more than 6\% of the TOI entries have already been flagged as FP so far, and likely much more will follow before the PLATO launch.} will eventually turn out to be genuine planets, but also because PLATO itself (both through its light curves and its follow-up program) can play a fundamental role in their validation and confirmation  \citep{Mantovan2022}.

From the 2024-08-20 release of the TESS Object of Interest database (TOI; \citealt{Guerrero2021}) we identified 824 entries in LOPS2 and matched them with the TESS Follow-up program (TFOP) data base, which includes a disposition flag (``TFOPWG'') reflecting the current status of each target according to the follow-up results.  After selecting only successfully vetted planetary candidates (disposition keyword: PC, APC or CP, meaning ``planetary candidate'', ``ambiguous planetary candidate'' and ``confirmed planet'', respectively) not already published we are left with 544 candidates hosted by 513 stars\footnote{The full list is available on \href{https://zenodo.org/records/14720127}{zenodo14720127}.} (Fig.~\ref{fig:lops2_trans}, right panel). We emphasize that these numbers include candidates with a very wide range of FPR, and that a quick inspection at the ``comment'' column of the TOI database reveals that a considerable fraction of APC entries could likely turn out as false positives. Nevertheless, an individual vetting of each TOI is beyond the scope of this paper.
Notably, 27 stars among the LOPS2 TOI host multiple candidates, a configuration which is known to significantly decrease the a-priori FPR. Three systems have multiplicity $N=3$: TOI-699, TOI-790, and TOI-2392 currently under active follow-up by the TFOP working group. 

It is worth mentioning that 27 among our list of TESS candidates are listed in the TOI table as having orbital periods longer than 100~d (they are visible at the right end of Fig.~\ref{fig:lops2_trans_prad}, right panel) or with no period at all. A closer inspection at the comment field reveals that almost all of them are so called ``monotransits", that is, candidates detected through a single event whose orbital period is unconstrained. In fact, an even larger number of TOI candidates, especially in the long-period end, are flagged as having an ambiguous period due to an insufficient orbital phase coverage by TESS. The almost uninterrupted 2-year coverage of LOPS2 will easily break the ambiguity and detect the correct orbital period \citep{Magliano2024}.

\subsubsection{Non-transiting planets}

\begin{figure*}
    \centering
    \includegraphics[width=0.9\columnwidth]{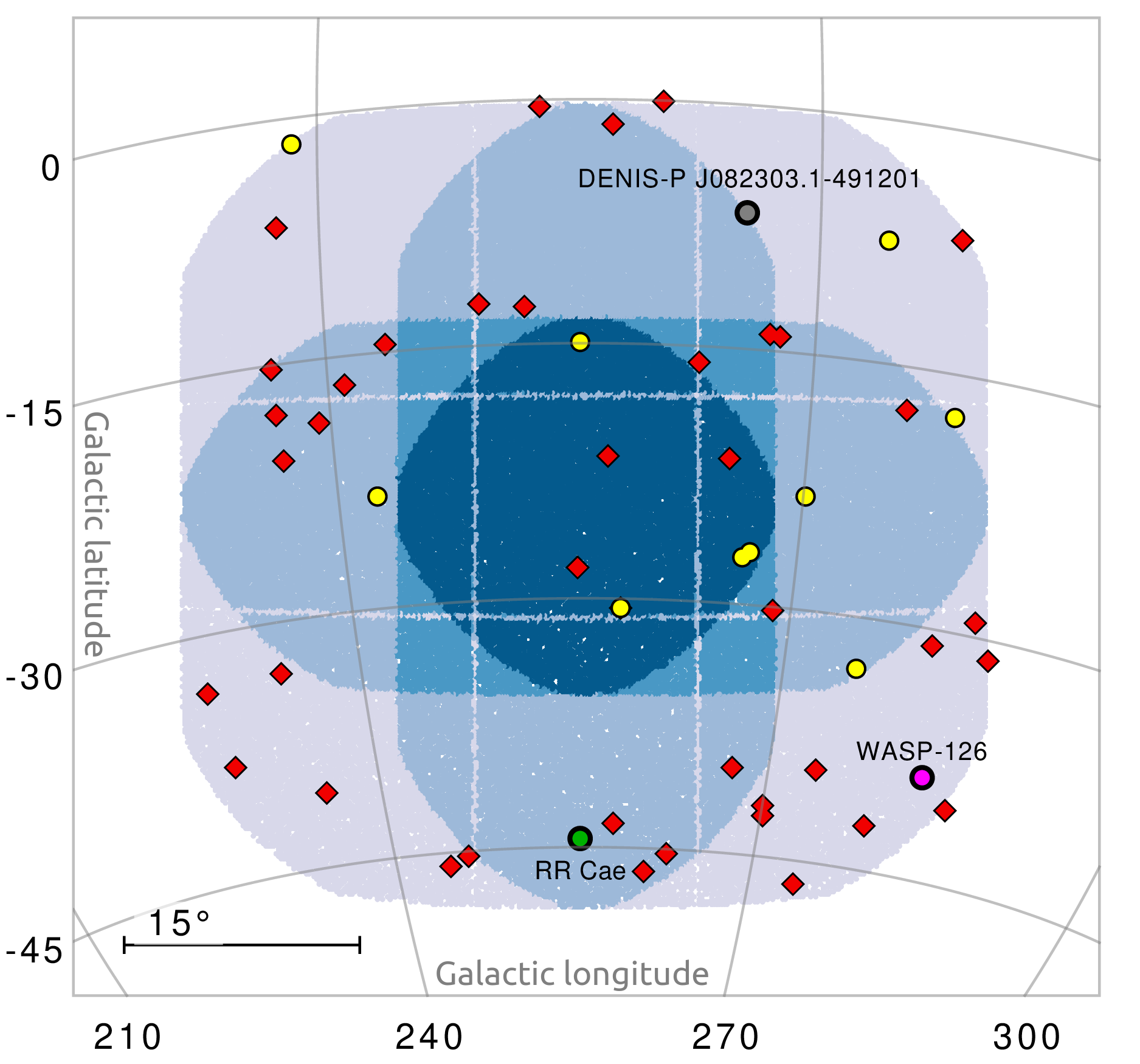}\hspace{5mm}
    \includegraphics[width=0.9\columnwidth]{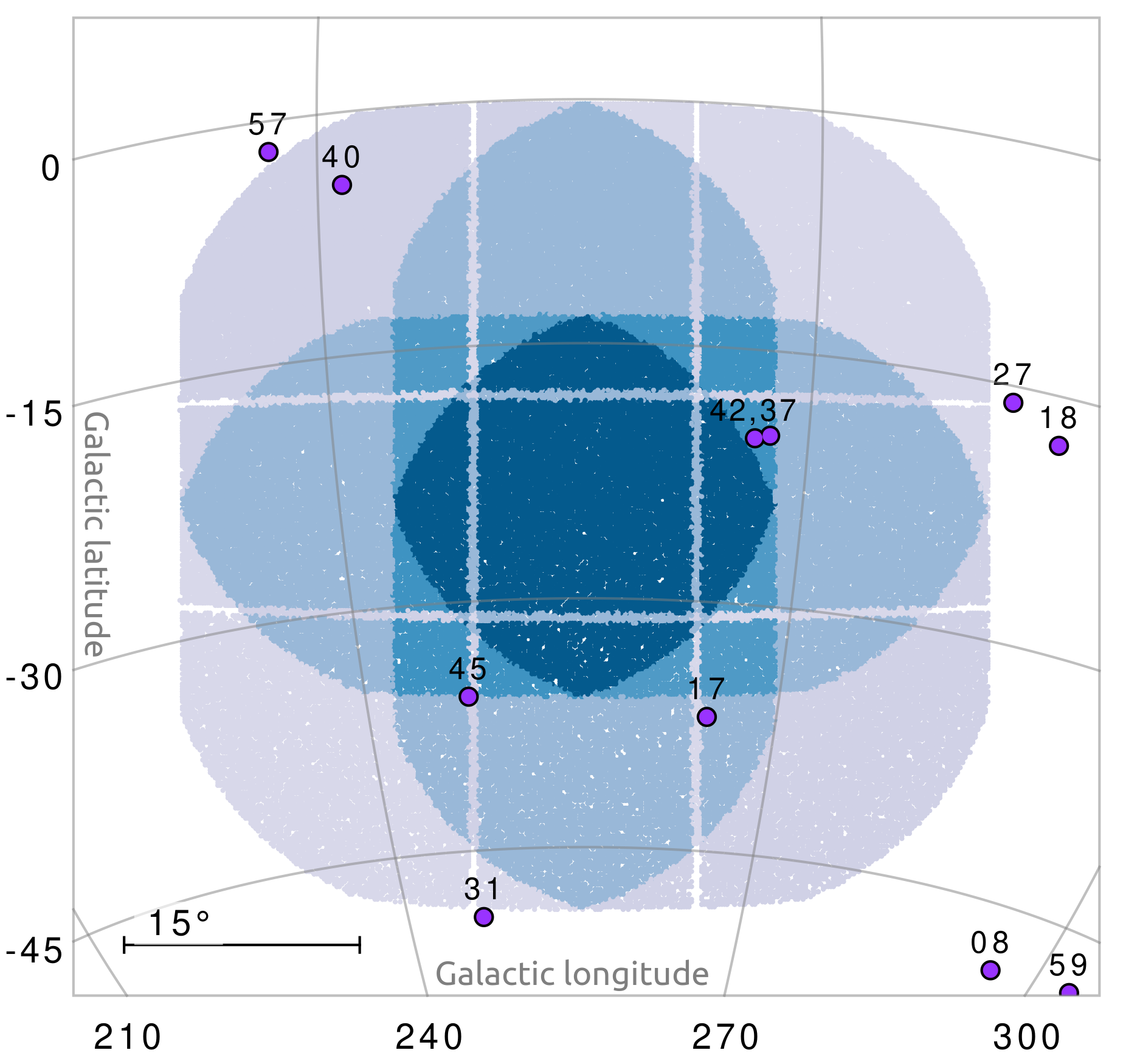}
    \caption{Known non-transiting planetary systems in LOPS2 (orthographic projection in galactic coordinates). \emph{Left panel:} Entries from the Exo-MerCat database (Section~\ref{sec:known}): planets discovered through RVs (red diamonds), direct imaging (yellow circles), astrometry (gray circle), and ETV (green circle). The location of the candidate TTV system WASP-126 is also marked with a magenta point. \emph{Right panel:} Candidate astrometric planets published by the \emph{Gaia} collaboration (purple circles). Each data point is labelled with its ASOI ID number.}
    \label{fig:lops2_nontrans}
\end{figure*}

As already mentioned, also non-transiting planets are worth observing with PLATO with the goal of better characterizing the stellar host or to discover additional transiting companions (or both). By cross-matching Exo-MerCat with the sky footprint of LOPS2, after excluding the transiting planets already discussed in the previous sections, we retain a total of 77 planets in 53 systems discovered through three different instrumental techniques (Fig.~\ref{fig:lops2_nontrans}, left plot):
\begin{itemize}
    \item \emph{Radial velocity:} sixty-four planets grouped in 41 planetary systems. Among the 15 multiple systems, HD~40307 \citep{Mayor2009} stands out as that with the highest multiplicity: at least four planets confirmed by independent analyses, plus two more controversial candidates\footnote{As a side note, the inclusion of HD~40307 and other bright RV systems in LOPS2 brings up the opportunity of planning a $\ge 2$~yr observing campaign where ultra-high-precision RV measurements can be combined with a simultaneous and almost uninterrupted space-based light curve to disentangle the planetary signal from stellar activity effects, following the approach developed by \citet{Lanza2011,Aigrain2012,Haywood2014}, among others.} \citep{Tuomi2013,Diaz2016}. A few RV systems are not counted here because they also host confirmed transiting planets, so they were included in the sample described in Section ~\ref{sec:transiting_planets}: they are TOI-500 and TOI-431 (both hosting a transiting super-Earth), L98-59 (with three transiting planets, \citealt{Kostov2019,Demangeon2021}), and TOI-1338A = BEBOP-1 \citep{Standing2023};
    \item \emph{Direct imaging:} eleven planets hosted by ten systems, among which $\beta$ Pic b \citep{Lagrange2010,Lagrange2019}, AB~Pic \citep{Chauvin2005}, HR~2562 \citep{Konopacky2016}, and WD~0806-661~b \citep{Luhman2011} stand out as the best characterized. All of them are very massive planets, close to or across the planet-BD transition;
    \item \emph{Eclipse Timing Variation} (ETV): RR~Cae ($V=14.4$), a post-common-envelope dM+WD eclipsing binary showing a $P\simeq 16$~years timing modulation claimed as due to the light-travel time effect from an outer $M_\mathrm{p}=3.4\pm 0.2$~$M_\mathrm{jup}$ giant planet \citep{Qian2012,Rattanamala2023}. PLATO will obviously be unable to sample the full phase of its signal, yet its uninterrupted series of $\sim 2400$ eclipses will constitute the most precise photometric data set on this target; 
    \item \emph{Astrometry:} A sub-stellar candidate orbiting a L1.5 ultracool dwarf, DENIS-P~J082303.1-491201 \citep{Sahlmann2013}. At $V\sim 19$ this target is likely too faint to be included in the PLATO target list.
\end{itemize}

It is also worth mentioning WASP-126c, a non-transiting outer companion of the transiting hot Jupiter WASP-126b claimed by \citet{Pearson2019} based on the Transit Timing Variation (TTV) analysis of TESS data, but unconfirmed by subsequent studies \citep{Macie2020}. If real, its $P\simeq 23$~d, 1-min wide timing modulation would be very easy to confirm or disprove by PLATO, even after a single quarter of photometry.

As anticipated in Paper I, the synergy between PLATO and the thousands of astrometric planets that will be discovered by \emph{Gaia} \citep{Perryman2014,Sozzetti2014} is not to be missed. According to the early estimate by \citet{Perryman2014}, 25-50 among those planets, mostly in the short orbital period tail ($P=2$-3~ yr), are expected to transit; even for the most favorable cases, the ephemeris will be accurate enough to pinpoint the transit with an error of a few weeks at best \citep{Sozzetti2023}, implying that ground-based campaigns will be largely ineffective or even unfeasible. This opens an exciting opportunity with PLATO, especially for the LOP phase. On top of this, the discovery of additional transiting companions on shorter orbits would be extremely interesting to investigate the architecture of such largely unexplored planetary systems.

The first official release of confirmed astrometric planets from \emph{Gaia} will be included in DR4, expected not before mid-2026. A preliminary list of candidates has been released in DR3, in the non-single-star (NSS) part \citep{Areanou2023}. Five targets, all M dwarfs, \emph{Gaia}-ASOI-017, -037, -040, -042, -045 will be monitored in LOPS2 (Fig.~\ref{fig:lops2_nontrans}, right panel). Only \emph{Gaia}-ASOI-037 ($V\simeq 13.7$), and -040 ($V\simeq 15.4$) are currently included in the PIC; the remaining three are fainter than the $V=16$ requirement set for the PLATO P4 sample. These will possibly be added on a special separate list if simulations will show that meaningful photometry can be extracted. 

\subsection{Star clusters and associations}

It is well known that star clusters provide us with an ideal laboratory to study the formation and evolutionary processes shaping the architecture of planetary systems \citep{Adibekyan2021}. Not only does it provide the opportunity to compare the property of planets hosted in different dynamical environments, also the stellar parameters (including age and chemical composition) of stars belonging to clusters can be measured or derived with a much better accuracy than for field stars \citep{Vejar2021}. 

\begin{figure*}
    \centering
    \includegraphics[width=0.9\columnwidth]{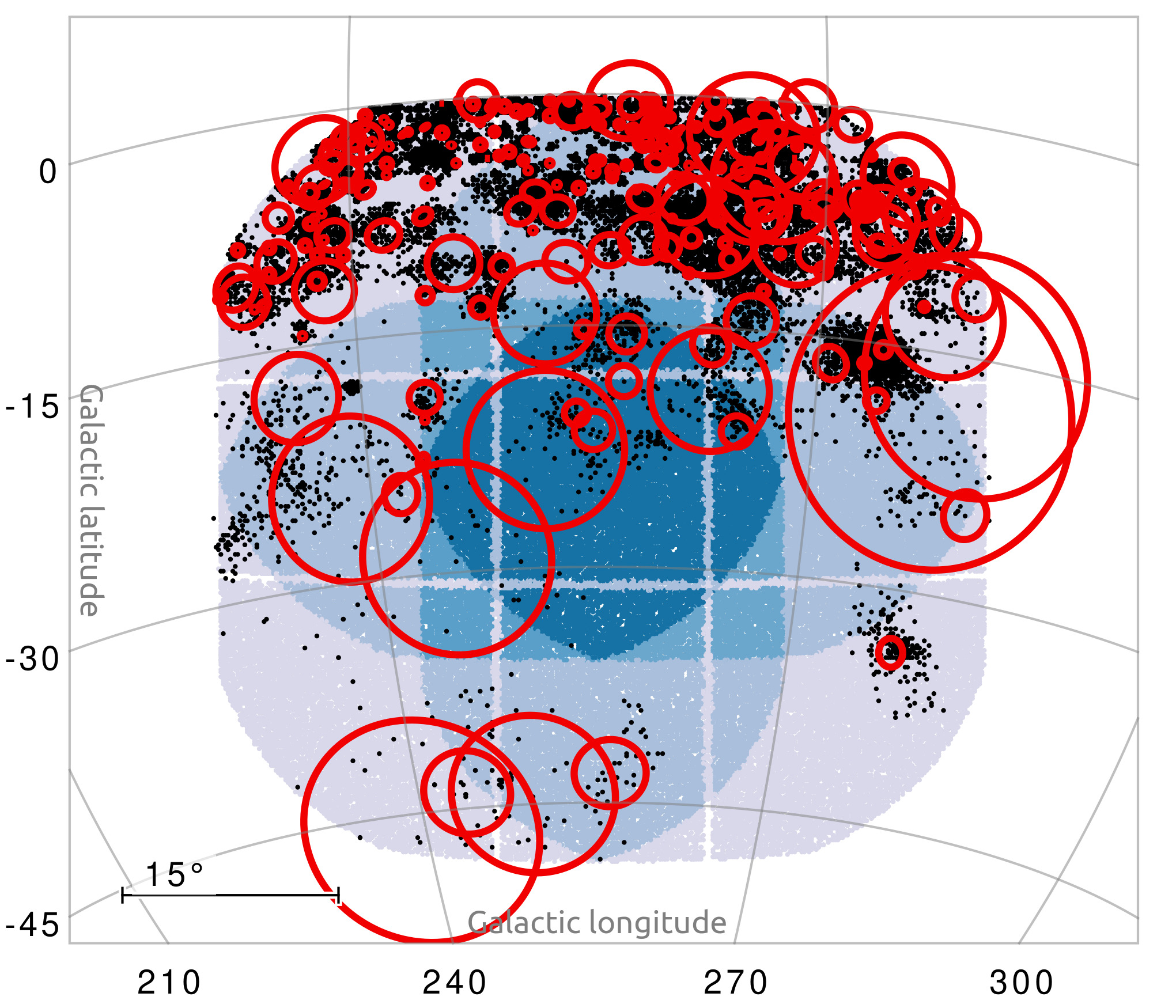}\hspace{5mm}
    \includegraphics[width=0.9\columnwidth]{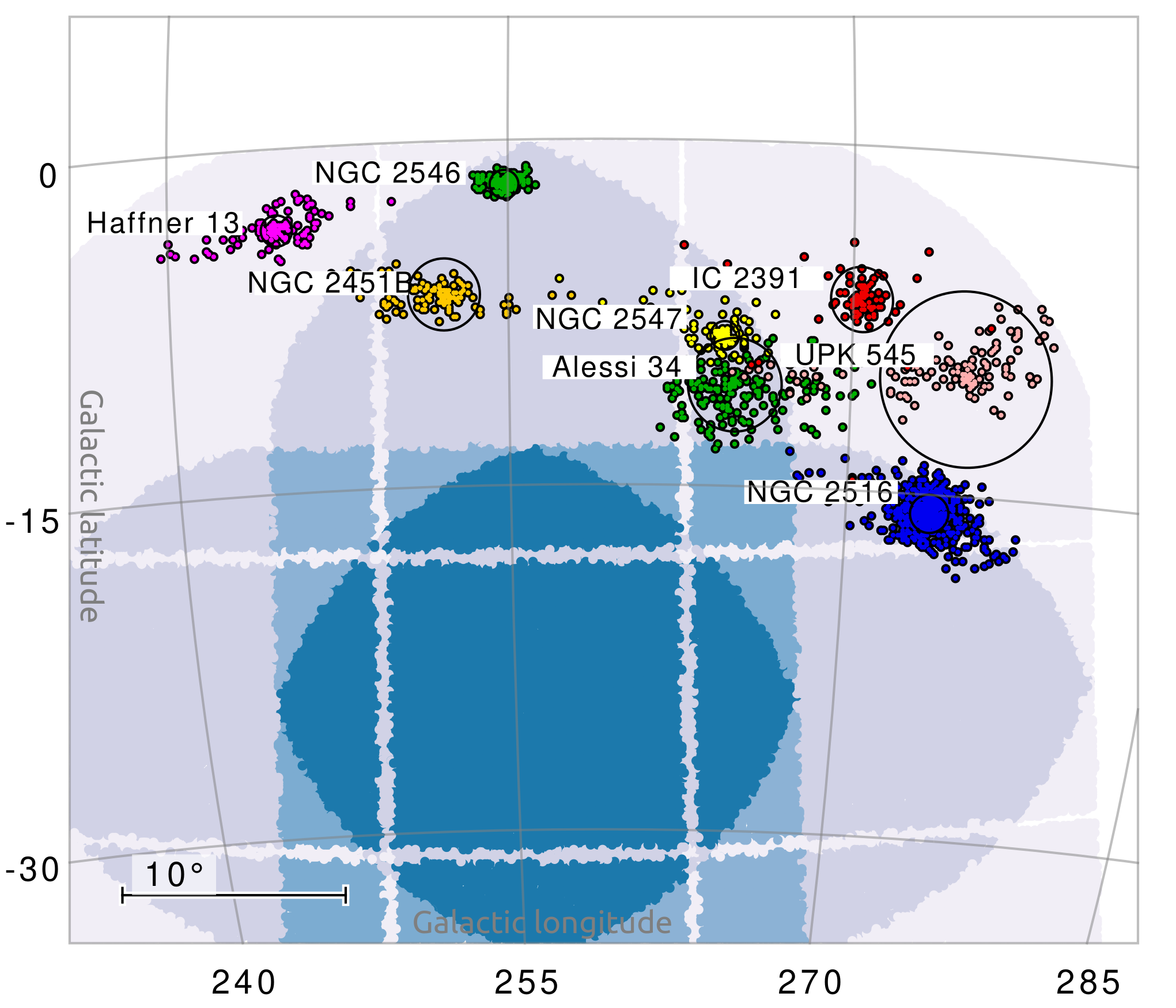}
    \caption{Open clusters and associations in LOPS2. \emph{Left panel:} Sky map of the 367 clusters identified by \citet{Hunt2023} at high confidence and with at least one member in LOPS2, plotted as red circles with radius \texttt{r50} (radius containing 50\% of the members within the tidal radius). All the 10\,682 members at $V<15$ are plotted as black points. \emph{Right panel:} Eight clusters in our sample having the largest number of $V<15$ targets of spectral type F5V. Each target is color coded with the same scheme as in Fig.~\ref{fig:lops2_clusters_cmd}.}
    \label{fig:lops2_clusters}
\end{figure*}

We adopted the catalog by \citet{Hunt2023} as starting point for our search, being the most recent and complete census of galactic star clusters and associations based on \emph{Gaia}~DR3 (Paper I was based on the work by \citealt{Cantat-Gaudin2020} instead). Following their prescription, we restricted our input list to a sub-sample of $4\,105$ ``high-quality'' clusters detected at high confidence from their initial sample of $7\,167$, by imposing $\texttt{CST}>5.0$ and $\texttt{CMDCl50}>0.5$. To identify clusters of possible interest for PLATO, we further selected just those having at least one $V<15$ member (the rough practical limit to detect transits around solar-type stars) in LOPS2. The positions of the resulting set of 367 clusters are plotted as red circles on the sky map of LOPS2 in the left panel of Fig.~\ref{fig:lops2_clusters}. As expected, there is a strong gradient as a function of galactic latitude, and the vast majority of clusters lie at $|b|<15^\circ$ from the plane.

\begin{figure*}
    \centering
    \includegraphics[width=1.0\columnwidth]{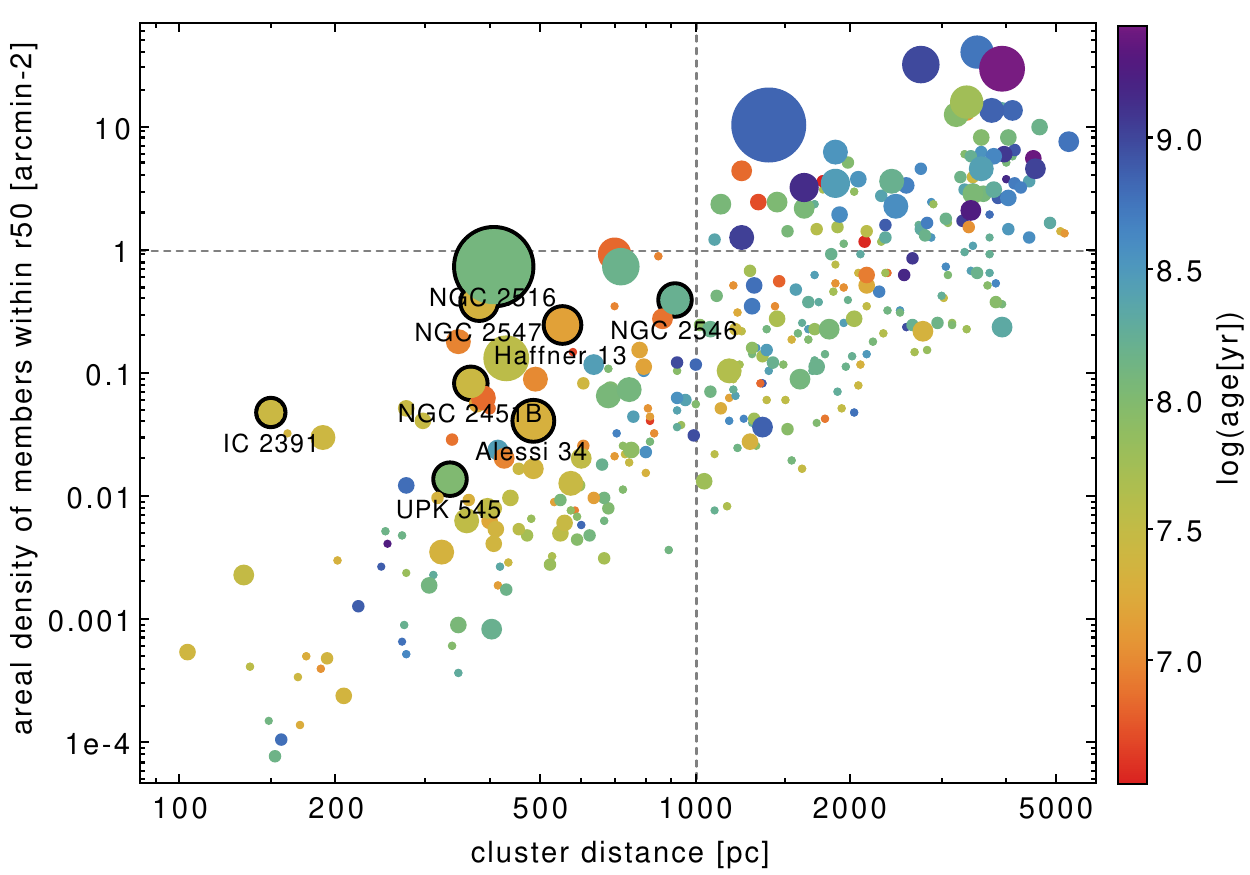}\hspace{3mm}
    \includegraphics[width=1.0\columnwidth]{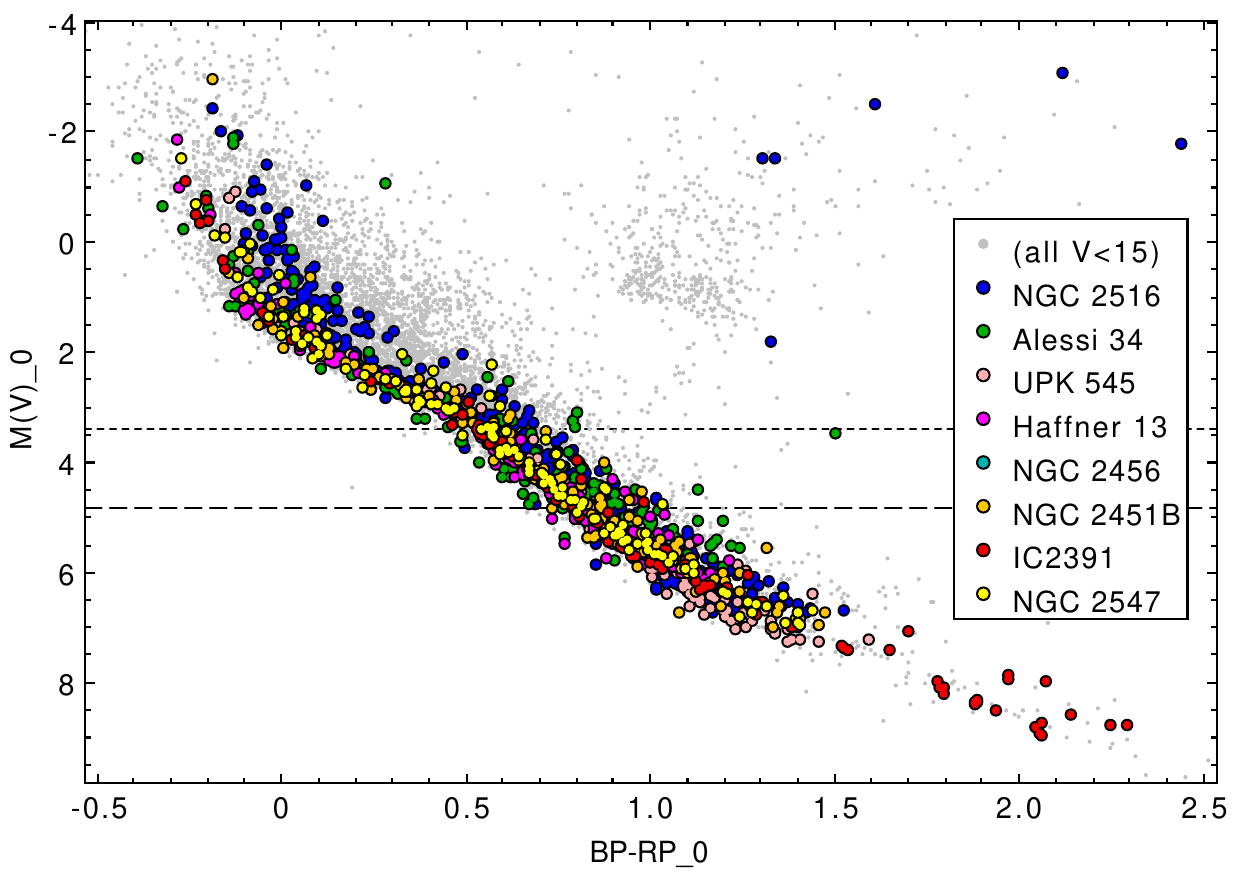}
    \caption{Open clusters and associations in LOPS2. \emph{Left panel:} the 367 clusters identified by \citet{Hunt2023} at high confidence and with at least one member in LOPS2, plotted as a function of their distance and areal density (as defined in the text). The logarithmic age of each cluster is color coded, while the point area is proportional to the number of $V<15$ stars. The eight clusters having the largest number of $V<15$ targets of spectral type F5V and later are labelled. \emph{Right panel:} Absolute and de-reddened color-magnitude diagram (CMD) for all the $10\,682$ $V<15$ stars in LOPS2 belonging to clusters (gray points). Members of the eight labeled clusters are plotted with larger circles and color coded as in the legend.}
    \label{fig:lops2_clusters_cmd}
\end{figure*}

As discussed in \citet{Rauer2024}, the main focus of the PLATO planet-hunting survey is on main-sequence stars later than F5V. If we neglect interstellar extinction and assume for the F5V/G2V spectral types an absolute magnitude $M(V)=3.37$ and 4.85, respectively (from \citealt{Pecaut2013}), we get an upper limit for the distance of 2100 and 1075~pc at a limiting magnitude of $V=15$. Translated at the target level and taking into account the best-fit extinction coefficient $\texttt{AV50}$ and distance $\texttt{dist50}$  from \citet{Hunt2023}, we can set the requirement
\begin{equation}
    V-\texttt{AV50}+5-5\log(\texttt{dist50}) < 3.37 \texttt{ AND } V < 15
\end{equation}
to get a final list of $3\,506$ targets of interest with spectral type later than F5V. Despite the large number of clusters in LOPS2, more than one third of those targets are hosted by just eight relatively nearby open clusters: NGC~2516, Alessi~34, UPK~545, Haffner~13, NGC~2546, NGC~2451B, IC~2391, NGC~2547, sorted by decreasing number of later than F5V, $V<15$ targets (from 543 to 77) and labelled in the left panel of Fig.~\ref{fig:lops2_clusters_cmd}. They are all located within 1000~pc from the Sun. Their average areal density within \texttt{r50} (defined as the radius containing half of the cluster members) is much smaller than 1~$\textrm{arcmin}^{-2}$ even for the densest ones, meaning none of them is critically crowded considering the $15''$ pixel size of PLATO, with 90\% of the flux within $3\times 3$~pixels, and the availability of sophisticated PSF or DIA techniques \citep{Nardiello2020,Montalto2020} when a target has been allocated an imagette.

All the mentioned clusters are much younger than the Sun, having estimated ages in the range of 15-150~Myr. This is confirmed by independent spectroscopic studies in the literature, but also by their de-reddened color-magnitude diagrams, showing (with only one exception, see below) a straight main sequence with no hint of evolved members (Fig.~\ref{fig:lops2_clusters_cmd}, right panel). Two clusters in particular stand out because of their properties and deserve some more discussion:
\begin{itemize}
    \item \emph{IC~2391} is by far the closest object in our sample at a distance of 150~pc. It is a sparse, young ($51\pm5$~Myr from the Li depletion boundary; \citealt{Randich2018}) solar-metallicity cluster ($[\mathrm{Fe}/\mathrm{H}]=-0.04\pm0.03$; \citealt{DeSilva2013}) with about 350 high-confidence members identified so far \citep{Nisak2022}. IC~2391 is also the only cluster in our short list for which late-K stars will be accessible by PLATO. Stellar activity at such young ages could be a limiting factor. For instance, the solar analog \emph{Gaia}~DR3~5318186221414047104 ($V=10.9$, $T_\mathrm{eff}=5720$~K, $R_\star = 0.96$~$R_\odot$, $\log g = 4.46$), which is a bona-fide member of IC~2391, shows (in TESS light curves) an obvious rotational modulation at $P_\mathrm{rot}\simeq 4$~d and with a 3\% semi-amplitude. The same star is also a moderately fast rotator, at $v\sin i = 9$~km/s \citep{DeSilva2013} requiring an extra effort to get RV confirmation in the rocky planet regime.
    \item \emph{NGC~2516}. The richest cluster of our sample ($\sim$2000 known members, of which 543 at $V<15$ later than F5V; $d=407$~pc), and the only one lying for the most part at $|b|>15^\circ$. At an age of $138\pm 40$~Myr \citep{Franciosini2022} it also is, together with NGC~2546, the oldest cluster of our sample. Metallicity is slightly super-solar, $[\mathrm{Fe}/\mathrm{H}]=+0.08\pm0.01$ according to \citet{Baratella2020}. Interestingly, NGC~2516 has also been proven to possess an extended halo of stars spanning up to 500~pc, or $20^\circ$ in the sky \citep{Bouma2021}. As a passing note, we also mention that the hot Jupiter TOI-1937A~b \citep{Yee2023} could possibly belong to this cluster, but the evidence about its membership has been inconclusive so far.
\end{itemize}

Being a particularly complex task, the selection, characterization and prioritization of the cluster stars to be included in the PIC as a special subset will constitute the main topic of a dedicated future paper.

Other stars in associations, especially young stellar objects (YSO) not yet on the main sequence, could be identified from specific catalogs, such as the Konkoly Optical YSO catalog (KYSO; \citealt{Marton2023}), a data base of bona-fide, mostly spectroscopically confirmed YSOs compiled by cross-matching \emph{Gaia}~DR3 with several existing catalogs. Among the $11\, 671$ KYSO entries, 431 fall on LOPS2. If we further restrict our sample to $V<15$, only 53 stars survive: 33 belong to the $\gamma$ Velorum cluster \citep{Franciosini2018}, 12 to the Gum Nebula, 5 to different regions/branches of the Vela molecular ridge. Two more stars are field Herbig Ae/Be stars, and one is the brightest star of the open cluster NGC~2362 \citep{Currie2009}, the only one to make it through our magnitude threshold. 

\subsection{Variable and binary stars}

Thanks to its very large area (2149~$\textrm{deg}^2$, that is, 5.2\% of the whole sky) and to the wide range in galactic latitude covered, LOPS2 will include a large number and variety of variable stars. A cross-match with the latest versions of the VSX (Variable Star indeX; \citealt{Watson2006}) and \citet{Gavras2023} catalogs yields $47\,356$ and $282\,366$ entries, respectively (with notable over-densities in the LMC and close to the galactic plane). The numbers decrease to $31\,211$ and $15\,932$, respectively, when we limit our sample at $V<15$. Among these we mention:
\begin{itemize}
    \item $7\,787$ eclipsing binaries (including 3017/2155 with explicit detached/contact-type classification);
    \item $6\,111$ long-period variables (LPV);
    \item 268 $\gamma$ Cassiopeiae stars;
    \item 185 chemically peculiar stars;
    \item 47 cataclysmic variables (CV);
    \item 700 $\delta$ Scuti pulsators;
    \item 651 Cepheids and 699 RR~Lyr;
    \item 29 slowly pulsating B stars (SPB);
    \item 17 $\beta$ Cephei stars (BCEP);
    \item seven $\gamma$ Doradus stars (GDOR).
\end{itemize}
It should be emphasized that some of these classes are interesting not just for stellar science, but also to plan more focused planet-hunting surveys. For instance, circumbinary planets represent a rare laboratory to challenge our theories on planetary formation and migration through a disk, and can be discovered both by detecting their transits \citep{Kostov2020},\footnote{The first (and so far only) circumbinary planet detected by TESS, TOI-1338 \citep{Kostov2020}, is by chance in LOPS2.} or by modeling their eclipse timing variations (ETV; \citealt{Goldberg2023,Brown2021}). Pulsators with intrinsically stable modes, such as $\delta$~Scuti and $\gamma$~Dor variables, may enable the detection of non-transiting planets on wide orbits through the pulsation timing technique, thanks to the unique combination of timing accuracy and temporal baseline of PLATO \citep{Vaulato2022}. Early-type stars with very specific chemical signatures such as $\lambda$~Bootis stars have been predicted since long to host a much higher fraction of giant planets with respect to ordinary field stars, although this has been recently disputed \citep{Saffe2021}.

Some wide-orbit detached eclipsing binaries (DEBs) in LOPS2 can be considered as benchmarks for stellar evolution studies, since their absolute radii and masses can be measured with an extremely high precision ($<3\%$; \citealt{Serenelli2021}). From the comprehensive list of 273 benchmark DEBs by \citet{Southworth2015}, 36 are within LOPS2, and 12 are brighter than $V=15$. Among them CV~Velorum ($V=6.69$) a B2.5V+B2.5V system \citep{Albrecht2014} known to show a significant misalignment between the orbital spin and the rotation axis of its components \citep{Marcussen2022}, stands out. 

Cataclysmic Variables (CVs) are compact binary systems, typically containing a white dwarf which is accreting material from a low-mass star through Roche lobe overflow. Many CVs show outbursts where they brighten by $\sim$2-5 mag on a recurrence time of a few weeks to many months, which is why setting a fixed magnitude limit as done above could be misleading. If we relax that constraint, there are 94 CVs in both VSX and \citet{Gavras2023} which lie in the LOPS2 field, many of which at quiescent are fainter than $V=18$ mag. However, VW~Hyi, which shows normal and superoutbursts, and IX~Vel which shows Z~Cam low states, are relatively bright ($V=12.5$ and $V=10.1$, respectively). Other CVs in the LOPS2 field include UW~Pic which is polar, where the white dwarf has a magnetic field strength of $\sim$20 MG. We expect that many CVs and other transients will be readily observable using PLATO during outbursts.

A census of all the SIMBAD objects in LOPS2 sorted by decreasing number of associated publications reveals, besides the objects already mentioned in the previous sections, other specific variable stars of interest:
\begin{itemize}
    \item $\gamma$~Doradus ($V\simeq 4.2$), the prototype of the $\gamma$~Dor class of variables \citep{Kaye1999};
    \item AI~Velorum ($V\simeq 6.70$), one of the most studied high-amplitude, double-mode $\delta$~Scuti pulsator;
    \item AB~Doradus, a pre-main-sequence quadruple system \citep{Guirado2011} known for the super-flaring activity of its primary component \citep{Schmitt2019}, and also the eponymous and brightest member of the AB~Dor moving group \citep{Zuckerman2004};
    \item $\gamma^2$~Velorum ($V\simeq 1.83$), a binary made of the closest and brightest known Wolf-Rayet (WR; \citealt{DeMarco2000}) star and a blue supergiant, belonging to the Vela~OB2 association \citep{Jeffries2014}. Another bright WR star is HR~2583 = HD~50896 ($V= 6.91$), a suspected binary system showing brightness variations still of unknown origin \citep{Flores2023};
    \item $\epsilon$~Canis Majoris, a binary with a very bright B2II component ($V\simeq 1.50$), the strongest EUV source in the whole sky;
    \item R~Doradus, an asymptotic giant branch (AGB) star with a 300~$R_\odot$ radius, probably the star with the largest apparent diameter as seen from the Sun \citep{Bedding1997}.
\end{itemize}

Aside from these few very bright variables and the cross-match with the catalog by \citet{Gavras2023}, we have also searched for confirmed non-radial pulsators in LOPS2. This was done from a combined \emph{Gaia}--TESS approach. \citet{HeyAerts2024} distilled about $60\,000$ variables using light curves from the first year of the TESS mission, starting from the original \emph{Gaia}~DR3 catalog of candidate main-sequence pulsators by \citet{DeRidder2023} relying on their stellar properties derived by \citet{Aerts2023}. \citet{HeyAerts2024} reclassified all the stars whose dominant frequency in the totally independent sparsely sampled \emph{Gaia}~DR3 light curves and the 30-min sampled TESS light curves are equal. All these stars have a dominant amplitude above 4~mmag in the \emph{Gaia} $G$ band.  Among the 6430 stars from \citet{HeyAerts2024} occurring in LOPS2, about 5000 are now classified as confirmed multi-periodic pulsators, while the others are rotational variables or eclipsing binaries. These non-radial pulsators in LOPS2 are split up into 1455 gravity-mode pulsators (of $\gamma\,$Dor or SPB type), 2079 $\delta\,$Sct pulsators, and 1449 hybrid pressure- and gravity mode pulsators. \citet{Mombarg2024} provide estimates of the (convective core) masses, radii, and relative ages for more than 10\,000 of the $\gamma\,$Dor, SPB, and hybrid pulsators classified by \citet{HeyAerts2024}. They did so by subjecting the pulsators' {\it Gaia\/} data to grid-based modelling based on rotating stellar models. These models were constructed such as to be compliant with the asteroseismology of rotating {\it Kepler\/} pulsators in the mass range from 1.3~$M_\odot$ and 9~$M_\odot$ from \citet{GangLi2020,Pedersen2021}. Some 3000 of the stars treated by \citet{Mombarg2024} occur in LOPS2. For about 300 among them Aerts et al.~(A\&A subm.) distilled the near-core rotation frequency from their identified dominant dipole prograde gravito-inertial mode as found consistently in both the \emph{Gaia}~DR3 and TESS light curves. Ongoing work on the full 5-year TESS light curves will increase the number of new pulsators with a measurement of the internal rotation frequency and this estimate will become more precise. Moreover, simulations with \texttt{PLATOSIM} \citep{Jannsen2024} reveal that detections of a dense spectrum of oscillation modes suitable for asteroseismology are expected for all these $\gamma\,$Dor and hybrid pulsators brighter than $V<14$ \citep{Jannsen2024b}, making them ideal calibration stars for the stellar science program of the mission.

\subsection{Synergy with TESS}\label{sec:tess}

\begin{figure*}
    \centering
    \includegraphics[width=0.9\columnwidth]{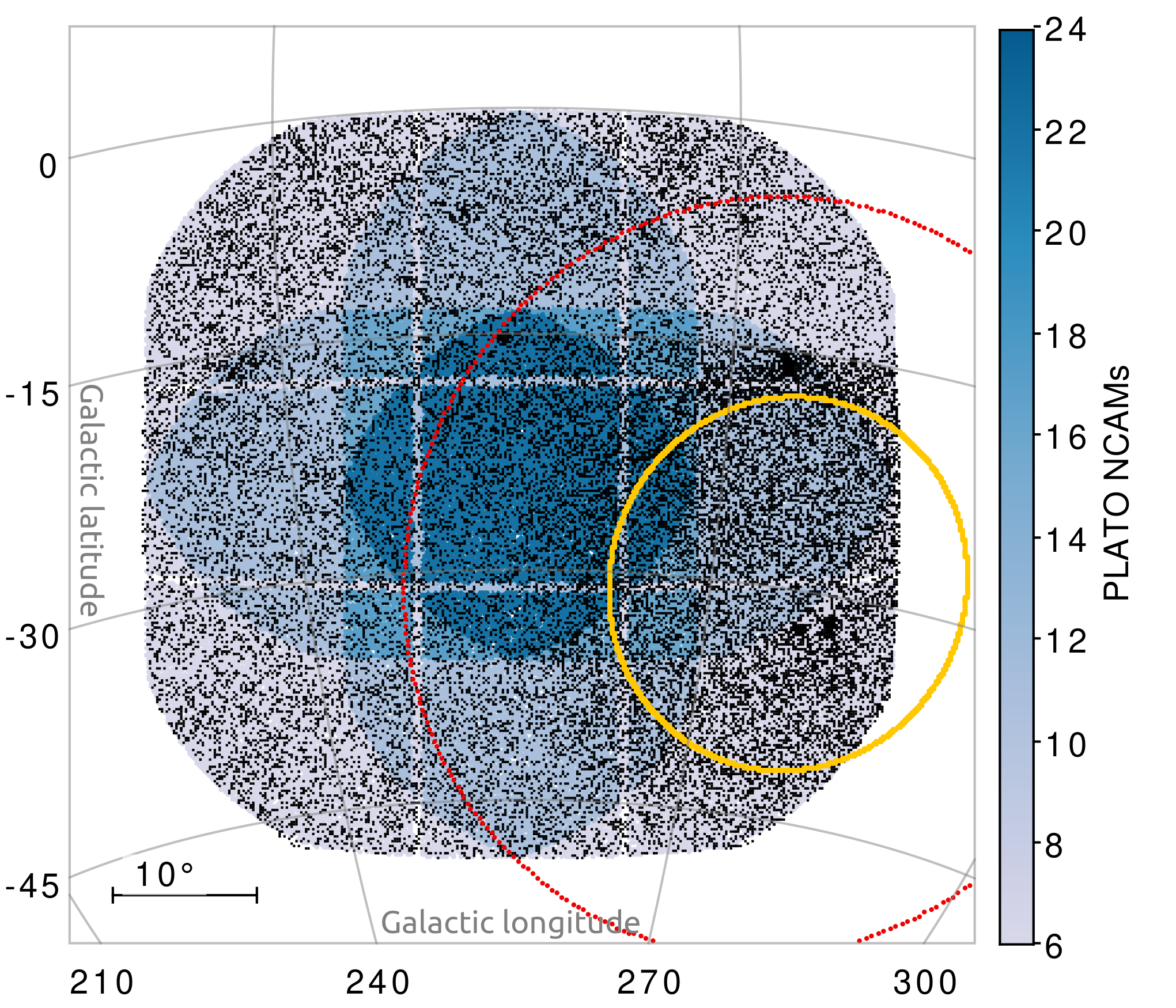}\hspace{5mm}
    \includegraphics[width=0.9\columnwidth]{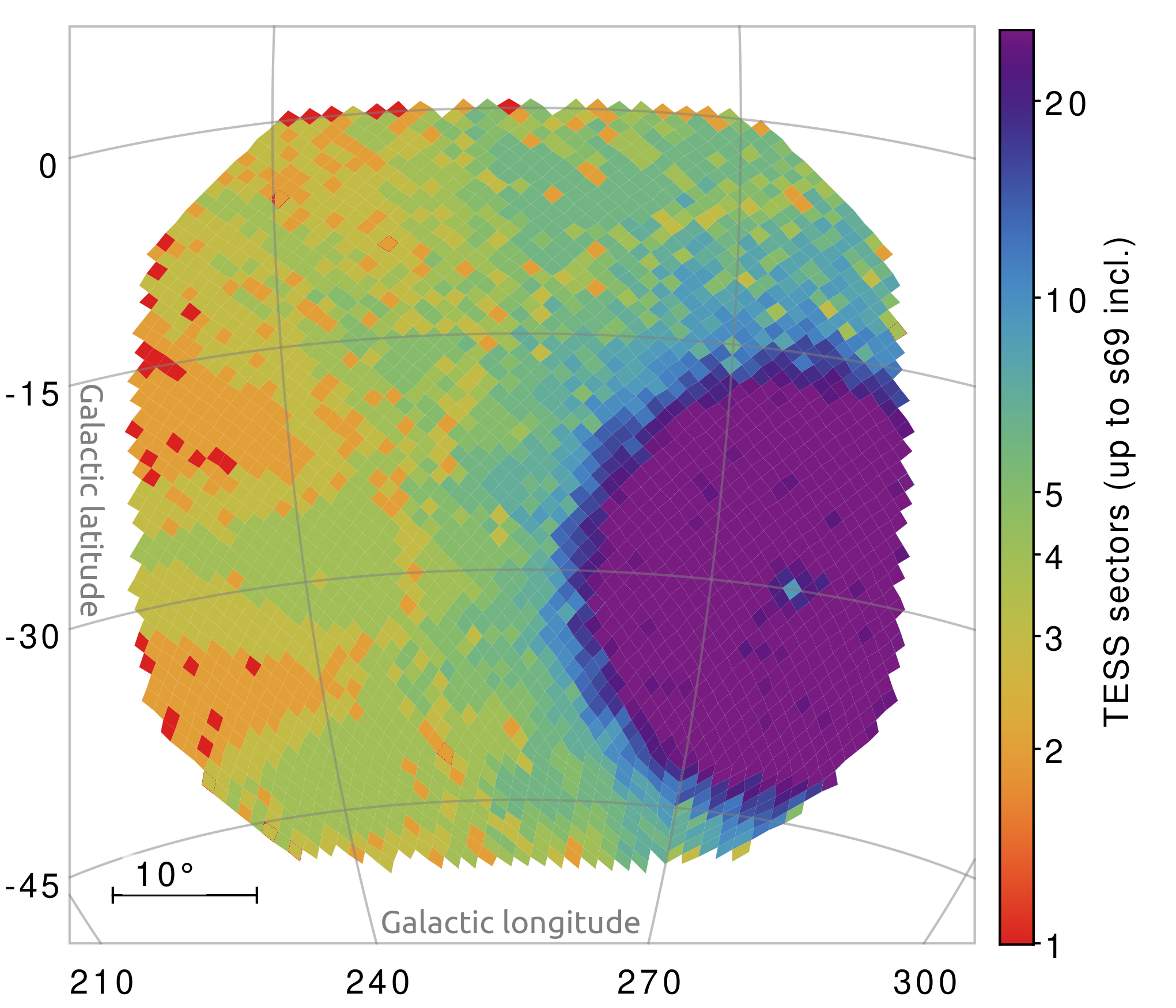}
    \caption{Distribution of observed TESS targets in LOPS2, orthographic projection in galactic coordinates. \emph{Left panel}: All $37\,910$ CTL targets within LOPS2 observed in short cadence by TESS from sector 1 to 82 included (black points), superimposed on the LOPS2 field. The yellow circle is the TESS southern CVZ (up to 35 sectors), the red circle is roughly the northern boundary of Camera 4, that is, the sky region where at least 4 TESS sectors are available.  \emph{Right panel:} Same but the TESS FFI coverage in terms of TESS sectors is color coded over an HEALPix level-6 grid.}
    \label{fig:lops2_ctl}
\end{figure*}

It is worth investigating the synergy with TESS not only in terms of known planets and planetary candidates, but also because the availability of long ($\ge$~28~d) and precise light curves can enable a useful characterization of the target stars in advance, including parameters which can play a role at the prioritization stage, such as rotational periods, activity levels, variability and/or binarity. 

Among the $519\,704$ unique TESS targets \citep{Stassun2018,Stassun2019} observed at the regular 2-min cadence up to Sector~82 included (i.e., August 2024, end of Cycle~6), $37\,910$ lie within the approximate footprint of LOPS2 with an average areal density spanning the 5-20 stars per $\textrm{deg}^2$ range (Fig.~\ref{fig:lops2_ctl}, left panel; the two most evident over-densities are centered on the LMC and NGC~2516). Among these, 55\% were observed for at least three TESS sectors (not always contiguous) and 15\% for at least 13 sectors, or one year cumulated (mostly within the $\beta\lesssim -78^\circ$ cap). Slightly more than 40\% of the available TESS short-cadence targets are not included in the current version of the PIC,\footnote{Being included in the PIC does not automatically imply being scheduled for observations or being included in the P1-P2-P4-P5 sample, even though the vast majority of them will do \citep{Montalto2021}.} mostly because their stellar parameters fall outside our $(T_\mathrm{eff}, R_\star)$ parameter space \citepalias{Montalto2021}, or, to a lesser extent, because they do not meet our magnitude requirements. All those remaining targets of scientific interest left out can be requested by the community through the PLATO GO program, if they are technically feasible. 

If we focus on the general sector coverage of the TESS FFIs instead (Fig.~\ref{fig:lops2_ctl}, right panel) we find that 25\% of the LOPS2 sky area has been already surveyed through at least 12 TESS sectors, 55\% through at least five sectors, and 88\% by three.

\subsection{Synergy with CHEOPS}

CHEOPS \citep{Benz2021} is an ESA S-class mission launched in 2019, equipped with a 30-cm reflecting telescope designed for ultra-high-precision photometry of individual targets in a single optical band \citep{Fortier2024}. Its main goal is the follow-up and characterization of known transiting planets, articulated over a wide range of science cases; see \citet{Benz2021} for a summary of the CHEOPS GTO program. Its 3.5-year primary mission ended in 2023, and currently CHEOPS is running its first mission extension until 2026, with a possible second extension foreseen until 2029. Given the scientific (and possibly temporal) overlap with the PLATO observations, it is worth investigating how much of LOPS2 can be accessed by CHEOPS.

\begin{figure*}
    \centering
    \includegraphics[width=0.9\columnwidth]{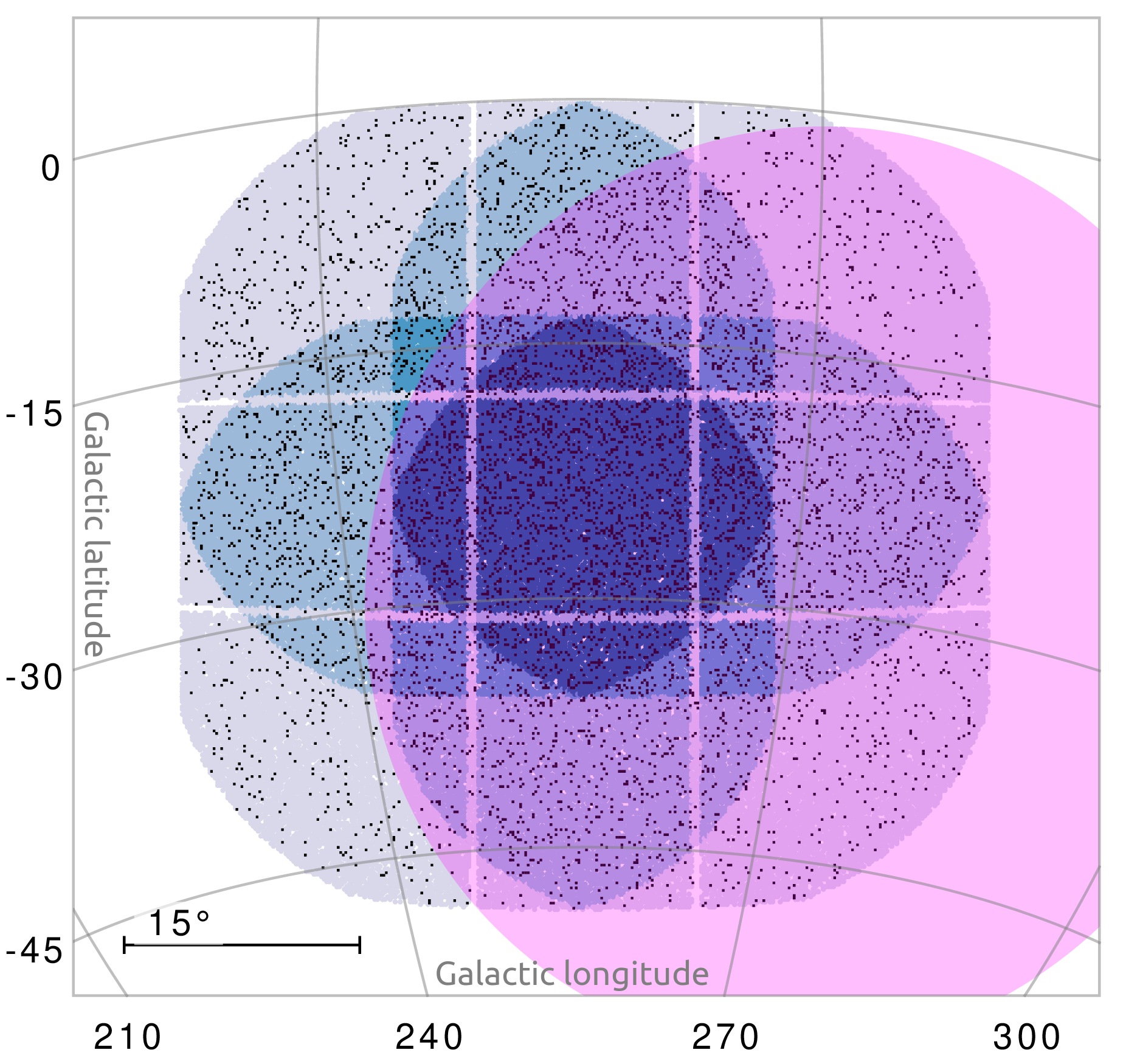}\hspace{5mm}
    \includegraphics[width=0.9\columnwidth]{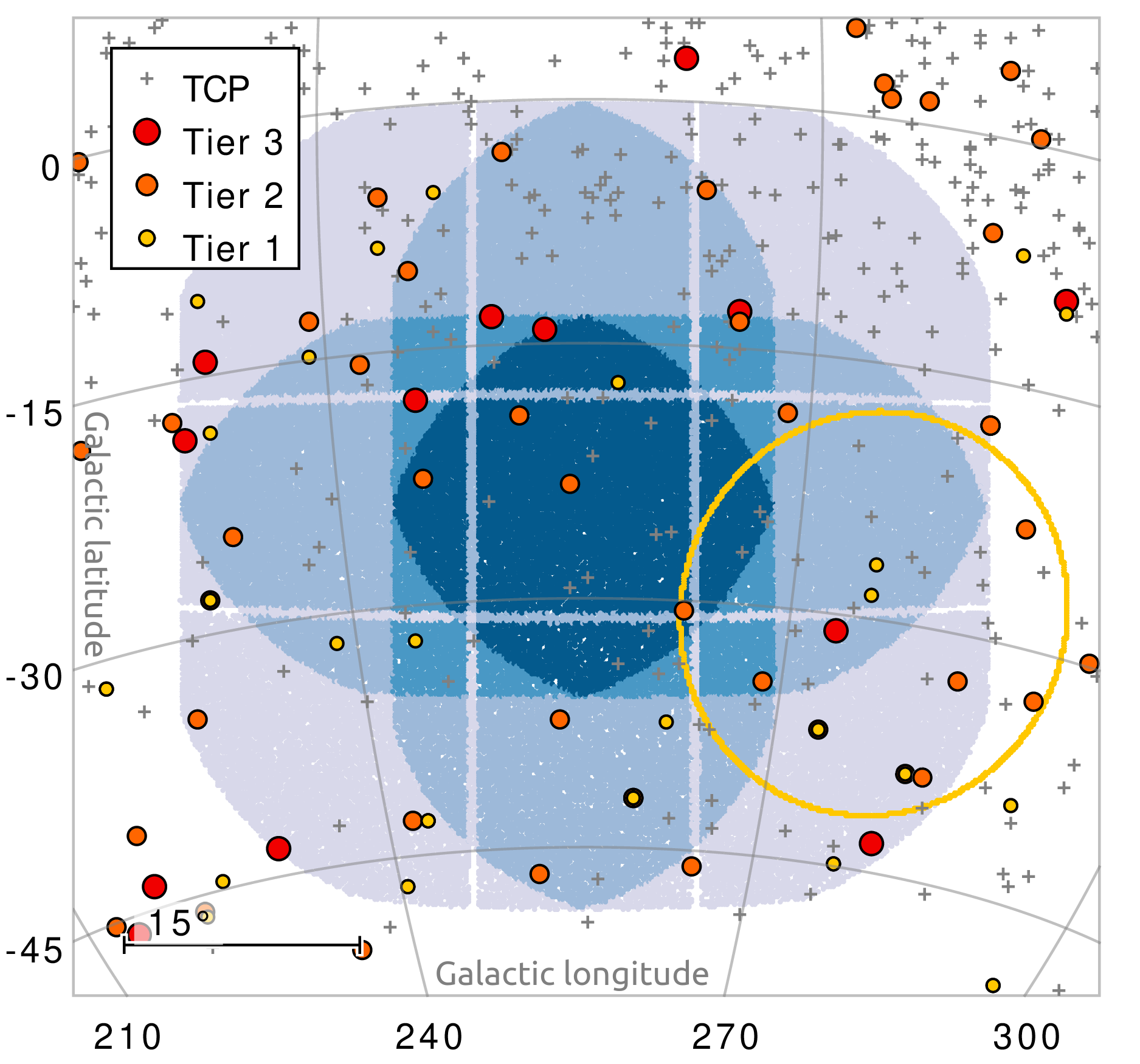}
    \caption{Overlap between CHEOPS/Ariel and LOPS2 (orthographic projection in galactic coordinates). \emph{Left panel:} Sky area forbidden to CHEOPS (in magenta) due to the Sun Exclusion Angle currently set at $120^\circ$ (corresponding to $\beta<-60^\circ$ for the southern ecliptic cap). P1 targets are plotted as black dots. \emph{Right panel:} Ariel targets from the current MRS. Confirmed targets belonging to the Ariel Tier 1/2/3 (see text for details) are plotted as yellow, orange and red dots, respectively, while the TESS candidate planets are plotted as gray crosses (see text for details). The yellow circle is the TESS CVZ.}
    \label{fig:lops2_cheops_ariel}
\end{figure*}

CHEOPS is not subject to a rigid scanning law such as TESS. Rather, it works on a flexible schedule, and its pointing ability is limited only by three avoidance angles to minimize scattered light from the Sun, the Moon and the Earth limb. The Sun Exclusion Angle (SEA) is currently set to $120^\circ$ (as throughout the whole nominal mission), and is the most limiting factor to reach large ecliptic latitudes such as those spanned by LOPS2. It is easy to see that a SEA of 120$^\circ$ implies that the whole $\beta < -60^\circ$ cap (the magenta area plotted in Fig.~\ref{fig:lops2_cheops_ariel}, left panel) is inaccessible to CHEOPS. In other words, only 33\% of the LOPS2 footprint can currently be pointed by CHEOPS, mostly in the six- and 12-camera regions and in any case for a very limited amount of days per year, because of the SEA limitation. The 24-NCAM region is completely inaccessible. Moreover, the fraction of accessible P1 targets (black dots in Fig.~\ref{fig:lops2_cheops_ariel}) is even lower (23\%), since P1 targets are more densely located in the inner regions (18 and 24 telescope) of LOPS. Should the CHEOPS observing strategy and/or the value of the avoidance angles be changed during the extended mission, the geometric overlap could increase by a significant amount; a relaxation of the SEA of a few degrees, for instance, would make part of the PLATO 24-camera region and its dense population of P1 targets accessible to CHEOPS.

\subsection{Synergy with Ariel}

Ariel is an ESA M-class mission currently in development, with a planned launch to L2 in 2029 \citep{Tinetti2018,Tinetti2021}, i.~e., with a significant temporal overlap with the PLATO nominal mission. Based on a 1-m class telescope equipped with two near-to-mid infrared low-resolution spectrographs (covering the 1.1-7.8~$\mu$m range) and three photometric channels also employed as fine guidance system (FGS; 0.5-1.1~$\mu$m), Ariel will monitor transiting planets through emission spectroscopy, transmission spectroscopy and phase curves, with repeated observations to gather the required SNR. Ariel targets will be observed through a four-``Tier'' approach: Tier 1 targets will undergo an extensive low-resolution reconnaissance survey, Tier 2 targets will be selected for a deeper survey at higher resolution, and Tier 3 is made of a relatively small number of benchmark planets to be explored for variability \citep{Edwards2022}. Tier 4 will be dedicated to phase curves and bespoke observations of targets of special interest, which are expected to be identified, for example, in Tier 1 or by space-based missions and ground-based surveys prior to Ariel.

The current Ariel target list, called Mission Reference Sample (MRS) is a living catalog, started by \citet{Edwards2022} and regularly updated.\footnote{\url{https://github.com/arielmission-space/Mission_Candidate_Sample}} The latest version (2024-07-09) contains 722 confirmed transiting planets, and 2025 candidates from the TESS TOI/ExoFOP database \citep{Guerrero2021}. The majority of MRS targets (58\%) are hot Jupiters. Also, nearly 90\% of the “confirmed planets” sub-sample have transits deeper than 1~mmag, i.~e., are easily detectable by TESS at high SNR (or in most cases, even from ground-based facilities). A cross-match with the LOPS2 footprint reveals that PLATO will observe 60 confirmed planets from the MRS (22 in Tier-1, 29 in Tier-2, nine in Tier-3) belonging to 49 host stars (Fig.~\ref{fig:lops2_cheops_ariel}, right panel).

Given the selection criteria of the MRS, PLATO will be a provider of new, additional targets for Ariel, mainly of a small (but scientifically very interesting) population of long-period transiting giants not (yet) discovered by TESS, or discovered as mono-transit events only \citep{Magliano2024}. Moreover, the PLATO versus Ariel synergy will be strong, for the already known and future targets, on the extremely accurate stellar parameters delivered by PLATO through asteroseismological analysis (and in particular, ages; \citealt{Goupil2024}), the in-depth characterization of stellar activity and rotational periods \citep{Breton2024}, the refinement of the planetary ephemeris for targets difficult to be followed up from the ground, the extension of the temporal baseline for TTV studies \citep{Borsato2022}, and the accurate planetary masses obtained by the PLATO follow-up program \citep{Rauer2024}. On the hot Jupiters belonging to the brighter tail of the MRS, PLATO will also be able to detect the planetary albedo and/or heat redistribution behavior through phase curves \citep{Shporer2017,Singh2019}, in a few cases possibly with color information thanks to the FCAM coverage.

\subsection{Synergy with JWST}

On a closing note, we also mention that the JWST southern CVZ at $\beta < -85^\circ$ \citep{Gardner2006} is fully enclosed in the LOPS2 footprint, unfolding another synergy with a space-based mission that is and will be crucial in the investigation of exoplanetary atmospheres. Further, JWST is able to monitor continuously for at least 200 days every southern source at $\beta < -45^\circ$, and this includes the whole LOPS2.

\section{Conclusions}\label{sec:conclusions}

In this work we presented the first field to be observed by PLATO during its LOP phase, and illustrated some of its astrophysical content relevant for planetary and stellar science. While the position of LOPS2 on the sky is now fixed, the PIC will continue to evolve and improve. The release of \emph{Gaia} DR4 will deliver much more accurate stellar parameters for all of our targets of interest, including all non-single star (NSS) solutions, individual epoch measurements, variability metrics, metallicities/abundances etc. based on the first 66 months of observations, and a large catalog of planets and planetary candidates discovered by \emph{Gaia} astrometry. As presented in this paper, LOPS2 also contains numerous highly relevant targets to address complementary science topics. In this context, the general community will be invited to propose observations in response to ESA announcements of opportunity for a Guest Observer program. The first open call is planned to be issued nine months prior to launch.

Meanwhile, effort will be spent on building the more effective target lists and defining criteria on how to allocate the PLATO telemetry resources (imagettes, centroids, light curves) among the selected targets. The selected field meets the mission scientific requirements on samples P1-P2-P4-P5 as defined by SciRD. In the near future, the PSWT will implement avprioritization metric able to define a subset of most valuable stars for the ground-based follow-up, called prime sample. For planets hosted by this sample, the PLATO mission will eventually provide planet candidate confirmation and high-resolution spectroscopy, with planetary mass measurements. The prioritization process will also allow prioritization of tPIC targets for imagettes acquisition, sampling timing, etc. This process, to be documented in a third paper of this series, does not separately consider the different samples. The prioritization metric is evaluated regardless of whether a star belongs to the P1, P2 or P5 sample.
For instance, a K3 subgiant at $V\lesssim 11$ and $R_\star \simeq 3$~$R_\odot$ can be included in P1 with a similar $\textrm{NSR}\simeq 45$~ppm with respect to a K3V star at $V\gtrsim 11$ and $R_\star \simeq 0.8$~$R_\odot$ belonging to P5, due to the magnitude constraint. However, detecting a true Earth analog during LOPS would be impossible in the former case, while perfectly within the reach\footnote{As a figure of reference, a photometric precision of 80~ppm in one hour enables the robust detection of a true Earth analog around a quiet Sun twin with three transits \citep{Rauer2024}.} of PLATO in the latter, so that this observation will be given higher priority.

In synthesis, the prioritization scheme will lead to the selection of the PLATO prime sample, a PIC subset of up to $20\,000$ stars to be observed during the LOP for which the PLATO Consortium is committed to do and deliver the follow-up observations. For that purpose, additional constraints (also to be documented in Paper III) other than the prioritization metric will be added to ensure that the targets are feasible with the available ground-based facilities and within the timeline of the PLATO operations.

\section*{Data availability}

The table with the confirmed transiting planets in LOPS2, the table with the TESS candidates in LOPS2 and the MOC regions of LOPS2 are publicly available on Zenodo: \href{https://zenodo.org/records/14720127}{zenodo14720127}.

\begin{acknowledgements}
We thank the referee for his valuable comments and suggestions. We thank Hans Deeg for his useful comments.\\
This work presents results from the European Space Agency (ESA) space mission PLATO. The PLATO payload, the PLATO Ground Segment and PLATO data processing are joint developments of ESA and the PLATO mission consortium (PMC). Funding for the PMC is provided at national levels, in particular by countries participating in the PLATO Multilateral Agreement (Austria, Belgium, Czech Republic, Denmark, France, Germany, Italy, Netherlands, Portugal, Spain, Sweden, Switzerland, Norway, and United Kingdom) and institutions from Brazil. Members of the PLATO Consortium can be found at \url{https://platomission.com/}. The ESA PLATO mission website is \url{https://www.cosmos.esa.int/plato}. We thank the teams working for PLATO for all their work.\\
CA and AT acknowledge support from the long-term structural Methusalem funding program by means of the project SOUL: Stellar evolution in full glory, KU Leuven grant METH/24/012.
JMMH is funded by Spanish MCIU/AEI/10.13039/501100011033 grant PID2019-107061GB-C61.
VN, GP, MM, SB, SD, VG, DM, LM, IP, LP, RR acknowledge support from PLATO ASI-INAF agreements n. 2022-28-HH.0\\
This research has made use of the SIMBAD database (operated at CDS, Strasbourg, France; \citealt{Wenger2000}), TOPCAT and STILTS \citep{Taylor2005,Taylor2006}, UCAC4-RPM \citep{Nascimbeni2016}, TEPCAT \citep{Southworth2011}. This research has made use of the Exoplanet Follow-up Observation Program (ExoFOP; DOI:10.26134/ExoFOP5) website, which is operated by the California Institute of Technology, under contract with the National Aeronautics and Space Administration under the Exoplanet Exploration Program.

\end{acknowledgements}

\bibliographystyle{aa}
\bibliography{biblio}

\begin{thebibliography}{162}
\expandafter\ifx\csname natexlab\endcsname\relax\def\natexlab#1{#1}\fi

\bibitem[{{Adibekyan} {et~al.}(2021){Adibekyan}, {Santos}, {Demangeon}, {Faria}, {Barros}, {Oshagh}, {Figueira}, {Delgado Mena}, {Sousa}, {Israelian}, {Campante}, \& {Hakobyan}}]{Adibekyan2021}
{Adibekyan}, V., {Santos}, N.~C., {Demangeon}, O.~D.~S., {et~al.} 2021, \aap, 649, A111

\bibitem[{{Aerts} {et~al.}(2023){Aerts}, {Molenberghs}, \& {De Ridder}}]{Aerts2023}
{Aerts}, C., {Molenberghs}, G., \& {De Ridder}, J. 2023, \aap, 672, A183

\bibitem[{{Aigrain} {et~al.}(2012){Aigrain}, {Pont}, \& {Zucker}}]{Aigrain2012}
{Aigrain}, S., {Pont}, F., \& {Zucker}, S. 2012, \mnras, 419, 3147

\bibitem[{{Albrecht} {et~al.}(2014){Albrecht}, {Winn}, {Torres}, {Fabrycky}, {Setiawan}, {Gillon}, {Jehin}, {Triaud}, {Queloz}, {Snellen}, \& {Eggleton}}]{Albrecht2014}
{Albrecht}, S., {Winn}, J.~N., {Torres}, G., {et~al.} 2014, \apj, 785, 83

\bibitem[{{Alei} {et~al.}(2020){Alei}, {Claudi}, {Bignamini}, \& {Molinaro}}]{Alei2020}
{Alei}, E., {Claudi}, R., {Bignamini}, A., \& {Molinaro}, M. 2020, Astronomy and Computing, 31, 100370

\bibitem[{{Anglada-Escude} {et~al.}(2014){Anglada-Escude}, {Arriagada}, {Tuomi}, {Zechmeister}, {Jenkins}, {Ofir}, {Dreizler}, {Gerlach}, {Marvin}, {Reiners}, {Jeffers}, {Butler}, {Vogt}, {Amado}, {Rodriguez-Lopez}, {Berdinas}, {Morin}, {Crane}, {Shectman}, {Thompson}, {Diaz}, {Rivera}, {Sarmiento}, \& {Jones}}]{Anglada2014}
{Anglada-Escude}, G., {Arriagada}, P., {Tuomi}, M., {et~al.} 2014, \mnras, 443, L89

\bibitem[{{Auvergne} {et~al.}(2009){Auvergne}, {Bodin}, {Boisnard}, {Buey}, {Chaintreuil}, {Epstein}, {Jouret}, {Lam-Trong}, {Levacher}, {Magnan}, {Perez}, {Plasson}, {Plesseria}, {Peter}, {Steller}, {Tiph{\`e}ne}, {Baglin}, {Agogu{\'e}}, {Appourchaux}, {Barbet}, {Beaufort}, {Bellenger}, {Berlin}, {Bernardi}, {Blouin}, {Boumier}, {Bonneau}, {Briet}, {Butler}, {Cautain}, {Chiavassa}, {Costes}, {Cuvilho}, {Cunha-Parro}, {de Oliveira Fialho}, {Decaudin}, {Defise}, {Djalal}, {Docclo}, {Drummond}, {Dupuis}, {Exil}, {Faur{\'e}}, {Gaboriaud}, {Gamet}, {Gavalda}, {Grolleau}, {Gueguen}, {Guivarc'h}, {Guterman}, {Hasiba}, {Huntzinger}, {Hustaix}, {Imbert}, {Jeanville}, {Johlander}, {Jorda}, {Journoud}, {Karioty}, {Kerjean}, {Lafond}, {Lapeyrere}, {Landiech}, {Larqu{\'e}}, {Laudet}, {Le Merrer}, {Leporati}, {Leruyet}, {Levieuge}, {Llebaria}, {Martin}, {Mazy}, {Mesnager}, {Michel}, {Moalic}, {Monjoin}, {Naudet}, {Neukirchner}, {Nguyen-Kim}, {Ollivier}, {Orcesi}, {Ottacher}, {Oulali}, {Parisot}, {Perruchot}, {Piacentino},
  {Pinheiro da Silva}, {Platzer}, {Pontet}, {Pradines}, {Quentin}, {Rohbeck}, {Rolland}, {Rollenhagen}, {Romagnan}, {Russ}, {Samadi}, {Schmidt}, {Schwartz}, {Sebbag}, {Smit}, {Sunter}, {Tello}, {Toulouse}, {Ulmer}, {Vandermarcq}, {Vergnault}, {Wallner}, {Waultier}, \& {Zanatta}}]{Auvergne2009}
{Auvergne}, M., {Bodin}, P., {Boisnard}, L., {et~al.} 2009, \aap, 506, 411

\bibitem[{{Baratella} {et~al.}(2020){Baratella}, {D'Orazi}, {Carraro}, {Desidera}, {Randich}, {Magrini}, {Adibekyan}, {Smiljanic}, {Spina}, {Tsantaki}, {Tautvai{\v{s}}ien{\.{e}}}, {Sousa}, {Jofr{\'e}}, {Jim{\'e}nez-Esteban}, {Delgado-Mena}, {Martell}, {Van der Swaelmen}, {Roccatagliata}, {Gilmore}, {Alfaro}, {Bayo}, {Bensby}, {Bragaglia}, {Franciosini}, {Gonneau}, {Heiter}, {Hourihane}, {Jeffries}, {Koposov}, {Morbidelli}, {Prisinzano}, {Sacco}, {Sbordone}, {Worley}, {Zaggia}, \& {Lewis}}]{Baratella2020}
{Baratella}, M., {D'Orazi}, V., {Carraro}, G., {et~al.} 2020, \aap, 634, A34

\bibitem[{{Barros} {et~al.}(2022){Barros}, {Demangeon}, {Alibert}, {Leleu}, {Adibekyan}, {Lovis}, {Bossini}, {Sousa}, {Hara}, {Bouchy}, {Lavie}, {Rodrigues}, {Gomes da Silva}, {Lillo-Box}, {Pepe}, {Tabernero}, {Zapatero Osorio}, {Sozzetti}, {Su{\'a}rez Mascare{\~n}o}, {Micela}, {Allende Prieto}, {Cristiani}, {Damasso}, {Di Marcantonio}, {Ehrenreich}, {Faria}, {Figueira}, {Gonz{\'a}lez Hern{\'a}ndez}, {Jenkins}, {Lo Curto}, {Martins}, {Micela}, {Nunes}, {Pall{\'e}}, {Santos}, {Rebolo}, {Seager}, {Twicken}, {Udry}, {Vanderspek}, \& {Winn}}]{Barros2022}
{Barros}, S.~C.~C., {Demangeon}, O.~D.~S., {Alibert}, Y., {et~al.} 2022, \aap, 665, A154

\bibitem[{{Bedding} {et~al.}(1997){Bedding}, {Zijlstra}, {von der Luhe}, {Robertson}, {Marson}, {Barton}, \& {Carter}}]{Bedding1997}
{Bedding}, T.~R., {Zijlstra}, A.~A., {von der Luhe}, O., {et~al.} 1997, \mnras, 286, 957

\bibitem[{{Bell} \& {Rodgers}(1964)}]{Bell1964}
{Bell}, R.~A. \& {Rodgers}, A.~W. 1964, The Observatory, 84, 29

\bibitem[{{Benz} {et~al.}(2021){Benz}, {Broeg}, {Fortier}, {Rando}, {Beck}, {Beck}, {Queloz}, {Ehrenreich}, {Maxted}, {Isaak}, {Billot}, {Alibert}, {Alonso}, {Ant{\'o}nio}, {Asquier}, {Bandy}, {B{\'a}rczy}, {Barrado}, {Barros}, {Baumjohann}, {Bekkelien}, {Bergomi}, {Biondi}, {Bonfils}, {Borsato}, {Brandeker}, {Busch}, {Cabrera}, {Cessa}, {Charnoz}, {Chazelas}, {Collier Cameron}, {Corral Van Damme}, {Cortes}, {Davies}, {Deleuil}, {Deline}, {Delrez}, {Demangeon}, {Demory}, {Erikson}, {Farinato}, {Fossati}, {Fridlund}, {Futyan}, {Gandolfi}, {Garcia Munoz}, {Gillon}, {Guterman}, {Gutierrez}, {Hasiba}, {Heng}, {Hernandez}, {Hoyer}, {Kiss}, {Kovacs}, {Kuntzer}, {Laskar}, {Lecavelier des Etangs}, {Lendl}, {L{\'o}pez}, {Lora}, {Lovis}, {L{\"u}ftinger}, {Magrin}, {Malvasio}, {Marafatto}, {Michaelis}, {de Miguel}, {Modrego}, {Munari}, {Nascimbeni}, {Olofsson}, {Ottacher}, {Ottensamer}, {Pagano}, {Palacios}, {Pall{\'e}}, {Peter}, {Piazza}, {Piotto}, {Pizarro}, {Pollaco}, {Ragazzoni}, {Ratti}, {Rauer}, {Ribas}, {Rieder},
  {Rohlfs}, {Safa}, {Salatti}, {Santos}, {Scandariato}, {S{\'e}gransan}, {Simon}, {Smith}, {Sordet}, {Sousa}, {Steller}, {Szab{\'o}}, {Szoke}, {Thomas}, {Tschentscher}, {Udry}, {Van Grootel}, {Viotto}, {Walter}, {Walton}, {Wildi}, \& {Wolter}}]{Benz2021}
{Benz}, W., {Broeg}, C., {Fortier}, A., {et~al.} 2021, Experimental Astronomy, 51, 109

\bibitem[{{B{\'e}trisey} {et~al.}(2023){B{\'e}trisey}, {Buldgen}, {Reese}, {Farnir}, {Dupret}, {Khan}, {Goupil}, {Eggenberger}, \& {Meynet}}]{Betrisey2023}
{B{\'e}trisey}, J., {Buldgen}, G., {Reese}, D.~R., {et~al.} 2023, \aap, 676, A10

\bibitem[{{B{\"o}rner} {et~al.}(2024){B{\"o}rner}, {Paproth}, {Cabrera}, {Pertenais}, {Rauer}, {Mas-Hesse}, {Pagano}, {Alvarez}, {Erikson}, {Grie{\ss}bach}, {Levillain}, {Magrin}, {Mogulsky}, {Niemi}, {Prod'homme}, {Regibo}, {De Ridder}, {Rockstein}, {Samadi}, {Serrano-Velarde}, {Smith}, {Verhoeve}, \& {Walton}}]{Borner2024}
{B{\"o}rner}, A., {Paproth}, C., {Cabrera}, J., {et~al.} 2024, Experimental Astronomy, 58, 1

\bibitem[{{Borsato} {et~al.}(2022){Borsato}, {Nascimbeni}, {Piotto}, \& {Szab{\'o}}}]{Borsato2022}
{Borsato}, L., {Nascimbeni}, V., {Piotto}, G., \& {Szab{\'o}}, G. 2022, Experimental Astronomy, 53, 635

\bibitem[{{Bortle} {et~al.}(2021){Bortle}, {Fausey}, {Ji}, {Dodson-Robinson}, {Ramirez Delgado}, \& {Gizis}}]{Bortle2021}
{Bortle}, A., {Fausey}, H., {Ji}, J., {et~al.} 2021, \aj, 161, 230

\bibitem[{{Borucki} {et~al.}(2010){Borucki}, {Koch}, {Basri}, {Batalha}, {Brown}, {Caldwell}, {Caldwell}, {Christensen-Dalsgaard}, {Cochran}, {DeVore}, {Dunham}, {Dupree}, {Gautier}, {Geary}, {Gilliland}, {Gould}, {Howell}, {Jenkins}, {Kondo}, {Latham}, {Marcy}, {Meibom}, {Kjeldsen}, {Lissauer}, {Monet}, {Morrison}, {Sasselov}, {Tarter}, {Boss}, {Brownlee}, {Owen}, {Buzasi}, {Charbonneau}, {Doyle}, {Fortney}, {Ford}, {Holman}, {Seager}, {Steffen}, {Welsh}, {Rowe}, {Anderson}, {Buchhave}, {Ciardi}, {Walkowicz}, {Sherry}, {Horch}, {Isaacson}, {Everett}, {Fischer}, {Torres}, {Johnson}, {Endl}, {MacQueen}, {Bryson}, {Dotson}, {Haas}, {Kolodziejczak}, {Van Cleve}, {Chandrasekaran}, {Twicken}, {Quintana}, {Clarke}, {Allen}, {Li}, {Wu}, {Tenenbaum}, {Verner}, {Bruhweiler}, {Barnes}, \& {Prsa}}]{Borucki2010}
{Borucki}, W.~J., {Koch}, D., {Basri}, G., {et~al.} 2010, Science, 327, 977

\bibitem[{{Bouma} {et~al.}(2021){Bouma}, {Curtis}, {Hartman}, {Winn}, \& {Bakos}}]{Bouma2021}
{Bouma}, L.~G., {Curtis}, J.~L., {Hartman}, J.~D., {Winn}, J.~N., \& {Bakos}, G.~{\'A}. 2021, \aj, 162, 197

\bibitem[{{Bourrier} {et~al.}(2020){Bourrier}, {Kitzmann}, {Kuntzer}, {Nascimbeni}, {Lendl}, {Lavie}, {Hoeijmakers}, {Pino}, {Ehrenreich}, {Heng}, {Allart}, {Cegla}, {Dumusque}, {Melo}, {Astudillo-Defru}, {Caldwell}, {Cretignier}, {Giles}, {Henze}, {Jenkins}, {Lovis}, {Murgas}, {Pepe}, {Ricker}, {Rose}, {Seager}, {Segransan}, {Su{\'a}rez-Mascare{\~n}o}, {Udry}, {Vanderspek}, \& {Wyttenbach}}]{Bourrier2020}
{Bourrier}, V., {Kitzmann}, D., {Kuntzer}, T., {et~al.} 2020, \aap, 637, A36

\bibitem[{{Bray} {et~al.}(2023){Bray}, {Kolb}, {Rowden}, {Farmer}, {B{\"o}rner}, \& {Kozhura}}]{Bray2023}
{Bray}, J.~C., {Kolb}, U., {Rowden}, P., {et~al.} 2023, \mnras, 518, 3637

\bibitem[{{Breton} {et~al.}(2024){Breton}, {Lanza}, {Messina}, {Pagano}, {Bugnet}, {Corsaro}, {Garc{\'\i}a}, {Mathur}, {Santos}, {Aigrain}, {Amard}, {Brun}, {Degott}, {Noraz}, {Palakkatharappil}, {Panetier}, {Strugarek}, {Belkacem}, {Goupil}, {Ouazzani}, {Philidet}, {Reni{\'e}}, \& {Roth}}]{Breton2024}
{Breton}, S.~N., {Lanza}, A.~F., {Messina}, S., {et~al.} 2024, \aap, 689, A229

\bibitem[{{Brown-Sevilla} {et~al.}(2021){Brown-Sevilla}, {Nascimbeni}, {Borsato}, {Tartaglia}, {Nardiello}, {Granata}, {Libralato}, {Damasso}, {Piotto}, {Pollacco}, {West}, {Colombo}, {Cunial}, {Piazza}, \& {Scaggiante}}]{Brown2021}
{Brown-Sevilla}, S.~B., {Nascimbeni}, V., {Borsato}, L., {et~al.} 2021, \mnras, 506, 2122

\bibitem[{{Cantat-Gaudin} {et~al.}(2020){Cantat-Gaudin}, {Anders}, {Castro-Ginard}, {Jordi}, {Romero-G{\'o}mez}, {Soubiran}, {Casamiquela}, {Tarricq}, {Moitinho}, {Vallenari}, {Bragaglia}, {Krone-Martins}, \& {Kounkel}}]{Cantat-Gaudin2020}
{Cantat-Gaudin}, T., {Anders}, F., {Castro-Ginard}, A., {et~al.} 2020, \aap, 640, A1

\bibitem[{{Cantat-Gaudin} {et~al.}(2019){Cantat-Gaudin}, {Jordi}, {Wright}, {Armstrong}, {Vallenari}, {Balaguer-N{\'u}{\~n}ez}, {Ramos}, {Bossini}, {Padoan}, {Pelkonen}, {Mapelli}, \& {Jeffries}}]{CantatGaudin2019}
{Cantat-Gaudin}, T., {Jordi}, C., {Wright}, N.~J., {et~al.} 2019, \aap, 626, A17

\bibitem[{{Chauvin} {et~al.}(2005){Chauvin}, {Lagrange}, {Zuckerman}, {Dumas}, {Mouillet}, {Song}, {Beuzit}, {Lowrance}, \& {Bessell}}]{Chauvin2005}
{Chauvin}, G., {Lagrange}, A.~M., {Zuckerman}, B., {et~al.} 2005, \aap, 438, L29

\bibitem[{{Cunha} {et~al.}(2021){Cunha}, {Roxburgh}, {Aguirre B{\o}rsen-Koch}, {Ball}, {Basu}, {Chaplin}, {Goupil}, {Nsamba}, {Ong}, {Reese}, {Verma}, {Belkacem}, {Campante}, {Christensen-Dalsgaard}, {Clara}, {Deheuvels}, {Monteiro}, {Noll}, {Ouazzani}, {R{\o}rsted}, {Stokholm}, \& {Winther}}]{Cunha2021}
{Cunha}, M.~S., {Roxburgh}, I.~W., {Aguirre B{\o}rsen-Koch}, V., {et~al.} 2021, \mnras, 508, 5864

\bibitem[{{Currie} {et~al.}(2009){Currie}, {Lada}, {Plavchan}, {Robitaille}, {Irwin}, \& {Kenyon}}]{Currie2009}
{Currie}, T., {Lada}, C.~J., {Plavchan}, P., {et~al.} 2009, \apj, 698, 1

\bibitem[{{Daylan} {et~al.}(2021){Daylan}, {G{\"u}nther}, {Mikal-Evans}, {Sing}, {Wong}, {Shporer}, {Niraula}, {de Wit}, {Koll}, {Parmentier}, {Fetherolf}, {Kane}, {Ricker}, {Vanderspek}, {Seager}, {Winn}, {Jenkins}, {Caldwell}, {Charbonneau}, {Henze}, {Paegert}, {Rinehart}, {Rose}, {Sha}, {Quintana}, \& {Villasenor}}]{Daylan2021}
{Daylan}, T., {G{\"u}nther}, M.~N., {Mikal-Evans}, T., {et~al.} 2021, \aj, 161, 131

\bibitem[{{de Jong} {et~al.}(2012){de Jong}, {Bellido-Tirado}, {Chiappini}, {Depagne}, {Haynes}, {Johl}, {Schnurr}, {Schwope}, {Walcher}, {Dionies}, {Haynes}, {Kelz}, {Kitaura}, {Lamer}, {Minchev}, {M{\"u}ller}, {Nuza}, {Olaya}, {Piffl}, {Popow}, {Steinmetz}, {Ural}, {Williams}, {Winkler}, {Wisotzki}, {Ansorge}, {Banerji}, {Gonzalez Solares}, {Irwin}, {Kennicutt}, {King}, {McMahon}, {Koposov}, {Parry}, {Sun}, {Walton}, {Finger}, {Iwert}, {Krumpe}, {Lizon}, {Vincenzo}, {Amans}, {Bonifacio}, {Cohen}, {Francois}, {Jagourel}, {Mignot}, {Royer}, {Sartoretti}, {Bender}, {Grupp}, {Hess}, {Lang-Bardl}, {Muschielok}, {B{\"o}hringer}, {Boller}, {Bongiorno}, {Brusa}, {Dwelly}, {Merloni}, {Nandra}, {Salvato}, {Pragt}, {Navarro}, {Gerlofsma}, {Roelfsema}, {Dalton}, {Middleton}, {Tosh}, {Boeche}, {Caffau}, {Christlieb}, {Grebel}, {Hansen}, {Koch}, {Ludwig}, {Quirrenbach}, {Sbordone}, {Seifert}, {Thimm}, {Trifonov}, {Helmi}, {Trager}, {Feltzing}, {Korn}, \& {Boland}}]{deJong2012}
{de Jong}, R.~S., {Bellido-Tirado}, O., {Chiappini}, C., {et~al.} 2012, in Society of Photo-Optical Instrumentation Engineers (SPIE) Conference Series, Vol. 8446, Ground-based and Airborne Instrumentation for Astronomy IV, ed. I.~S. {McLean}, S.~K. {Ramsay}, \& H.~{Takami}, 84460T

\bibitem[{{De Marco} {et~al.}(2000){De Marco}, {Schmutz}, {Crowther}, {Hillier}, {Dessart}, {de Koter}, \& {Schweickhardt}}]{DeMarco2000}
{De Marco}, O., {Schmutz}, W., {Crowther}, P.~A., {et~al.} 2000, \aap, 358, 187

\bibitem[{{De Silva} {et~al.}(2013){De Silva}, {D'Orazi}, {Melo}, {Torres}, {Gieles}, {Quast}, \& {Sterzik}}]{DeSilva2013}
{De Silva}, G.~M., {D'Orazi}, V., {Melo}, C., {et~al.} 2013, \mnras, 431, 1005

\bibitem[{{Deeg} {et~al.}(2023){Deeg}, {Georgieva}, {Nowak}, {Persson}, {Cale}, {Murgas}, {Pall{\'e}}, {Godoy-Rivera}, {Dai}, {Ciardi}, {Murphy}, {Beck}, {Burke}, {Cabrera}, {Carleo}, {Cochran}, {Collins}, {Csizmadia}, {El Mufti}, {Fridlund}, {Fukui}, {Gandolfi}, {Garc{\'\i}a}, {Guenther}, {Guerra}, {Grziwa}, {Isaacson}, {Isogai}, {Jenkins}, {K{\'a}bath}, {Korth}, {Lam}, {Latham}, {Luque}, {Lund}, {Livingston}, {Mathis}, {Mathur}, {Narita}, {Orell-Miquel}, {Osborne}, {Parviainen}, {Plavchan}, {Redfield}, {Rodriguez}, {Schwarz}, {Seager}, {Smith}, {Van Eylen}, {Van Zandt}, {Winn}, \& {Ziegler}}]{Deeg2023}
{Deeg}, H.~J., {Georgieva}, I.~Y., {Nowak}, G., {et~al.} 2023, \aap, 677, A12

\bibitem[{{Delrez} {et~al.}(2016){Delrez}, {Santerne}, {Almenara}, {Anderson}, {Collier-Cameron}, {D{\'\i}az}, {Gillon}, {Hellier}, {Jehin}, {Lendl}, {Maxted}, {Neveu-VanMalle}, {Pepe}, {Pollacco}, {Queloz}, {S{\'e}gransan}, {Smalley}, {Smith}, {Triaud}, {Udry}, {Van Grootel}, \& {West}}]{Delrez2016}
{Delrez}, L., {Santerne}, A., {Almenara}, J.~M., {et~al.} 2016, \mnras, 458, 4025

\bibitem[{{Demangeon} {et~al.}(2021){Demangeon}, {Zapatero Osorio}, {Alibert}, {Barros}, {Adibekyan}, {Tabernero}, {Antoniadis-Karnavas}, {Camacho}, {Su{\'a}rez Mascare{\~n}o}, {Oshagh}, {Micela}, {Sousa}, {Lovis}, {Pepe}, {Rebolo}, {Cristiani}, {Santos}, {Allart}, {Allende Prieto}, {Bossini}, {Bouchy}, {Cabral}, {Damasso}, {Di Marcantonio}, {D'Odorico}, {Ehrenreich}, {Faria}, {Figueira}, {G{\'e}nova Santos}, {Haldemann}, {Hara}, {Gonz{\'a}lez Hern{\'a}ndez}, {Lavie}, {Lillo-Box}, {Lo Curto}, {Martins}, {M{\'e}gevand}, {Mehner}, {Molaro}, {Nunes}, {Pall{\'e}}, {Pasquini}, {Poretti}, {Sozzetti}, \& {Udry}}]{Demangeon2021}
{Demangeon}, O.~D.~S., {Zapatero Osorio}, M.~R., {Alibert}, Y., {et~al.} 2021, \aap, 653, A41

\bibitem[{{D{\'\i}az} {et~al.}(2016){D{\'\i}az}, {S{\'e}gransan}, {Udry}, {Lovis}, {Pepe}, {Dumusque}, {Marmier}, {Alonso}, {Benz}, {Bouchy}, {Coffinet}, {Collier Cameron}, {Deleuil}, {Figueira}, {Gillon}, {Lo Curto}, {Mayor}, {Mordasini}, {Motalebi}, {Moutou}, {Pollacco}, {Pompei}, {Queloz}, {Santos}, \& {Wyttenbach}}]{Diaz2016}
{D{\'\i}az}, R.~F., {S{\'e}gransan}, D., {Udry}, S., {et~al.} 2016, \aap, 585, A134

\bibitem[{{Dodson-Robinson} {et~al.}(2011){Dodson-Robinson}, {Beichman}, {Carpenter}, \& {Bryden}}]{DobsonRobinson2011}
{Dodson-Robinson}, S.~E., {Beichman}, C.~A., {Carpenter}, J.~M., \& {Bryden}, G. 2011, \aj, 141, 11

\bibitem[{{Edwards} \& {Tinetti}(2022)}]{Edwards2022}
{Edwards}, B. \& {Tinetti}, G. 2022, \aj, 164, 15

\bibitem[{{Eschen} {et~al.}(2024){Eschen}, {Bayliss}, {Wilson}, {Kunimoto}, {Pelisoli}, \& {Rodel}}]{Eschen2024}
{Eschen}, Y. N.~E., {Bayliss}, D., {Wilson}, T.~G., {et~al.} 2024, \mnras, 535, 1778

\bibitem[{{Fermiano} {et~al.}(2024){Fermiano}, {Saito}, {Ivanov}, {Caceres}, {Almeida}, {Aires}, {Beamin}, {Minniti}, {Ferreira}, {Andrade}, {Borges}, {de Almeida}, {Jablonski}, \& {Schlindwein}}]{Fermiano2024}
{Fermiano}, V., {Saito}, R.~K., {Ivanov}, V.~D., {et~al.} 2024, \aap, 690, L7

\bibitem[{{Fernique} {et~al.}(2014){Fernique}, {Boch}, {Donaldson}, {Durand}, {O'Mullane}, {Reinecke}, \& {Taylor}}]{Fernique2014}
{Fernique}, P., {Boch}, T., {Donaldson}, T., {et~al.} 2014, {MOC - HEALPix Multi-Order Coverage map Version 1.0}, IVOA Recommendation 02 June 2014

\bibitem[{{Flores} {et~al.}(2023){Flores}, {Hillier}, \& {Dessart}}]{Flores2023}
{Flores}, B.~L., {Hillier}, D.~J., \& {Dessart}, L. 2023, \mnras, 518, 5001

\bibitem[{{Fortier} {et~al.}(2024){Fortier}, {Simon}, {Broeg}, {Olofsson}, {Deline}, {Wilson}, {Maxted}, {Brandeker}, {Collier Cameron}, {Beck}, {Bekkelien}, {Billot}, {Bonfanti}, {Bruno}, {Cabrera}, {Delrez}, {Demory}, {Futyan}, {Flor{\'e}n}, {G{\"u}nther}, {Heitzmann}, {Hoyer}, {Isaak}, {Sousa}, {Stalport}, {Turin}, {Verhoeve}, {Akinsanmi}, {Alibert}, {Alonso}, {B{\'a}nhidi}, {B{\'a}rczy}, {Barrado}, {Barros}, {Baumjohann}, {Baycroft}, {Beck}, {Benz}, {B{\'\i}r{\'o}}, {B{\'o}di}, {Bonfils}, {Borsato}, {Charnoz}, {Cseh}, {Csizmadia}, {Cs{\'a}nyi}, {Cubillos}, {Davies}, {Davis}, {Deleuil}, {Demangeon}, {Derekas}, {Dransfield}, {Ducrot}, {Ehrenreich}, {Erikson}, {Fari{\~n}a}, {Fossati}, {Fridlund}, {Gandolfi}, {Garai}, {Garcia}, {Gillon}, {G{\'o}mez Maqueo Chew}, {G{\'o}mez-Mu{\~n}oz}, {Granata}, {G{\"u}del}, {Guterman}, {Heged{\"u}s}, {Helling}, {Jehin}, {Kalup}, {Kilkenny}, {Kiss}, {Kriskovics}, {Lam}, {Laskar}, {Lecavelier des Etangs}, {Lendl}, {Lopez Pina}, {Luntzer}, {Magrin}, {Miller}, {Modrego
  Contreras}, {Mordasini}, {Munari}, {Murray}, {Nascimbeni}, {Ottacher}, {Ottensamer}, {Pagano}, {P{\'a}l}, {Pall{\'e}}, {Pasetti}, {Pedersen}, {Peter}, {Petrucci}, {Piotto}, {Pizarro-Rubio}, {Pollacco}, {Pribulla}, {Queloz}, {Ragazzoni}, {Rando}, {Rauer}, {Ribas}, {Sabin}, {Santos}, {Scandariato}, {Schanche}, {Schroffenegger}, {Scutt}, {Sebastian}, {S{\'e}gransan}, {Seli}, {Smith}, {Southworth}, {Standing}, {Szab{\'o}}, {Szak{\'a}ts}, {Thomas}, {Timmermans}, {Triaud}, {Udry}, {Van Grootel}, {Venturini}, {Villaver}, {Vink{\'o}}, {Walton}, {Wells}, \& {Wolter}}]{Fortier2024}
{Fortier}, A., {Simon}, A.~E., {Broeg}, C., {et~al.} 2024, \aap, 687, A302

\bibitem[{{Franciosini} {et~al.}(2018){Franciosini}, {Sacco}, {Jeffries}, {Damiani}, {Roccatagliata}, {Fedele}, \& {Randich}}]{Franciosini2018}
{Franciosini}, E., {Sacco}, G.~G., {Jeffries}, R.~D., {et~al.} 2018, \aap, 616, L12

\bibitem[{{Franciosini} {et~al.}(2022){Franciosini}, {Tognelli}, {Degl'Innocenti}, {Prada Moroni}, {Randich}, {Sacco}, {Magrini}, {Pancino}, {Lanzafame}, {Smiljanic}, {Prisinzano}, {Sanna}, {Roccatagliata}, {Bonito}, {de Laverny}, {Guti{\'e}rrez Albarr{\'a}n}, {Montes}, {Jim{\'e}nez-Esteban}, {Gilmore}, {Bergemann}, {Carraro}, {Damiani}, {Gonneau}, {Hourihane}, {Morbidelli}, {Worley}, \& {Zaggia}}]{Franciosini2022}
{Franciosini}, E., {Tognelli}, E., {Degl'Innocenti}, S., {et~al.} 2022, \aap, 659, A85

\bibitem[{{Fuhrmann} {et~al.}(2011){Fuhrmann}, {Chini}, {Hoffmeister}, \& {Stahl}}]{Fuhrmann2011}
{Fuhrmann}, K., {Chini}, R., {Hoffmeister}, V.~H., \& {Stahl}, O. 2011, \mnras, 416, 391

\bibitem[{{Gaia Collaboration} {et~al.}(2023{\natexlab{a}}){Gaia Collaboration}, {Arenou}, {Babusiaux}, {Barstow}, {Faigler}, {Jorissen}, {Kervella}, {Mazeh}, {Mowlavi}, {Panuzzo}, \& et~al.}]{Areanou2023}
{Gaia Collaboration}, {Arenou}, F., {Babusiaux}, C., {et~al.} 2023{\natexlab{a}}, \aap, 674, A34

\bibitem[{{Gaia Collaboration} {et~al.}(2023{\natexlab{b}}){Gaia Collaboration}, {De Ridder}, {Ripepi}, {Aerts}, {Palaversa}, \& {Eyer}}]{DeRidder2023}
{Gaia Collaboration}, {De Ridder}, J., {Ripepi}, V., {et~al.} 2023{\natexlab{b}}, \aap, 674, A36

\bibitem[{{Gardner} {et~al.}(2006){Gardner}, {Mather}, {Clampin}, {Doyon}, {Greenhouse}, {Hammel}, {Hutchings}, {Jakobsen}, {Lilly}, {Long}, {Lunine}, {McCaughrean}, {Mountain}, {Nella}, {Rieke}, {Rieke}, {Rix}, {Smith}, {Sonneborn}, {Stiavelli}, {Stockman}, {Windhorst}, \& {Wright}}]{Gardner2006}
{Gardner}, J.~P., {Mather}, J.~C., {Clampin}, M., {et~al.} 2006, \ssr, 123, 485

\bibitem[{{Gavras} {et~al.}(2023){Gavras}, {Rimoldini}, {Nienartowicz}, {de Fombelle}, {Holl}, {{\'A}brah{\'a}m}, {Audard}, {Carnerero}, {Clementini}, {De Ridder}, {Distefano}, {Garcia-Lario}, {Garofalo}, {K{\'o}sp{\'a}l}, {Kruszy{\'n}ska}, {Kun}, {Lecoeur-Ta{\"\i}bi}, {Marton}, {Mazeh}, {Mowlavi}, {Raiteri}, {Ripepi}, {Szabados}, {Zucker}, \& {Eyer}}]{Gavras2023}
{Gavras}, P., {Rimoldini}, L., {Nienartowicz}, K., {et~al.} 2023, \aap, 674, A22

\bibitem[{{Ge} {et~al.}(2022){Ge}, {Zhang}, {Zang}, {Deng}, {Mao}, {Xie}, {Liu}, {Zhou}, {Willis}, {Huang}, {Howell}, {Feng}, {Zhu}, {Yao}, {Liu}, {Aizawa}, {Zhu}, {Li}, {Ma}, {Ye}, {Yu}, {Xiang}, {Yu}, {Liu}, {Yang}, {Wang}, {Shi}, {Fang}, {Zong}, {Liu}, {Zhang}, {Zhang}, {El-Badry}, {Shen}, {Tam}, {Hu}, {Yang}, {Zou}, {Wu}, {Lei}, {Wei}, {Wu}, {Sun}, {Wang}, {Zhang}, {Xu}, {Yang}, {Li}, {Xiang}, {Wang}, {Wang}, {Zhang}, {Jia}, {Yuan}, {Zhang}, {Xuesong Wang}, {Gan}, {Wang}, {Zhao}, {Liu}, {Wei}, {Kang}, {Yang}, {Qi}, {Liu}, {Zhang}, {Zhu}, {Zhou}, {Zhang}, {Yu}, {Zhang}, {Li}, {Tang}, {Wang}, {Wang}, {Li}, {Cheng}, {Shen}, {Li}, {Pan}, {Yang}, {Gao}, {Song}, {Wang}, {Zhang}, {Chen}, {Wang}, {Zhang}, {Wang}, {Zeng}, {Zheng}, {Zhu}, {Guo}, {Zhang}, {Li}, {Wen}, {Feng}, {Chen}, {Chen}, {Han}, {Yang}, {Wang}, {Duan}, {Huang}, {Liang}, {Bi}, {Gai}, {Ge}, {Guo}, {Huang}, {Li}, {Li}, {Li}, {Yuxi}, {Lu}, {Rix}, {Shi}, {Song}, {Tang}, {Ting}, {Wu}, {Wu}, {Yang}, {Yin}, {Gould}, {Lee}, {Dong}, {Yee}, {Shvartzvald},
  {Yang}, {Kuang}, {Zhang}, {Liao}, {Qi}, {Yang}, {Zhang}, {Jiang}, {Ou}, {Li}, {Beck}, {Bedding}, {Campante}, {Chaplin}, {Christensen-Dalsgaard}, {Garc{\'\i}a}, {Gaulme}, {Gizon}, {Hekker}, {Huber}, {Khanna}, {Li}, {Mathur}, {Miglio}, {Mosser}, {Ong}, {Santos}, {Stello}, {Bowman}, {Lares-Martiz}, {Murphy}, {Niu}, {Ma}, {Moln{\'a}r}, {Fu}, {De Cat}, {Su}, \& {consortium}}]{Ge2022}
{Ge}, J., {Zhang}, H., {Zang}, W., {et~al.} 2022, arXiv e-prints, arXiv:2206.06693

\bibitem[{{Gentile Fusillo} {et~al.}(2021){Gentile Fusillo}, {Tremblay}, {Cukanovaite}, {Vorontseva}, {Lallement}, {Hollands}, {G{\"a}nsicke}, {Burdge}, {McCleery}, \& {Jordan}}]{GentileFusillo2021}
{Gentile Fusillo}, N.~P., {Tremblay}, P.~E., {Cukanovaite}, E., {et~al.} 2021, \mnras, 508, 3877

\bibitem[{{Giacalone} {et~al.}(2022){Giacalone}, {Dressing}, {Hedges}, {Kostov}, {Collins}, {Jensen}, {Yahalomi}, {Bieryla}, {Ciardi}, {Howell}, {Lillo-Box}, {Barkaoui}, {Winters}, {Matthews}, {Livingston}, {Quinn}, {Safonov}, {Cadieux}, {Furlan}, {Crossfield}, {Mandell}, {Gilbert}, {Kruse}, {Quintana}, {Ricker}, {Seager}, {Winn}, {Jenkins}, {Duffy Adkins}, {Baker}, {Barclay}, {Barrado}, {Batalha}, {Belinski}, {Benkhaldoun}, {Buchhave}, {Cacciapuoti}, {Charbonneau}, {Chontos}, {Christiansen}, {Cloutier}, {Collins}, {Conti}, {Cutting}, {Dixon}, {Doyon}, {Mufti}, {Esparza-Borges}, {Essack}, {Fukui}, {Gan}, {Gary}, {Ghachoui}, {Gillon}, {Girardin}, {Glidden}, {Gonzales}, {Guerra}, {Horch}, {He{\l}miniak}, {Howard}, {Huber}, {Irwin}, {Isopi}, {Jehin}, {Kagetani}, {Kane}, {Kawauchi}, {Kielkopf}, {Lewin}, {Luker}, {Lund}, {Mallia}, {Mao}, {Massey}, {Matson}, {Mireles}, {Mori}, {Murgas}, {Narita}, {O'Dwyer}, {Petigura}, {Polanski}, {Pozuelos}, {Palle}, {Parviainen}, {Plavchan}, {Relles}, {Robertson}, {Rose},
  {Rowden}, {Roy}, {Savel}, {Schlieder}, {Schnaible}, {Schwarz}, {Sefako}, {Selezneva}, {Skinner}, {Stockdale}, {Strakhov}, {Tan}, {Torres}, {Tronsgaard}, {Twicken}, {Vermilion}, {Waite}, {Walter}, {Wang}, {Ziegler}, \& {Zou}}]{Giacalone2022}
{Giacalone}, S., {Dressing}, C.~D., {Hedges}, C., {et~al.} 2022, \aj, 163, 99

\bibitem[{{Gilbert} {et~al.}(2020){Gilbert}, {Barclay}, {Schlieder}, {Quintana}, {Hord}, {Kostov}, {Lopez}, {Rowe}, {Hoffman}, {Walkowicz}, {Silverstein}, {Rodriguez}, {Vanderburg}, {Suissa}, {Airapetian}, {Clement}, {Raymond}, {Mann}, {Kruse}, {Lissauer}, {Col{\'o}n}, {Kopparapu}, {Kreidberg}, {Zieba}, {Collins}, {Quinn}, {Howell}, {Ziegler}, {Vrijmoet}, {Adams}, {Arney}, {Boyd}, {Brande}, {Burke}, {Cacciapuoti}, {Chance}, {Christiansen}, {Covone}, {Daylan}, {Dineen}, {Dressing}, {Essack}, {Fauchez}, {Galgano}, {Howe}, {Kaltenegger}, {Kane}, {Lam}, {Lee}, {Lewis}, {Logsdon}, {Mandell}, {Monsue}, {Mullally}, {Mullally}, {Paudel}, {Pidhorodetska}, {Plavchan}, {Reyes}, {Rinehart}, {Rojas-Ayala}, {Smith}, {Stassun}, {Tenenbaum}, {Vega}, {Villanueva}, {Wolf}, {Youngblood}, {Ricker}, {Vanderspek}, {Latham}, {Seager}, {Winn}, {Jenkins}, {Bakos}, {Brice{\~n}o}, {Ciardi}, {Cloutier}, {Conti}, {Couperus}, {Di Sora}, {Eisner}, {Everett}, {Gan}, {Hartman}, {Henry}, {Isopi}, {Jao}, {Jensen}, {Law}, {Mallia}, {Matson},
  {Shappee}, {Le Wood}, \& {Winters}}]{Gilbert2020}
{Gilbert}, E.~A., {Barclay}, T., {Schlieder}, J.~E., {et~al.} 2020, \aj, 160, 116

\bibitem[{{Gilbert} {et~al.}(2023){Gilbert}, {Vanderburg}, {Rodriguez}, {Hord}, {Clement}, {Barclay}, {Quintana}, {Schlieder}, {Kane}, {Jenkins}, {Twicken}, {Kunimoto}, {Vanderspek}, {Arney}, {Charbonneau}, {G{\"u}nther}, {Huang}, {Isopi}, {Kostov}, {Kristiansen}, {Latham}, {Mallia}, {Mamajek}, {Mireles}, {Quinn}, {Ricker}, {Schulte}, {Seager}, {Suissa}, {Winn}, {Youngblood}, \& {Zapparata}}]{Gilbert2023}
{Gilbert}, E.~A., {Vanderburg}, A., {Rodriguez}, J.~E., {et~al.} 2023, \apjl, 944, L35

\bibitem[{{Goldberg} {et~al.}(2023){Goldberg}, {Fabrycky}, {Martin}, {Albrecht}, {Deeg}, \& {Nowak}}]{Goldberg2023}
{Goldberg}, M., {Fabrycky}, D., {Martin}, D.~V., {et~al.} 2023, \mnras, 525, 4628

\bibitem[{{Goupil} {et~al.}(2024){Goupil}, {Catala}, {Samadi}, {Belkacem}, {Ouazzani}, {Reese}, {Appourchaux}, {Mathur}, {Cabrera}, {B{\"o}rner}, {Paproth}, {Moedas}, {Verma}, {Lebreton}, {Deal}, {Ballot}, {Chaplin}, {Christensen-Dalsgaard}, {Cunha}, {Lanza}, {Miglio}, {Morel}, {Serenelli}, {Mosser}, {Creevey}, {Moya}, {Garcia}, {Nielsen}, \& {Hatt}}]{Goupil2024}
{Goupil}, M.~J., {Catala}, C., {Samadi}, R., {et~al.} 2024, \aap, 683, A78

\bibitem[{{Grandjean} {et~al.}(2023){Grandjean}, {Lagrange}, {Meunier}, {Chauvin}, {Borgniet}, {Desidera}, {Galland}, {Kiefer}, {Messina}, {Iglesias}, {Nicholson}, {Pantoja}, {Rubini}, {Sedaghati}, {Sterzik}, \& {Zicher}}]{Grandjean2023}
{Grandjean}, A., {Lagrange}, A.~M., {Meunier}, N., {et~al.} 2023, \aap, 669, A12

\bibitem[{{Guerrero} {et~al.}(2021){Guerrero}, {Seager}, {Huang}, {Vanderburg}, {Garcia Soto}, {Mireles}, {Hesse}, {Fong}, {Glidden}, {Shporer}, {Latham}, {Collins}, {Quinn}, {Burt}, {Dragomir}, {Crossfield}, {Vanderspek}, {Fausnaugh}, {Burke}, {Ricker}, {Daylan}, {Essack}, {G{\"u}nther}, {Osborn}, {Pepper}, {Rowden}, {Sha}, {Villanueva}, {Yahalomi}, {Yu}, {Ballard}, {Batalha}, {Berardo}, {Chontos}, {Dittmann}, {Esquerdo}, {Mikal-Evans}, {Jayaraman}, {Krishnamurthy}, {Louie}, {Mehrle}, {Niraula}, {Rackham}, {Rodriguez}, {Rowden}, {Sousa-Silva}, {Watanabe}, {Wong}, {Zhan}, {Zivanovic}, {Christiansen}, {Ciardi}, {Swain}, {Lund}, {Mullally}, {Fleming}, {Rodriguez}, {Boyd}, {Quintana}, {Barclay}, {Col{\'o}n}, {Rinehart}, {Schlieder}, {Clampin}, {Jenkins}, {Twicken}, {Caldwell}, {Coughlin}, {Henze}, {Lissauer}, {Morris}, {Rose}, {Smith}, {Tenenbaum}, {Ting}, {Wohler}, {Bakos}, {Bean}, {Berta-Thompson}, {Bieryla}, {Bouma}, {Buchhave}, {Butler}, {Charbonneau}, {Doty}, {Ge}, {Holman}, {Howard}, {Kaltenegger}, {Kane},
  {Kjeldsen}, {Kreidberg}, {Lin}, {Minsky}, {Narita}, {Paegert}, {P{\'a}l}, {Palle}, {Sasselov}, {Spencer}, {Sozzetti}, {Stassun}, {Torres}, {Udry}, \& {Winn}}]{Guerrero2021}
{Guerrero}, N.~M., {Seager}, S., {Huang}, C.~X., {et~al.} 2021, \apjs, 254, 39

\bibitem[{{Guirado} {et~al.}(2011){Guirado}, {Marcaide}, {Mart{\'\i}-Vidal}, {Le Bouquin}, {Close}, {Cotton}, \& {Montalb{\'a}n}}]{Guirado2011}
{Guirado}, J.~C., {Marcaide}, J.~M., {Mart{\'\i}-Vidal}, I., {et~al.} 2011, \aap, 533, A106

\bibitem[{{Haywood} {et~al.}(2014){Haywood}, {Collier Cameron}, {Queloz}, {Barros}, {Deleuil}, {Fares}, {Gillon}, {Lanza}, {Lovis}, {Moutou}, {Pepe}, {Pollacco}, {Santerne}, {S{\'e}gransan}, \& {Unruh}}]{Haywood2014}
{Haywood}, R.~D., {Collier Cameron}, A., {Queloz}, D., {et~al.} 2014, \mnras, 443, 2517

\bibitem[{{Heitzmann} {et~al.}(2023){Heitzmann}, {Zhou}, {Quinn}, {Huang}, {Dong}, {Bouma}, {Dawson}, {Marsden}, {Wright}, {Petit}, {Collins}, {Barkaoui}, {Wittenmyer}, {Gillen}, {Brahm}, {Hobson}, {Hellier}, {Ziegler}, {Brice{\~n}o}, {Law}, {Mann}, {Howell}, {Gnilka}, {Littlefield}, {Latham}, {Lissauer}, {Newton}, {Krolikowski}, {Kerr}, {Rampalli}, {Douglas}, {Eisner}, {Guedj}, {Sun}, {Smit}, {Huten}, {Eschweiler}, {Abe}, {Guillot}, {Ricker}, {Vanderspek}, {Seager}, {Jenkins}, {Ting}, {Winn}, {Ciardi}, {Vanderburg}, {Burke}, {Rodriguez}, \& {Daylan}}]{Heitzmann2023}
{Heitzmann}, A., {Zhou}, G., {Quinn}, S.~N., {et~al.} 2023, \aj, 165, 121

\bibitem[{{Heller} {et~al.}(2022){Heller}, {Harre}, \& {Samadi}}]{Heller2022}
{Heller}, R., {Harre}, J.-V., \& {Samadi}, R. 2022, \aap, 665, A11

\bibitem[{{Hey} \& {Aerts}(2024)}]{HeyAerts2024}
{Hey}, D. \& {Aerts}, C. 2024, \aap, 688, A93

\bibitem[{{Hobson} {et~al.}(2021){Hobson}, {Brahm}, {Jord{\'a}n}, {Espinoza}, {Kossakowski}, {Henning}, {Rojas}, {Schlecker}, {Sarkis}, {Trifonov}, {Thorngren}, {Binnenfeld}, {Shahaf}, {Zucker}, {Ricker}, {Latham}, {Seager}, {Winn}, {Jenkins}, {Addison}, {Bouchy}, {Bowler}, {Briegal}, {Bryant}, {Collins}, {Daylan}, {Grieves}, {Horner}, {Huang}, {Kane}, {Kielkopf}, {McLean}, {Mengel}, {Nielsen}, {Okumura}, {Jones}, {Plavchan}, {Shporer}, {Smith}, {Tilbrook}, {Tinney}, {Twicken}, {Udry}, {Unger}, {West}, {Wittenmyer}, {Wohler}, {Torres}, \& {Wright}}]{Hobson2021}
{Hobson}, M.~J., {Brahm}, R., {Jord{\'a}n}, A., {et~al.} 2021, \aj, 161, 235

\bibitem[{{Hoffleit} \& {Warren}(1995)}]{Hoffleit1995}
{Hoffleit}, D. \& {Warren}, W.~H., J. 1995, VizieR Online Data Catalog, V/50

\bibitem[{{Howell} {et~al.}(2014){Howell}, {Sobeck}, {Haas}, {Still}, {Barclay}, {Mullally}, {Troeltzsch}, {Aigrain}, {Bryson}, {Caldwell}, {Chaplin}, {Cochran}, {Huber}, {Marcy}, {Miglio}, {Najita}, {Smith}, {Twicken}, \& {Fortney}}]{Howell2014}
{Howell}, S.~B., {Sobeck}, C., {Haas}, M., {et~al.} 2014, \pasp, 126, 398

\bibitem[{{Hunt} \& {Reffert}(2023)}]{Hunt2023}
{Hunt}, E.~L. \& {Reffert}, S. 2023, \aap, 673, A114

\bibitem[{{Ivezi{\'c}} {et~al.}(2019){Ivezi{\'c}}, {Kahn}, {Tyson}, {Abel}, {Acosta}, {Allsman}, {Alonso}, {AlSayyad}, {Anderson}, {Andrew}, \& et~al.}]{Ivezic2019}
{Ivezi{\'c}}, {\v{Z}}., {Kahn}, S.~M., {Tyson}, J.~A., {et~al.} 2019, \apj, 873, 111

\bibitem[{{Jannsen} {et~al.}(2024{\natexlab{a}}){Jannsen}, {De Ridder}, {Seynaeve}, {Regibo}, {Huygen}, {Royer}, {Paproth}, {Grie{\ss}bach}, {Samadi}, {Reese}, {Pertenais}, {Grolleau}, {Heller}, {Niemi}, {Cabrera}, {B{\"o}rner}, {Aigrain}, {McCormac}, {Verhoeve}, {Astier}, {Kutrowski}, {Vandenbussche}, {Tkachenko}, \& {Aerts}}]{Jannsen2024}
{Jannsen}, N., {De Ridder}, J., {Seynaeve}, D., {et~al.} 2024{\natexlab{a}}, \aap, 681, A18

\bibitem[{{Jannsen} {et~al.}(2024{\natexlab{b}}){Jannsen}, {Tkachenko}, {Royer}, {De Ridder}, {Seynaeve}, {Aerts}, {Aigrain}, {Plachy}, {Bodi}, {Uzundag}, {Bowman}, {Fritzewski}, {IJspeert}, {Li}, {Pedersen}, {Vanrespaille}, \& {Van Reeth}}]{Jannsen2024b}
{Jannsen}, N., {Tkachenko}, A., {Royer}, P., {et~al.} 2024{\natexlab{b}}, arXiv e-prints, arXiv:2412.10508

\bibitem[{{Jeffries} {et~al.}(2014){Jeffries}, {Jackson}, {Cottaar}, {Koposov}, {Lanzafame}, {Meyer}, {Prisinzano}, {Randich}, {Sacco}, {Brugaletta}, {Caramazza}, {Damiani}, {Franciosini}, {Frasca}, {Gilmore}, {Feltzing}, {Micela}, {Alfaro}, {Bensby}, {Pancino}, {Recio-Blanco}, {de Laverny}, {Lewis}, {Magrini}, {Morbidelli}, {Costado}, {Jofr{\'e}}, {Klutsch}, {Lind}, \& {Maiorca}}]{Jeffries2014}
{Jeffries}, R.~D., {Jackson}, R.~J., {Cottaar}, M., {et~al.} 2014, \aap, 563, A94

\bibitem[{{Johns} {et~al.}(2012){Johns}, {McCarthy}, {Raybould}, {Bouchez}, {Farahani}, {Filgueira}, {Jacoby}, {Shectman}, \& {Sheehan}}]{Johns2012}
{Johns}, M., {McCarthy}, P., {Raybould}, K., {et~al.} 2012, in Society of Photo-Optical Instrumentation Engineers (SPIE) Conference Series, Vol. 8444, Ground-based and Airborne Telescopes IV, ed. L.~M. {Stepp}, R.~{Gilmozzi}, \& H.~J. {Hall}, 84441H

\bibitem[{{Jontof-Hutter} {et~al.}(2021{\natexlab{a}}){Jontof-Hutter}, {Lissauer}, \& {Rowe}}]{JontofHutter2021b}
{Jontof-Hutter}, D., {Lissauer}, J., \& {Rowe}, J. 2021{\natexlab{a}}, in Plato Mission Conference 2021. Presentations and posters of the online PLATO Mission Conference 2021, 11

\bibitem[{{Jontof-Hutter} {et~al.}(2021{\natexlab{b}}){Jontof-Hutter}, {Wolfgang}, {Ford}, {Lissauer}, {Fabrycky}, \& {Rowe}}]{JontofHutter2021a}
{Jontof-Hutter}, D., {Wolfgang}, A., {Ford}, E.~B., {et~al.} 2021{\natexlab{b}}, \aj, 161, 246

\bibitem[{{Kasting} {et~al.}(1993){Kasting}, {Whitmire}, \& {Reynolds}}]{Kasting1993}
{Kasting}, J.~F., {Whitmire}, D.~P., \& {Reynolds}, R.~T. 1993, \icarus, 101, 108

\bibitem[{{Kaye} {et~al.}(1999){Kaye}, {Handler}, {Krisciunas}, {Poretti}, \& {Zerbi}}]{Kaye1999}
{Kaye}, A.~B., {Handler}, G., {Krisciunas}, K., {Poretti}, E., \& {Zerbi}, F.~M. 1999, \pasp, 111, 840

\bibitem[{{Konopacky} {et~al.}(2016){Konopacky}, {Rameau}, {Duch{\^e}ne}, {Filippazzo}, {Giorla Godfrey}, {Marois}, {Nielsen}, {Pueyo}, {Rafikov}, {Rice}, {Wang}, {Ammons}, {Bailey}, {Barman}, {Bulger}, {Bruzzone}, {Chilcote}, {Cotten}, {Dawson}, {De Rosa}, {Doyon}, {Esposito}, {Fitzgerald}, {Follette}, {Goodsell}, {Graham}, {Greenbaum}, {Hibon}, {Hung}, {Ingraham}, {Kalas}, {Lafreni{\`e}re}, {Larkin}, {Macintosh}, {Maire}, {Marchis}, {Marley}, {Matthews}, {Metchev}, {Millar-Blanchaer}, {Oppenheimer}, {Palmer}, {Patience}, {Perrin}, {Poyneer}, {Rajan}, {Rantakyr{\"o}}, {Savransky}, {Schneider}, {Sivaramakrishnan}, {Song}, {Soummer}, {Thomas}, {Wallace}, {Ward-Duong}, {Wiktorowicz}, \& {Wolff}}]{Konopacky2016}
{Konopacky}, Q.~M., {Rameau}, J., {Duch{\^e}ne}, G., {et~al.} 2016, \apjl, 829, L4

\bibitem[{{Kostov} {et~al.}(2020){Kostov}, {Orosz}, {Feinstein}, {Welsh}, {Cukier}, {Haghighipour}, {Quarles}, {Martin}, {Montet}, {Torres}, {Triaud}, {Barclay}, {Boyd}, {Briceno}, {Cameron}, {Correia}, {Gilbert}, {Gill}, {Gillon}, {Haqq-Misra}, {Hellier}, {Dressing}, {Fabrycky}, {Furesz}, {Jenkins}, {Kane}, {Kopparapu}, {Hod{\v{z}}i{\'c}}, {Latham}, {Law}, {Levine}, {Li}, {Lintott}, {Lissauer}, {Mann}, {Mazeh}, {Mardling}, {Maxted}, {Eisner}, {Pepe}, {Pepper}, {Pollacco}, {Quinn}, {Quintana}, {Rowe}, {Ricker}, {Rose}, {Seager}, {Santerne}, {S{\'e}gransan}, {Short}, {Smith}, {Standing}, {Tokovinin}, {Trifonov}, {Turner}, {Twicken}, {Udry}, {Vanderspek}, {Winn}, {Wolf}, {Ziegler}, {Ansorge}, {Barnet}, {Bergeron}, {Huten}, {Pappa}, \& {van der Straeten}}]{Kostov2020}
{Kostov}, V.~B., {Orosz}, J.~A., {Feinstein}, A.~D., {et~al.} 2020, \aj, 159, 253

\bibitem[{{Kostov} {et~al.}(2019){Kostov}, {Schlieder}, {Barclay}, {Quintana}, {Col{\'o}n}, {Brande}, {Collins}, {Feinstein}, {Hadden}, {Kane}, {Kreidberg}, {Kruse}, {Lam}, {Matthews}, {Montet}, {Pozuelos}, {Stassun}, {Winters}, {Ricker}, {Vanderspek}, {Latham}, {Seager}, {Winn}, {Jenkins}, {Afanasev}, {Armstrong}, {Arney}, {Boyd}, {Barentsen}, {Barkaoui}, {Batalha}, {Beichman}, {Bayliss}, {Burke}, {Burdanov}, {Cacciapuoti}, {Carson}, {Charbonneau}, {Christiansen}, {Ciardi}, {Clampin}, {Collins}, {Conti}, {Coughlin}, {Covone}, {Crossfield}, {Delrez}, {Domagal-Goldman}, {Dressing}, {Ducrot}, {Essack}, {Everett}, {Fauchez}, {Foreman-Mackey}, {Gan}, {Gilbert}, {Gillon}, {Gonzales}, {Hamann}, {Hedges}, {Hocutt}, {Hoffman}, {Horch}, {Horne}, {Howell}, {Hynes}, {Ireland}, {Irwin}, {Isopi}, {Jensen}, {Jehin}, {Kaltenegger}, {Kielkopf}, {Kopparapu}, {Lewis}, {Lopez}, {Lissauer}, {Mann}, {Mallia}, {Mandell}, {Matson}, {Mazeh}, {Monsue}, {Moran}, {Moran}, {Morley}, {Morris}, {Muirhead}, {Mukai}, {Mullally}, {Mullally},
  {Murray}, {Narita}, {Palle}, {Pidhorodetska}, {Quinn}, {Relles}, {Rinehart}, {Ritsko}, {Rodriguez}, {Rowden}, {Rowe}, {Sebastian}, {Sefako}, {Shahaf}, {Shporer}, {Ta{\~n}{\'o}n Reyes}, {Tenenbaum}, {Ting}, {Twicken}, {van Belle}, {Vega}, {Volosin}, {Walkowicz}, \& {Youngblood}}]{Kostov2019}
{Kostov}, V.~B., {Schlieder}, J.~E., {Barclay}, T., {et~al.} 2019, \aj, 158, 32

\bibitem[{{Kotoneva} {et~al.}(2005){Kotoneva}, {Innanen}, {Dawson}, {Wood}, \& {De Robertis}}]{Kotoneva2005}
{Kotoneva}, E., {Innanen}, K., {Dawson}, P.~C., {Wood}, P.~R., \& {De Robertis}, M.~M. 2005, \aap, 438, 957

\bibitem[{{Kunimoto} {et~al.}(2023){Kunimoto}, {Bryson}, {Daylan}, {Lissauer}, {Matesic}, {Mullally}, \& {Rowe}}]{Kunimoto2023}
{Kunimoto}, M., {Bryson}, S., {Daylan}, T., {et~al.} 2023, Research Notes of the American Astronomical Society, 7, 7

\bibitem[{{Lagrange} {et~al.}(2010){Lagrange}, {Bonnefoy}, {Chauvin}, {Apai}, {Ehrenreich}, {Boccaletti}, {Gratadour}, {Rouan}, {Mouillet}, {Lacour}, \& {Kasper}}]{Lagrange2010}
{Lagrange}, A.~M., {Bonnefoy}, M., {Chauvin}, G., {et~al.} 2010, Science, 329, 57

\bibitem[{{Lagrange} {et~al.}(2019){Lagrange}, {Meunier}, {Rubini}, {Keppler}, {Galland}, {Chapellier}, {Michel}, {Balona}, {Beust}, {Guillot}, {Grandjean}, {Borgniet}, {M{\'e}karnia}, {Wilson}, {Kiefer}, {Bonnefoy}, {Lillo-Box}, {Pantoja}, {Jones}, {Iglesias}, {Rodet}, {Diaz}, {Zapata}, {Abe}, \& {Schmider}}]{Lagrange2019}
{Lagrange}, A.~M., {Meunier}, N., {Rubini}, P., {et~al.} 2019, Nature Astronomy, 3, 1135

\bibitem[{{Lanza} {et~al.}(2011){Lanza}, {Boisse}, {Bouchy}, {Bonomo}, \& {Moutou}}]{Lanza2011}
{Lanza}, A.~F., {Boisse}, I., {Bouchy}, F., {Bonomo}, A.~S., \& {Moutou}, C. 2011, \aap, 533, A44

\bibitem[{{Li} {et~al.}(2020){Li}, {Van Reeth}, {Bedding}, {Murphy}, {Antoci}, {Ouazzani}, \& {Barbara}}]{GangLi2020}
{Li}, G., {Van Reeth}, T., {Bedding}, T.~R., {et~al.} 2020, \mnras, 491, 3586

\bibitem[{{Lorenzo-Oliveira} {et~al.}(2018){Lorenzo-Oliveira}, {Freitas}, {Mel{\'e}ndez}, {Bedell}, {Ram{\'\i}rez}, {Bean}, {Asplund}, {Spina}, {Dreizler}, {Alves-Brito}, \& {Casagrande}}]{LorenzoOliveira2018}
{Lorenzo-Oliveira}, D., {Freitas}, F.~C., {Mel{\'e}ndez}, J., {et~al.} 2018, \aap, 619, A73

\bibitem[{{Luhman} {et~al.}(2011){Luhman}, {Burgasser}, \& {Bochanski}}]{Luhman2011}
{Luhman}, K.~L., {Burgasser}, A.~J., \& {Bochanski}, J.~J. 2011, \apjl, 730, L9

\bibitem[{{Maciejewski}(2020)}]{Macie2020}
{Maciejewski}, G. 2020, \actaa, 70, 181

\bibitem[{{Magliano} {et~al.}(2024){Magliano}, {Covone}, {Nascimbeni}, {Inno}, {Vines}, {Kostov}, {Fiscale}, {Granata}, {Montalto}, {Pagano}, {Piotto}, \& {Saggese}}]{Magliano2024}
{Magliano}, C., {Covone}, G., {Nascimbeni}, V., {et~al.} 2024, \mnras, 528, 2851

\bibitem[{{Mantovan} {et~al.}(2022){Mantovan}, {Montalto}, {Piotto}, {Wilson}, {Collier Cameron}, {Majidi}, {Borsato}, {Granata}, \& {Nascimbeni}}]{Mantovan2022}
{Mantovan}, G., {Montalto}, M., {Piotto}, G., {et~al.} 2022, \mnras, 516, 4432

\bibitem[{{Marconi} {et~al.}(2022){Marconi}, {Abreu}, {Adibekyan}, {Alberti}, {Albrecht}, {Alcaniz}, {Aliverti}, {Allende Prieto}, {Alvarado G{\'o}mez}, {Amado}, \& et~al.}]{Marconi2022}
{Marconi}, A., {Abreu}, M., {Adibekyan}, V., {et~al.} 2022, in Society of Photo-Optical Instrumentation Engineers (SPIE) Conference Series, Vol. 12184, Ground-based and Airborne Instrumentation for Astronomy IX, ed. C.~J. {Evans}, J.~J. {Bryant}, \& K.~{Motohara}, 1218424

\bibitem[{{Marcussen} \& {Albrecht}(2022)}]{Marcussen2022}
{Marcussen}, M.~L. \& {Albrecht}, S.~H. 2022, \apj, 933, 227

\bibitem[{{Marton} {et~al.}(2023){Marton}, {{\'A}brah{\'a}m}, {Rimoldini}, {Audard}, {Kun}, {Nagy}, {K{\'o}sp{\'a}l}, {Szabados}, {Holl}, {Gavras}, {Mowlavi}, {Nienartowicz}, {de Fombelle}, {Lecoeur-Ta{\"\i}bi}, {Karbevska}, {Lario}, \& {Eyer}}]{Marton2023}
{Marton}, G., {{\'A}brah{\'a}m}, P., {Rimoldini}, L., {et~al.} 2023, \aap, 674, A21

\bibitem[{{Matuszewski} {et~al.}(2023){Matuszewski}, {Nettelmann}, {Cabrera}, {B{\"o}rner}, \& {Rauer}}]{Matuszewski2023}
{Matuszewski}, F., {Nettelmann}, N., {Cabrera}, J., {B{\"o}rner}, A., \& {Rauer}, H. 2023, \aap, 677, A133

\bibitem[{{Mayor} {et~al.}(2003){Mayor}, {Pepe}, {Queloz}, {Bouchy}, {Rupprecht}, {Lo Curto}, {Avila}, {Benz}, {Bertaux}, {Bonfils}, {Dall}, {Dekker}, {Delabre}, {Eckert}, {Fleury}, {Gilliotte}, {Gojak}, {Guzman}, {Kohler}, {Lizon}, {Longinotti}, {Lovis}, {Megevand}, {Pasquini}, {Reyes}, {Sivan}, {Sosnowska}, {Soto}, {Udry}, {van Kesteren}, {Weber}, \& {Weilenmann}}]{Mayor2003}
{Mayor}, M., {Pepe}, F., {Queloz}, D., {et~al.} 2003, The Messenger, 114, 20

\bibitem[{{Mayor} {et~al.}(2009){Mayor}, {Udry}, {Lovis}, {Pepe}, {Queloz}, {Benz}, {Bertaux}, {Bouchy}, {Mordasini}, \& {Segransan}}]{Mayor2009}
{Mayor}, M., {Udry}, S., {Lovis}, C., {et~al.} 2009, \aap, 493, 639

\bibitem[{{Mazeh} {et~al.}(2016){Mazeh}, {Holczer}, \& {Faigler}}]{Mazeh2016}
{Mazeh}, T., {Holczer}, T., \& {Faigler}, S. 2016, \aap, 589, A75

\bibitem[{{Mombarg} {et~al.}(2024){Mombarg}, {Aerts}, {Van Reeth}, \& {Hey}}]{Mombarg2024}
{Mombarg}, J. S.~G., {Aerts}, C., {Van Reeth}, T., \& {Hey}, D. 2024, \aap, 691, A131

\bibitem[{{Montalto} {et~al.}(2020){Montalto}, {Borsato}, {Granata}, {Lacedelli}, {Malavolta}, {Manthopoulou}, {Nardiello}, {Nascimbeni}, \& {Piotto}}]{Montalto2020}
{Montalto}, M., {Borsato}, L., {Granata}, V., {et~al.} 2020, \mnras, 498, 1726

\bibitem[{{Montalto} {et~al.}(2021){Montalto}, {Piotto}, {Marrese}, {Nascimbeni, V.}, {Prisinzano, L.}, {Granata, V.}, {Marinoni, S.}, {Desidera, S.}, {Ortolani, S.}, {Aerts, C.}, {Alei, E.}, {Altavilla, G.}, {Benatti, S.}, {B\"orner, A.}, {Cabrera, J.}, {Claudi, R.}, {Deleuil, M.}, {Fabrizio, M.}, {Gizon, L.}, {Goupil, M. J.}, {Heras, A. M.}, {Magrin, D.}, {Malavolta, L.}, {Mas-Hesse, J. M.}, {Pagano, I.}, {Paproth, C.}, {Pertenais, M.}, {Pollacco, D.}, {Ragazzoni, R.}, {Ramsay, G.}, {Rauer, H.}, \& {Udry, S.}}]{Montalto2021}
{Montalto}, M., {Piotto}, G., {Marrese}, P.~M., {et~al.} 2021, A\&A, 653, A98

\bibitem[{{Nardiello} {et~al.}(2020){Nardiello}, {Piotto}, {Deleuil}, {Malavolta}, {Montalto}, {Bedin}, {Borsato}, {Granata}, {Libralato}, \& {Manthopoulou}}]{Nardiello2020}
{Nardiello}, D., {Piotto}, G., {Deleuil}, M., {et~al.} 2020, \mnras, 495, 4924

\bibitem[{{Nascimbeni} {et~al.}(2022){Nascimbeni}, {Piotto}, {B{\"o}rner}, {Montalto}, {Marrese}, {Cabrera}, {Marinoni}, {Aerts}, {Altavilla}, {Benatti}, {Claudi}, {Deleuil}, {Desidera}, {Fabrizio}, {Gizon}, {Goupil}, {Granata}, {Heras}, {Magrin}, {Malavolta}, {Mas-Hesse}, {Ortolani}, {Pagano}, {Pollacco}, {Prisinzano}, {Ragazzoni}, {Ramsay}, {Rauer}, \& {Udry}}]{Nascimbeni2022}
{Nascimbeni}, V., {Piotto}, G., {B{\"o}rner}, A., {et~al.} 2022, \aap, 658, A31

\bibitem[{{Nascimbeni} {et~al.}(2016){Nascimbeni}, {Piotto}, {Ortolani}, {Giuffrida}, {Marrese}, {Magrin}, {Ragazzoni}, {Pagano}, {Rauer}, {Cabrera}, {Pollacco}, {Heras}, {Deleuil}, {Gizon}, \& {Granata}}]{Nascimbeni2016}
{Nascimbeni}, V., {Piotto}, G., {Ortolani}, S., {et~al.} 2016, \mnras, 463, 4210

\bibitem[{{Newton} {et~al.}(2021){Newton}, {Mann}, {Kraus}, {Livingston}, {Vanderburg}, {Curtis}, {Thao}, {Hawkins}, {Wood}, {Rizzuto}, {Soubkiou}, {Tofflemire}, {Zhou}, {Crossfield}, {Pearce}, {Collins}, {Conti}, {Tan}, {Villeneuva}, {Spencer}, {Dragomir}, {Quinn}, {Jensen}, {Collins}, {Stockdale}, {Cloutier}, {Hellier}, {Benkhaldoun}, {Ziegler}, {Brice{\~n}o}, {Law}, {Benneke}, {Christiansen}, {Gorjian}, {Kane}, {Kreidberg}, {Morales}, {Werner}, {Twicken}, {Levine}, {Ciardi}, {Guerrero}, {Hesse}, {Quintana}, {Shiao}, {Smith}, {Torres}, {Ricker}, {Vanderspek}, {Seager}, {Winn}, {Jenkins}, \& {Latham}}]{Newton2021}
{Newton}, E.~R., {Mann}, A.~W., {Kraus}, A.~L., {et~al.} 2021, \aj, 161, 65

\bibitem[{{Nisak} {et~al.}(2022){Nisak}, {White}, {Yep}, {Henry}, {Paredes}, {James}, \& {Jao}}]{Nisak2022}
{Nisak}, A.~H., {White}, R.~J., {Yep}, A., {et~al.} 2022, \aj, 163, 278

\bibitem[{{O'Brien} {et~al.}(2023){O'Brien}, {Tremblay}, {Gentile Fusillo}, {Hollands}, {G{\"a}nsicke}, {Koester}, {Pelisoli}, {Cukanovaite}, {Cunningham}, {Doyle}, {Elms}, {Farihi}, {Hermes}, {Holberg}, {Jordan}, {Klein}, {Kleinman}, {Manser}, {De Martino}, {Marsh}, {McCleery}, {Melis}, {Nitta}, {Parsons}, {Raddi}, {Rebassa-Mansergas}, {Schreiber}, {Silvotti}, {Steeghs}, {Toloza}, {Toonen}, {Torres}, {Weinberger}, \& {Zuckerman}}]{Obrien2023}
{O'Brien}, M.~W., {Tremblay}, P.~E., {Gentile Fusillo}, N.~P., {et~al.} 2023, \mnras, 518, 3055

\bibitem[{{Osborn} {et~al.}(2021){Osborn}, {Armstrong}, {Cale}, {Brahm}, {Wittenmyer}, {Dai}, {Crossfield}, {Bryant}, {Adibekyan}, {Cloutier}, {Collins}, {Delgado Mena}, {Fridlund}, {Hellier}, {Howell}, {King}, {Lillo-Box}, {Otegi}, {Sousa}, {Stassun}, {Matthews}, {Ziegler}, {Ricker}, {Vanderspek}, {Latham}, {Seager}, {Winn}, {Jenkins}, {Acton}, {Addison}, {Anderson}, {Ballard}, {Barrado}, {Barros}, {Batalha}, {Bayliss}, {Barclay}, {Benneke}, {Berberian}, {Bouchy}, {Bowler}, {Brice{\~n}o}, {Burke}, {Burleigh}, {Casewell}, {Ciardi}, {Collins}, {Cooke}, {Demangeon}, {D{\'\i}az}, {Dorn}, {Dragomir}, {Dressing}, {Dumusque}, {Espinoza}, {Figueira}, {Fulton}, {Furlan}, {Gaidos}, {Geneser}, {Gill}, {Goad}, {Gonzales}, {Gorjian}, {G{\"u}nther}, {Helled}, {Henderson}, {Henning}, {Hogan}, {Hojjatpanah}, {Horner}, {Howard}, {Hoyer}, {Huber}, {Isaacson}, {Jenkins}, {Jensen}, {Jord{\'a}n}, {Kane}, {Kidwell}, {Kielkopf}, {Law}, {Lendl}, {Lund}, {Matson}, {Mann}, {McCormac}, {Mengel}, {Morales}, {Nielsen}, {Okumura},
  {Osborn}, {Petigura}, {Plavchan}, {Pollacco}, {Quintana}, {Raynard}, {Robertson}, {Rose}, {Roy}, {Reefe}, {Santerne}, {Santos}, {Sarkis}, {Schlieder}, {Schwarz}, {Scott}, {Shporer}, {Smith}, {Stibbard}, {Stockdale}, {Str{\o}m}, {Twicken}, {Tan}, {Tanner}, {Teske}, {Tilbrook}, {Tinney}, {Udry}, {Villase{\~n}or}, {Vines}, {Wang}, {Weiss}, {West}, {Wheatley}, {Wright}, {Zhang}, \& {Zohrabi}}]{Osborn2021}
{Osborn}, A., {Armstrong}, D.~J., {Cale}, B., {et~al.} 2021, \mnras, 507, 2782

\bibitem[{{Pearce} {et~al.}(2022){Pearce}, {Launhardt}, {Ostermann}, {Kennedy}, {Gennaro}, {Booth}, {Krivov}, {Cugno}, {Henning}, {Quirrenbach}, {Barcucci}, {Matthews}, {Ruh}, \& {Stone}}]{Pearce2022}
{Pearce}, T.~D., {Launhardt}, R., {Ostermann}, R., {et~al.} 2022, \aap, 659, A135

\bibitem[{{Pearson}(2019)}]{Pearson2019}
{Pearson}, K.~A. 2019, \aj, 158, 243

\bibitem[{{Pecaut} \& {Mamajek}(2013)}]{Pecaut2013}
{Pecaut}, M.~J. \& {Mamajek}, E.~E. 2013, \apjs, 208, 9

\bibitem[{{Pedersen} {et~al.}(2021){Pedersen}, {Aerts}, {P{\'a}pics}, {Michielsen}, {Gebruers}, {Rogers}, {Molenberghs}, {Burssens}, {Garcia}, \& {Bowman}}]{Pedersen2021}
{Pedersen}, M.~G., {Aerts}, C., {P{\'a}pics}, P.~I., {et~al.} 2021, Nature Astronomy, 5, 715

\bibitem[{{Pepe} {et~al.}(2021){Pepe}, {Cristiani}, {Rebolo}, {Santos}, {Dekker}, {Cabral}, {Di Marcantonio}, {Figueira}, {Lo Curto}, {Lovis}, {Mayor}, {M{\'e}gevand}, {Molaro}, {Riva}, {Zapatero Osorio}, {Amate}, {Manescau}, {Pasquini}, {Zerbi}, {Adibekyan}, {Abreu}, {Affolter}, {Alibert}, {Aliverti}, {Allart}, {Allende Prieto}, {{\'A}lvarez}, {Alves}, {Avila}, {Baldini}, {Bandy}, {Barros}, {Benz}, {Bianco}, {Borsa}, {Bourrier}, {Bouchy}, {Broeg}, {Calderone}, {Cirami}, {Coelho}, {Conconi}, {Coretti}, {Cumani}, {Cupani}, {D'Odorico}, {Damasso}, {Deiries}, {Delabre}, {Demangeon}, {Dumusque}, {Ehrenreich}, {Faria}, {Fragoso}, {Genolet}, {Genoni}, {G{\'e}nova Santos}, {Gonz{\'a}lez Hern{\'a}ndez}, {Hughes}, {Iwert}, {Kerber}, {Knudstrup}, {Landoni}, {Lavie}, {Lillo-Box}, {Lizon}, {Maire}, {Martins}, {Mehner}, {Micela}, {Modigliani}, {Monteiro}, {Monteiro}, {Moschetti}, {Murphy}, {Nunes}, {Oggioni}, {Oliveira}, {Oshagh}, {Pall{\'e}}, {Pariani}, {Poretti}, {Rasilla}, {Rebord{\~a}o}, {Redaelli}, {Santana Tschudi},
  {Santin}, {Santos}, {S{\'e}gransan}, {Schmidt}, {Segovia}, {Sosnowska}, {Sozzetti}, {Sousa}, {Span{\`o}}, {Su{\'a}rez Mascare{\~n}o}, {Tabernero}, {Tenegi}, {Udry}, \& {Zanutta}}]{Pepe2021}
{Pepe}, F., {Cristiani}, S., {Rebolo}, R., {et~al.} 2021, \aap, 645, A96

\bibitem[{{Perryman} {et~al.}(2014){Perryman}, {Hartman}, {Bakos}, \& {Lindegren}}]{Perryman2014}
{Perryman}, M., {Hartman}, J., {Bakos}, G.~{\'A}., \& {Lindegren}, L. 2014, \apj, 797, 14

\bibitem[{{Pertenais} {et~al.}(2022){Pertenais}, {Ammler-von Eiff}, {Burresi}, {Cabrera}, {Farinato}, {Gorius}, {Huygen}, {Magrin}, {Martin Garcia}, {Munari}, {Regibo}, {Royer}, {Vandenbussche}, \& {van Kempen}}]{Pertenais2022}
{Pertenais}, M., {Ammler-von Eiff}, M., {Burresi}, M., {et~al.} 2022, in Society of Photo-Optical Instrumentation Engineers (SPIE) Conference Series, Vol. 12180, Space Telescopes and Instrumentation 2022: Optical, Infrared, and Millimeter Wave, ed. L.~E. {Coyle}, S.~{Matsuura}, \& M.~D. {Perrin}, 121804M

\bibitem[{{Pertenais} {et~al.}(2021){Pertenais}, {Cabrera}, {Paproth}, {Boerner}, {Griessbach}, {Mogulski}, \& {Rauer}}]{Pertenais2021}
{Pertenais}, M., {Cabrera}, J., {Paproth}, C., {et~al.} 2021, in Society of Photo-Optical Instrumentation Engineers (SPIE) Conference Series, Vol. 11852, Society of Photo-Optical Instrumentation Engineers (SPIE) Conference Series, 118524Y

\bibitem[{{Pope} {et~al.}(2019){Pope}, {White}, {Farr}, {Yu}, {Greklek-McKeon}, {Huber}, {Aerts}, {Aigrain}, {Bedding}, {Boyajian}, {Creevey}, \& {Hogg}}]{Pope2019}
{Pope}, B. J.~S., {White}, T.~R., {Farr}, W.~M., {et~al.} 2019, \apjs, 245, 8

\bibitem[{{Porterfield} {et~al.}(2016){Porterfield}, {Coe}, {Gonzaga}, {Anderson}, \& {Grogin}}]{Porterfield2016}
{Porterfield}, B., {Coe}, D., {Gonzaga}, S., {Anderson}, J., \& {Grogin}, N. 2016, {Here Be Dragons: Characterization of ACS/WFC Scattered Light Anomalies}, Instrument Science Report ACS 2016-6, 16 pages

\bibitem[{{Qian} {et~al.}(2012){Qian}, {Liu}, {Zhu}, {Dai}, {Fern{\'a}ndez Laj{\'u}s}, \& {Baume}}]{Qian2012}
{Qian}, S.~B., {Liu}, L., {Zhu}, L.~Y., {et~al.} 2012, \mnras, 422, L24

\bibitem[{{Randich} {et~al.}(2018){Randich}, {Tognelli}, {Jackson}, {Jeffries}, {Degl'Innocenti}, {Pancino}, {Re Fiorentin}, {Spagna}, {Sacco}, {Bragaglia}, {Magrini}, {Prada Moroni}, {Alfaro}, {Franciosini}, {Morbidelli}, {Roccatagliata}, {Bouy}, {Bravi}, {Jim{\'e}nez-Esteban}, {Jordi}, {Zari}, {Tautvai{\v{s}}iene}, {Drazdauskas}, {Mikolaitis}, {Gilmore}, {Feltzing}, {Vallenari}, {Bensby}, {Koposov}, {Korn}, {Lanzafame}, {Smiljanic}, {Bayo}, {Carraro}, {Costado}, {Heiter}, {Hourihane}, {Jofr{\'e}}, {Lewis}, {Monaco}, {Prisinzano}, {Sbordone}, {Sousa}, {Worley}, \& {Zaggia}}]{Randich2018}
{Randich}, S., {Tognelli}, E., {Jackson}, R., {et~al.} 2018, \aap, 612, A99

\bibitem[{{Rattanamala} {et~al.}(2023){Rattanamala}, {Awiphan}, {Komonjinda}, {Phriksee}, {Sappankum}, {A-thano}, {Chitchak}, {Rittipruk}, {Sawangwit}, {Poshyachinda}, {Reichart}, \& {Haislip}}]{Rattanamala2023}
{Rattanamala}, R., {Awiphan}, S., {Komonjinda}, S., {et~al.} 2023, \mnras, 523, 5086

\bibitem[{{Rauer} {et~al.}(2024){Rauer}, {Aerts}, {Cabrera}, {Deleuil}, {Erikson}, {Gizon}, {Goupil}, {Heras}, {Lorenzo-Alvarez}, {Marliani}, \& et~al.}]{Rauer2024}
{Rauer}, H., {Aerts}, C., {Cabrera}, J., {et~al.} 2024, arXiv e-prints, arXiv:2406.05447

\bibitem[{{Reyl{\'e}} {et~al.}(2021){Reyl{\'e}}, {Jardine}, {Fouqu{\'e}}, {Caballero}, {Smart}, \& {Sozzetti}}]{Reyle2021}
{Reyl{\'e}}, C., {Jardine}, K., {Fouqu{\'e}}, P., {et~al.} 2021, \aap, 650, A201

\bibitem[{{Ricker} {et~al.}(2015){Ricker}, {Winn}, {Vanderspek}, {Latham}, {Bakos}, {Bean}, {Berta-Thompson}, {Brown}, {Buchhave}, {Butler}, {Butler}, {Chaplin}, {Charbonneau}, {Christensen-Dalsgaard}, {Clampin}, {Deming}, {Doty}, {De Lee}, {Dressing}, {Dunham}, {Endl}, {Fressin}, {Ge}, {Henning}, {Holman}, {Howard}, {Ida}, {Jenkins}, {Jernigan}, {Johnson}, {Kaltenegger}, {Kawai}, {Kjeldsen}, {Laughlin}, {Levine}, {Lin}, {Lissauer}, {MacQueen}, {Marcy}, {McCullough}, {Morton}, {Narita}, {Paegert}, {Palle}, {Pepe}, {Pepper}, {Quirrenbach}, {Rinehart}, {Sasselov}, {Sato}, {Seager}, {Sozzetti}, {Stassun}, {Sullivan}, {Szentgyorgyi}, {Torres}, {Udry}, \& {Villasenor}}]{Ricker2015}
{Ricker}, G.~R., {Winn}, J.~N., {Vanderspek}, R., {et~al.} 2015, Journal of Astronomical Telescopes, Instruments, and Systems, 1, 014003

\bibitem[{{Robertson} {et~al.}(2015){Robertson}, {Roy}, \& {Mahadevan}}]{Robertson2015}
{Robertson}, P., {Roy}, A., \& {Mahadevan}, S. 2015, \apjl, 805, L22

\bibitem[{{Rodr{\'\i}guez Mart{\'\i}nez} {et~al.}(2020){Rodr{\'\i}guez Mart{\'\i}nez}, {Gaudi}, {Rodriguez}, {Zhou}, {Labadie-Bartz}, {Quinn}, {Penev}, {Tan}, {Latham}, {Paredes}, {Kielkopf}, {Addison}, {Wright}, {Teske}, {Howell}, {Ciardi}, {Ziegler}, {Stassun}, {Johnson}, {Eastman}, {Siverd}, {Beatty}, {Bouma}, {Bedding}, {Pepper}, {Winn}, {Lund}, {Villanueva}, {Stevens}, {Jensen}, {Kilby}, {Crane}, {Tokovinin}, {Everett}, {Tinney}, {Fausnaugh}, {Cohen}, {Bayliss}, {Bieryla}, {Cargile}, {Collins}, {Conti}, {Col{\'o}n}, {Curtis}, {Depoy}, {Evans}, {Feliz}, {Gregorio}, {Rothenberg}, {James}, {Joner}, {Kuhn}, {Manner}, {Khakpash}, {Marshall}, {McLeod}, {Penny}, {Reed}, {Relles}, {Stephens}, {Stockdale}, {Trueblood}, {Trueblood}, {Yao}, {Zambelli}, {Vanderspek}, {Seager}, {Jenkins}, {Henry}, {James}, {Jao}, {Wang}, {Butler}, {Thompson}, {Shectman}, {Wittenmyer}, {Bowler}, {Horner}, {Kane}, {Mengel}, {Morton}, {Okumura}, {Plavchan}, {Zhang}, {Scott}, {Matson}, {Mann}, {Dragomir}, {G{\"u}nther}, {Ting},
  {Glidden}, \& {Quintana}}]{Rodriguez2020}
{Rodr{\'\i}guez Mart{\'\i}nez}, R., {Gaudi}, B.~S., {Rodriguez}, J.~E., {et~al.} 2020, \aj, 160, 111

\bibitem[{{Saffe} {et~al.}(2021){Saffe}, {Miquelarena}, {Alacoria}, {Flores}, {Jaque Arancibia}, {Calvo}, {Mart{\'\i}n Girardi}, {Grosso}, \& {Collado}}]{Saffe2021}
{Saffe}, C., {Miquelarena}, P., {Alacoria}, J., {et~al.} 2021, \aap, 647, A49

\bibitem[{{Sahlmann} {et~al.}(2013){Sahlmann}, {Lazorenko}, {S{\'e}gransan}, {Mart{\'\i}n}, {Queloz}, {Mayor}, \& {Udry}}]{Sahlmann2013}
{Sahlmann}, J., {Lazorenko}, P.~F., {S{\'e}gransan}, D., {et~al.} 2013, \aap, 556, A133

\bibitem[{{Sanderson} {et~al.}(2022){Sanderson}, {Bonsor}, \& {Mustill}}]{Sanderson2022}
{Sanderson}, H., {Bonsor}, A., \& {Mustill}, A. 2022, \mnras, 517, 5835

\bibitem[{{Schmitt} {et~al.}(2019){Schmitt}, {Ioannidis}, {Robrade}, {Czesla}, \& {Schneider}}]{Schmitt2019}
{Schmitt}, J.~H.~M.~M., {Ioannidis}, P., {Robrade}, J., {Czesla}, S., \& {Schneider}, P.~C. 2019, \aap, 628, A79

\bibitem[{{Serenelli} {et~al.}(2021){Serenelli}, {Weiss}, {Aerts}, {Angelou}, {Baroch}, {Bastian}, {Beck}, {Bergemann}, {Bestenlehner}, {Czekala}, {Elias-Rosa}, {Escorza}, {Van Eylen}, {Feuillet}, {Gandolfi}, {Gieles}, {Girardi}, {Lebreton}, {Lodieu}, {Martig}, {Miller Bertolami}, {Mombarg}, {Morales}, {Moya}, {Nsamba}, {Pavlovski}, {Pedersen}, {Ribas}, {Schneider}, {Silva Aguirre}, {Stassun}, {Tolstoy}, {Tremblay}, \& {Zwintz}}]{Serenelli2021}
{Serenelli}, A., {Weiss}, A., {Aerts}, C., {et~al.} 2021, \aapr, 29, 4

\bibitem[{{Serrano} {et~al.}(2022){Serrano}, {Gandolfi}, {Mustill}, {Barrag{\'a}n}, {Korth}, {Dai}, {Redfield}, {Fridlund}, {Lam}, {D{\'\i}az}, {Grziwa}, {Collins}, {Livingston}, {Cochran}, {Hellier}, {Bellomo}, {Trifonov}, {Rodler}, {Alarcon}, {Jenkins}, {Latham}, {Ricker}, {Seager}, {Vanderspeck}, {Winn}, {Albrecht}, {Collins}, {Csizmadia}, {Daylan}, {Deeg}, {Esposito}, {Fausnaugh}, {Georgieva}, {Goffo}, {Guenther}, {Hatzes}, {Howell}, {Jensen}, {Luque}, {Mann}, {Murgas}, {Osborne}, {Palle}, {Persson}, {Rowden}, {Rudat}, {Smith}, {Twicken}, {Van Eylen}, \& {Ziegler}}]{Serrano2022}
{Serrano}, L.~M., {Gandolfi}, D., {Mustill}, A.~J., {et~al.} 2022, Nature Astronomy, 6, 736

\bibitem[{{Shporer}(2017)}]{Shporer2017}
{Shporer}, A. 2017, \pasp, 129, 072001

\bibitem[{{Silverstein} {et~al.}(2024){Silverstein}, {Barclay}, {Schlieder}, {Collins}, {Schwarz}, {Hord}, {Rowe}, {Kruse}, {Astudillo-Defru}, {Bonfils}, {Caldwell}, {Charbonneau}, {Cloutier}, {Collins}, {Daylan}, {Fong}, {Jenkins}, {Kunimoto}, {McDermott}, {Murgas}, {Palle}, {Ricker}, {Seager}, {Shporer}, {Tey}, {Vanderspek}, \& {Winn}}]{Silverstein2024}
{Silverstein}, M.~L., {Barclay}, T., {Schlieder}, J.~E., {et~al.} 2024, \aj, 167, 255

\bibitem[{{Singh} {et~al.}(2019){Singh}, {Scandariato}, \& {Pagano}}]{Singh2019}
{Singh}, V., {Scandariato}, G., \& {Pagano}, I. 2019, \mnras, 486, 5867

\bibitem[{{Southworth}(2011)}]{Southworth2011}
{Southworth}, J. 2011, \mnras, 417, 2166

\bibitem[{{Southworth}(2015)}]{Southworth2015}
{Southworth}, J. 2015, in Astronomical Society of the Pacific Conference Series, Vol. 496, Living Together: Planets, Host Stars and Binaries, ed. S.~M. {Rucinski}, G.~{Torres}, \& M.~{Zejda}, 164

\bibitem[{{Sozzetti} {et~al.}(2014){Sozzetti}, {Giacobbe}, {Lattanzi}, {Micela}, {Morbidelli}, \& {Tinetti}}]{Sozzetti2014}
{Sozzetti}, A., {Giacobbe}, P., {Lattanzi}, M.~G., {et~al.} 2014, \mnras, 437, 497

\bibitem[{{Sozzetti} {et~al.}(2023){Sozzetti}, {Giacobbe}, {Lattanzi}, \& {Pinamonti}}]{Sozzetti2023}
{Sozzetti}, A., {Giacobbe}, P., {Lattanzi}, M.~G., \& {Pinamonti}, M. 2023, \mnras, 520, 1748

\bibitem[{{Standing} {et~al.}(2023){Standing}, {Sairam}, {Martin}, {Triaud}, {Correia}, {Coleman}, {Baycroft}, {Kunovac}, {Boisse}, {Cameron}, {Dransfield}, {Faria}, {Gillon}, {Hara}, {Hellier}, {Howard}, {Lane}, {Mardling}, {Maxted}, {Miller}, {Nelson}, {Orosz}, {Pepe}, {Santerne}, {Sebastian}, {Udry}, \& {Welsh}}]{Standing2023}
{Standing}, M.~R., {Sairam}, L., {Martin}, D.~V., {et~al.} 2023, Nature Astronomy, 7, 702

\bibitem[{{Stassun} {et~al.}(2019){Stassun}, {Oelkers}, {Paegert}, {Torres}, {Pepper}, {De Lee}, {Collins}, {Latham}, {Muirhead}, {Chittidi}, {Rojas-Ayala}, {Fleming}, {Rose}, {Tenenbaum}, {Ting}, {Kane}, {Barclay}, {Bean}, {Brassuer}, {Charbonneau}, {Ge}, {Lissauer}, {Mann}, {McLean}, {Mullally}, {Narita}, {Plavchan}, {Ricker}, {Sasselov}, {Seager}, {Sharma}, {Shiao}, {Sozzetti}, {Stello}, {Vanderspek}, {Wallace}, \& {Winn}}]{Stassun2019}
{Stassun}, K.~G., {Oelkers}, R.~J., {Paegert}, M., {et~al.} 2019, \aj, 158, 138

\bibitem[{{Stassun} {et~al.}(2018){Stassun}, {Oelkers}, {Pepper}, {Paegert}, {De Lee}, {Torres}, {Latham}, {Charpinet}, {Dressing}, {Huber}, {Kane}, {L{\'e}pine}, {Mann}, {Muirhead}, {Rojas-Ayala}, {Silvotti}, {Fleming}, {Levine}, \& {Plavchan}}]{Stassun2018}
{Stassun}, K.~G., {Oelkers}, R.~J., {Pepper}, J., {et~al.} 2018, \aj, 156, 102

\bibitem[{{Taylor}(2005)}]{Taylor2005}
{Taylor}, M.~B. 2005, in Astronomical Society of the Pacific Conference Series, Vol. 347, Astronomical Data Analysis Software and Systems XIV, ed. P.~{Shopbell}, M.~{Britton}, \& R.~{Ebert}, 29

\bibitem[{{Taylor}(2006)}]{Taylor2006}
{Taylor}, M.~B. 2006, in Astronomical Society of the Pacific Conference Series, Vol. 351, Astronomical Data Analysis Software and Systems XV, ed. C.~{Gabriel}, C.~{Arviset}, D.~{Ponz}, \& S.~{Enrique}, 666

\bibitem[{{Teske} {et~al.}(2021){Teske}, {Wang}, {Wolfgang}, {Gan}, {Plotnykov}, {Armstrong}, {Butler}, {Cale}, {Crane}, {Howard}, {Jensen}, {Law}, {Shectman}, {Plavchan}, {Valencia}, {Vanderburg}, {Ricker}, {Vanderspek}, {Latham}, {Seager}, {Winn}, {Jenkins}, {Adibekyan}, {Barrado}, {Barros}, {Benkhaldoun}, {Brown}, {Bryant}, {Burt}, {Caldwell}, {Charbonneau}, {Cloutier}, {Collins}, {Collins}, {Colon}, {Conti}, {Demangeon}, {Eastman}, {Elmufti}, {Feng}, {Flowers}, {Guerrero}, {Hojjatpanah}, {Irwin}, {Isopi}, {Lillo-Box}, {Mallia}, {Massey}, {Mori}, {Mullally}, {Narita}, {Nishiumi}, {Osborn}, {Paegert}, {de Leon}, {Quinn}, {Reefe}, {Schwarz}, {Shporer}, {Soubkiou}, {Sousa}, {Stockdale}, {Str{\o}m}, {Tan}, {Tang}, {Tenenbaum}, {Wheatley}, {Wittrock}, {Yahalomi}, \& {Zohrabi}}]{Teske2021}
{Teske}, J., {Wang}, S.~X., {Wolfgang}, A., {et~al.} 2021, \apjs, 256, 33

\bibitem[{{Tinetti} {et~al.}(2018){Tinetti}, {Drossart}, {Eccleston}, {Hartogh}, {Heske}, {Leconte}, {Micela}, {Ollivier}, {Pilbratt}, {Puig}, \& et~al.}]{Tinetti2018}
{Tinetti}, G., {Drossart}, P., {Eccleston}, P., {et~al.} 2018, Experimental Astronomy, 46, 135

\bibitem[{{Tinetti} {et~al.}(2021){Tinetti}, {Eccleston}, {Haswell}, {Lagage}, {Leconte}, {L{\"u}ftinger}, {Micela}, {Min}, {Pilbratt}, {Puig}, \& et~al.}]{Tinetti2021}
{Tinetti}, G., {Eccleston}, P., {Haswell}, C., {et~al.} 2021, arXiv e-prints, arXiv:2104.04824

\bibitem[{{Trifonov} {et~al.}(2019){Trifonov}, {Rybizki}, \& {K{\"u}rster}}]{Trifonov2019}
{Trifonov}, T., {Rybizki}, J., \& {K{\"u}rster}, M. 2019, \aap, 622, L7

\bibitem[{{Tuomi} {et~al.}(2013){Tuomi}, {Anglada-Escud{\'e}}, {Gerlach}, {Jones}, {Reiners}, {Rivera}, {Vogt}, \& {Butler}}]{Tuomi2013}
{Tuomi}, M., {Anglada-Escud{\'e}}, G., {Gerlach}, E., {et~al.} 2013, \aap, 549, A48

\bibitem[{{Vach} {et~al.}(2022){Vach}, {Quinn}, {Vanderburg}, {Kane}, {Collins}, {Kraus}, {Zhou}, {Medina}, {Schwarz}, {Collins}, {Conti}, {Stockdale}, {Massey}, {Suarez}, {Guillot}, {Mekarnia}, {Abe}, {Dransfield}, {Crouzet}, {Triaud}, {Schmider}, {Agabi}, {Buttu}, {Hellier}, {Furlan}, {Gnilka}, {Howell}, {Ziegler}, {Brice{\~n}o}, {Law}, {Mann}, {Rudat}, {Colon}, {Rose}, {Kunimoto}, {G{\"u}nther}, {Charbonneau}, {Ciardi}, {Ricker}, {Vanderspek}, {Latham}, {Seager}, {Winn}, \& {Jenkins}}]{Vach2022}
{Vach}, S., {Quinn}, S.~N., {Vanderburg}, A., {et~al.} 2022, \aj, 164, 71

\bibitem[{{Vaulato} {et~al.}(2022){Vaulato}, {Nascimbeni}, \& {Piotto}}]{Vaulato2022}
{Vaulato}, V., {Nascimbeni}, V., \& {Piotto}, G. 2022, \aap, 668, A110

\bibitem[{{Vejar} {et~al.}(2021){Vejar}, {Schuler}, \& {Stassun}}]{Vejar2021}
{Vejar}, G., {Schuler}, S.~C., \& {Stassun}, K.~G. 2021, \apj, 919, 100

\bibitem[{{Verhoeve} {et~al.}(2016){Verhoeve}, {Prod'homme}, {Oosterbroek}, {Duvet}, {Beaufort}, {Blommaert}, {Butler}, {Heijnen}, {Lemmel}, {van der Luijt}, {Smit}, \& {Visser}}]{Verhoeve2016}
{Verhoeve}, P., {Prod'homme}, T., {Oosterbroek}, T., {et~al.} 2016, in Society of Photo-Optical Instrumentation Engineers (SPIE) Conference Series, Vol. 9915, High Energy, Optical, and Infrared Detectors for Astronomy VII, ed. A.~D. {Holland} \& J.~{Beletic}, 99150Z

\bibitem[{{Watson} {et~al.}(2006){Watson}, {Henden}, \& {Price}}]{Watson2006}
{Watson}, C.~L., {Henden}, A.~A., \& {Price}, A. 2006, Society for Astronomical Sciences Annual Symposium, 25, 47

\bibitem[{{Wenger} {et~al.}(2000){Wenger}, {Ochsenbein}, {Egret}, {Dubois}, {Bonnarel}, {Borde}, {Genova}, {Jasniewicz}, {Lalo{\"e}}, {Lesteven}, \& {Monier}}]{Wenger2000}
{Wenger}, M., {Ochsenbein}, F., {Egret}, D., {et~al.} 2000, \aaps, 143, 9

\bibitem[{{West} {et~al.}(2019){West}, {Gillen}, {Bayliss}, {Burleigh}, {Delrez}, {G{\"u}nther}, {Hodgkin}, {Jackman}, {Jenkins}, {King}, {McCormac}, {Nielsen}, {Raynard}, {Smith}, {Soto}, {Turner}, {Wheatley}, {Almleaky}, {Armstrong}, {Belardi}, {Bouchy}, {Briegal}, {Burdanov}, {Cabrera}, {Casewell}, {Chaushev}, {Chazelas}, {Chote}, {Cooke}, {Csizmadia}, {Ducrot}, {Eigm{\"u}ller}, {Erikson}, {Foxell}, {G{\"a}nsicke}, {Gillon}, {Goad}, {Jehin}, {Lambert}, {Longstaff}, {Louden}, {Moyano}, {Murray}, {Pollacco}, {Queloz}, {Rauer}, {Sohy}, {Thompson}, {Udry}, {Walker}, \& {Watson}}]{West2019}
{West}, R.~G., {Gillen}, E., {Bayliss}, D., {et~al.} 2019, \mnras, 486, 5094

\bibitem[{{White} {et~al.}(2017){White}, {Pope}, {Antoci}, {P{\'a}pics}, {Aerts}, {Gies}, {Gordon}, {Huber}, {Schaefer}, {Aigrain}, {Albrecht}, {Barclay}, {Barentsen}, {Beck}, {Bedding}, {Fredslund Andersen}, {Grundahl}, {Howell}, {Ireland}, {Murphy}, {Nielsen}, {Silva Aguirre}, \& {Tuthill}}]{White2017}
{White}, T.~R., {Pope}, B.~J.~S., {Antoci}, V., {et~al.} 2017, \mnras, 471, 2882

\bibitem[{{Wilson} {et~al.}(2021){Wilson}, {Gibson}, {Lothringer}, {Sing}, {Mikal-Evans}, {de Mooij}, {Nikolov}, \& {Watson}}]{Wilson2021}
{Wilson}, J., {Gibson}, N.~P., {Lothringer}, J.~D., {et~al.} 2021, \mnras, 503, 4787

\bibitem[{{Yee} {et~al.}(2023){Yee}, {Winn}, {Hartman}, {Bouma}, {Zhou}, {Quinn}, {Latham}, {Bieryla}, {Rodriguez}, {Collins}, {Alfaro}, {Barkaoui}, {Beard}, {Belinski}, {Benkhaldoun}, {Benni}, {Bernacki}, {Boyle}, {Butler}, {Caldwell}, {Chontos}, {Christiansen}, {Ciardi}, {Collins}, {Conti}, {Crane}, {Daylan}, {Dressing}, {Eastman}, {Essack}, {Evans}, {Everett}, {Fajardo-Acosta}, {For{\'e}s-Toribio}, {Furlan}, {Ghachoui}, {Gillon}, {Hellier}, {Helm}, {Howard}, {Howell}, {Isaacson}, {Jehin}, {Jenkins}, {Jensen}, {Kielkopf}, {Laloum}, {Leonhardes-Barboza}, {Lewin}, {Logsdon}, {Lubin}, {Lund}, {MacDougall}, {Mann}, {Maslennikova}, {Massey}, {McLeod}, {Mu{\~n}oz}, {Newman}, {Orlov}, {Plavchan}, {Popowicz}, {Pozuelos}, {Pritchard}, {Radford}, {Reefe}, {Ricker}, {Rudat}, {Safonov}, {Schwarz}, {Schweiker}, {Scott}, {Seager}, {Shectman}, {Stockdale}, {Tan}, {Teske}, {Thomas}, {Timmermans}, {Vanderspek}, {Vermilion}, {Watanabe}, {Weiss}, {West}, {Van Zandt}, {Zejmo}, \& {Ziegler}}]{Yee2023}
{Yee}, S.~W., {Winn}, J.~N., {Hartman}, J.~D., {et~al.} 2023, \apjs, 265, 1

\bibitem[{{Zakhozhay} {et~al.}(2022){Zakhozhay}, {Launhardt}, {M{\"u}ller}, {Brems}, {Eigenthaler}, {Gennaro}, {Hempel}, {Hempel}, {Henning}, {Kennedy}, {Kim}, {K{\"u}rster}, {Lachaume}, {Manerikar}, {Patel}, {Pavlov}, {Reffert}, \& {Trifonov}}]{Zakhozhay2022}
{Zakhozhay}, O.~V., {Launhardt}, R., {M{\"u}ller}, A., {et~al.} 2022, \aap, 667, A63

\bibitem[{{Zhou} {et~al.}(2019){Zhou}, {Bakos}, {Bayliss}, {Bento}, {Bhatti}, {Brahm}, {Csubry}, {Espinoza}, {Hartman}, {Henning}, {Jord{\'a}n}, {Mancini}, {Penev}, {Rabus}, {Sarkis}, {Suc}, {de Val-Borro}, {Rodriguez}, {Osip}, {Kedziora-Chudczer}, {Bailey}, {Tinney}, {Durkan}, {L{\'a}z{\'a}r}, {Papp}, \& {S{\'a}ri}}]{Zhou2019}
{Zhou}, G., {Bakos}, G.~{\'A}., {Bayliss}, D., {et~al.} 2019, \aj, 157, 31

\bibitem[{{Zuckerman} {et~al.}(2004){Zuckerman}, {Song}, \& {Bessell}}]{Zuckerman2004}
{Zuckerman}, B., {Song}, I., \& {Bessell}, M.~S. 2004, \apjl, 613, L65

\bibitem[{{Zwintz}(2024)}]{Zwinze2024}
{Zwintz}, K. 2024, in EAS2024, 1836

\end{thebibliography}

\appendix

\section{The LOPS2 footprint}
\label{sec:favorite}

A software tool to check whether a given line of sight passes through the PLATO fields will be made available in the near future by ESA with support from the PLATO Consortium. Meanwhile, for any non-critical purpose, the footprint of LOPS2 can be reasonably approximated by the intersection between a spherical circle with $r=28.1^\circ$ and a ``square'' made with arcs of great circles (cf.~Fig.~\ref{fov}). The corresponding TOPCAT expression is:
\begin{flalign}
&\texttt{inSkyEllipse(l,b,255.9375,-24.62432,28.1,} \nonumber \\ & \texttt{28.1,0) \&\& inSkyPolygon(l,b,288.6075,-44.1,} \\ & \texttt{223.2675,-44.1,233.1675,-0.25,278.7075,-0.25)} \nonumber
\end{flalign}
where $(l, b)$ are the galactic coordinates of the target. This expression can be translated into any other language of choice since it contains the coordinates of the five pivot points involved (the center of the circles plus the four vertexes of the square). 

A more accurate representation of LOPS2, also including the NCAM sub-regions, can be built by approximating the field on a level-9 HEALPix grid, and then converting the footprints into  IVOA\footnote{International Virtual Observatory Alliance, \url{https://ivoa.net/}}-compliant Multi-Order Coverage (MOC) regions \citep{Fernique2014}, through the \texttt{mocpy} code\footnote{\url{https://cds-astro.github.io/mocpy/}}. The MOC regions for 6, 12, 18, 24 NCAMs and the underlying HEALPix grid are available online\footnote{\href{https://zenodo.org/records/14720127}{zenodo14720127}}. We emphasize that the actual sky coverage could be slightly different due to several factors (the in-orbit optical performances, the relative co-alignment of the cameras, the pointing accuracy, etc.) so all these regions should be used with some caution especially when a target of interest is close the the external boundaries or to the inner gaps.


\section{Additional tables and figures}

\begin{table}[h!]\centering\small
\caption{Glossary of acronyms used throughout this article.}
\begin{tabular}{lp{6cm}}
\hline
Acronym & Description \\
\hline\hline
asPIC & All-sky PLATO Input Catalog \citepalias{Montalto2021}\\
CVZ & Continuous Viewing Zone \\
DEB & Detached Eclipsing Binary\\
DIA & Difference Image Analysis \\
EB & Eclipsing Binary \\
FPR & False Positive Rate \\
FCAM & PLATO Fast camera \\
FOV & Field Of View \\
GC & Globular Cluster \\
GO & PLATO Guest Observing program\\
HJ & hot Jupiter \\
HZ & Habitable Zone \\
LMC & Large Magellanic Cloud\\
LOP & Long-duration Observation Phase \\
LOPN & LOP field North\\
LOPN1 & Current LOPN proposal (this work) \\
LOPS & LOP field South\\
LOPS1 & Current LOPS proposal (this work) \\
\citetalias{Montalto2021} & \citet{Montalto2021} \\
\citetalias{Nascimbeni2022} & \citet{Nascimbeni2022} \\
NCAM & PLATO Normal camera \\
NSR & Noise-to-Signal ratio \\
ORM & Observatorio Roque de Los Muchachos \\
OC & Open Cluster \\
P/L & PLATO Payload \\
PF & PLATO field \\
PIC & PLATO Input Catalog \citepalias{Montalto2021} \\
PLATO & PLAnetary Transits and Oscillations of stars \citep{Rauer2024} \\PPT & PLATO Performance Team \\
PSF & Point Spread Function \\
RV & Radial velocity \\
SNR & Signal-to-Noise ratio \\
SOP & Step and stare Observation Phase \\
SciRD & PLATO Scientific Requirements Document \\
SRJD & PLATO Scientific Requirements Justification Document \\
SWT & ESA PLATO Science Working Team \\
TCP & TESS candidate planets \\
TTV & Transit timing variations \\
WJ & warm Jupiter \\
\hline
\end{tabular}\label{table:glossary}
\end{table}

\begin{figure*}[t]
    \centering
    \includegraphics[width=0.97\columnwidth]{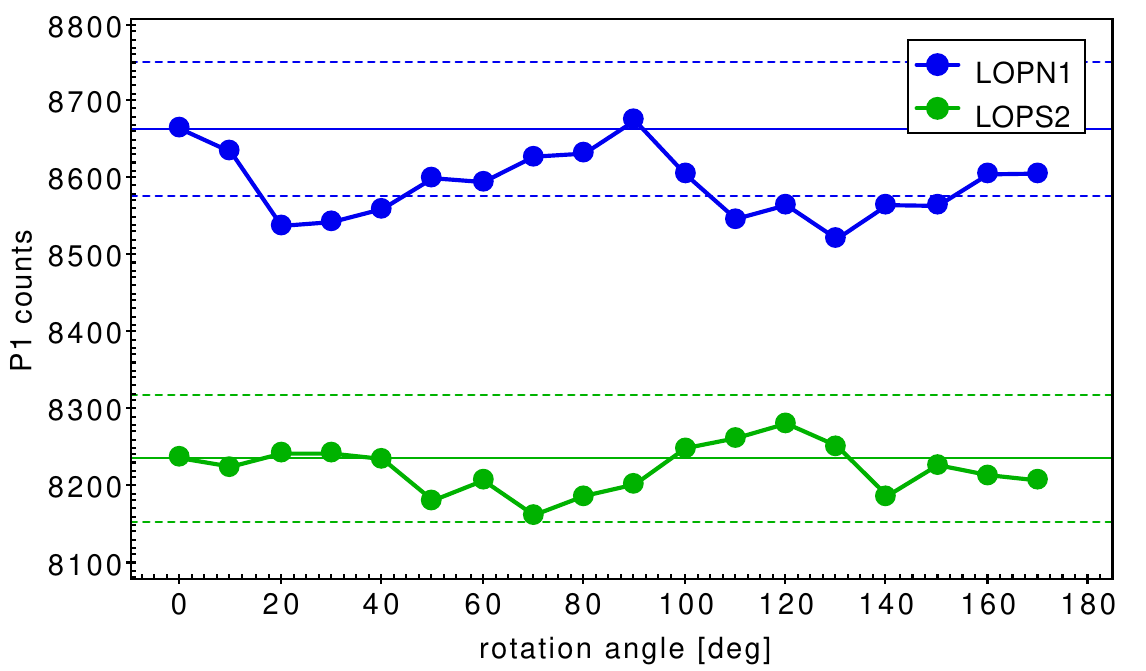}\hspace{5mm}
    \includegraphics[width=0.97\columnwidth]{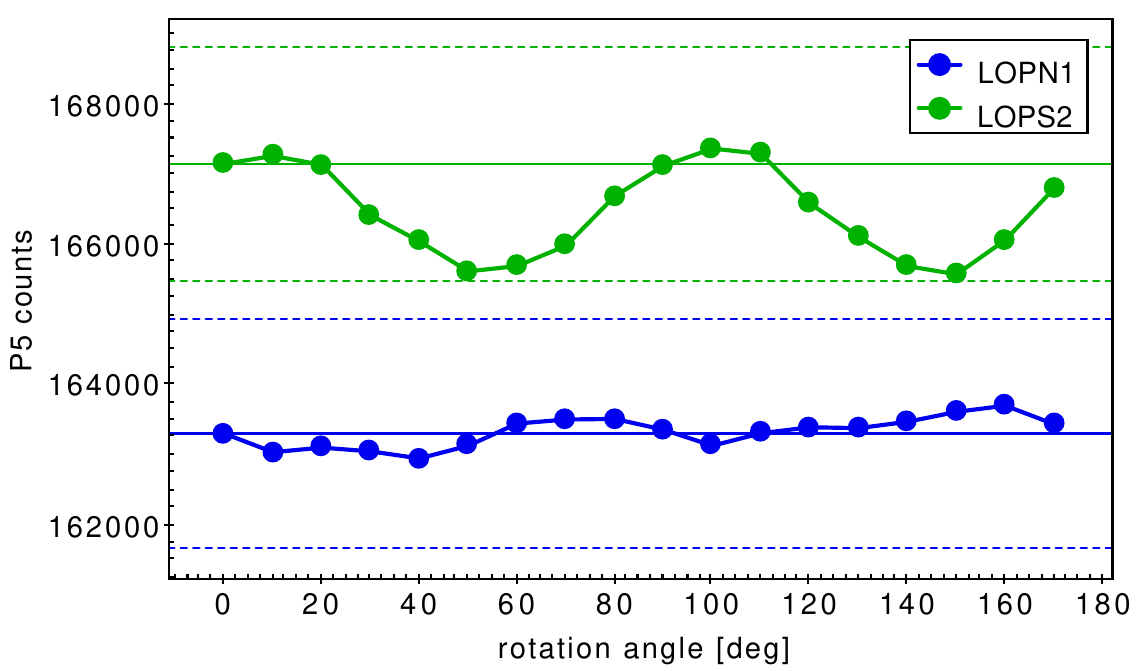}\\ 
    \caption{Optimization of the PLATO field rotation angle. The number of P1 (left panel) and P5 (right panel) targets are shown as a function of the rotation angle, for hypothetical PLATO fields centered at the LOPN1 (blue points and lines) or LOPS2 (green points and lines) coordinates. The continuous and dashed horizontal lines mark, for reference, the value of P1/P5 counts at $\varphi = 0^\circ$ and a $\pm 1\%$ deviation from it, respectively (see Section~\ref{sec:rotation} for more details).}
    \label{fig:rotation_counts}
\end{figure*}

\begin{figure*}[p]
    \centering\vspace{3mm}
    \includegraphics[height=0.54\columnwidth]{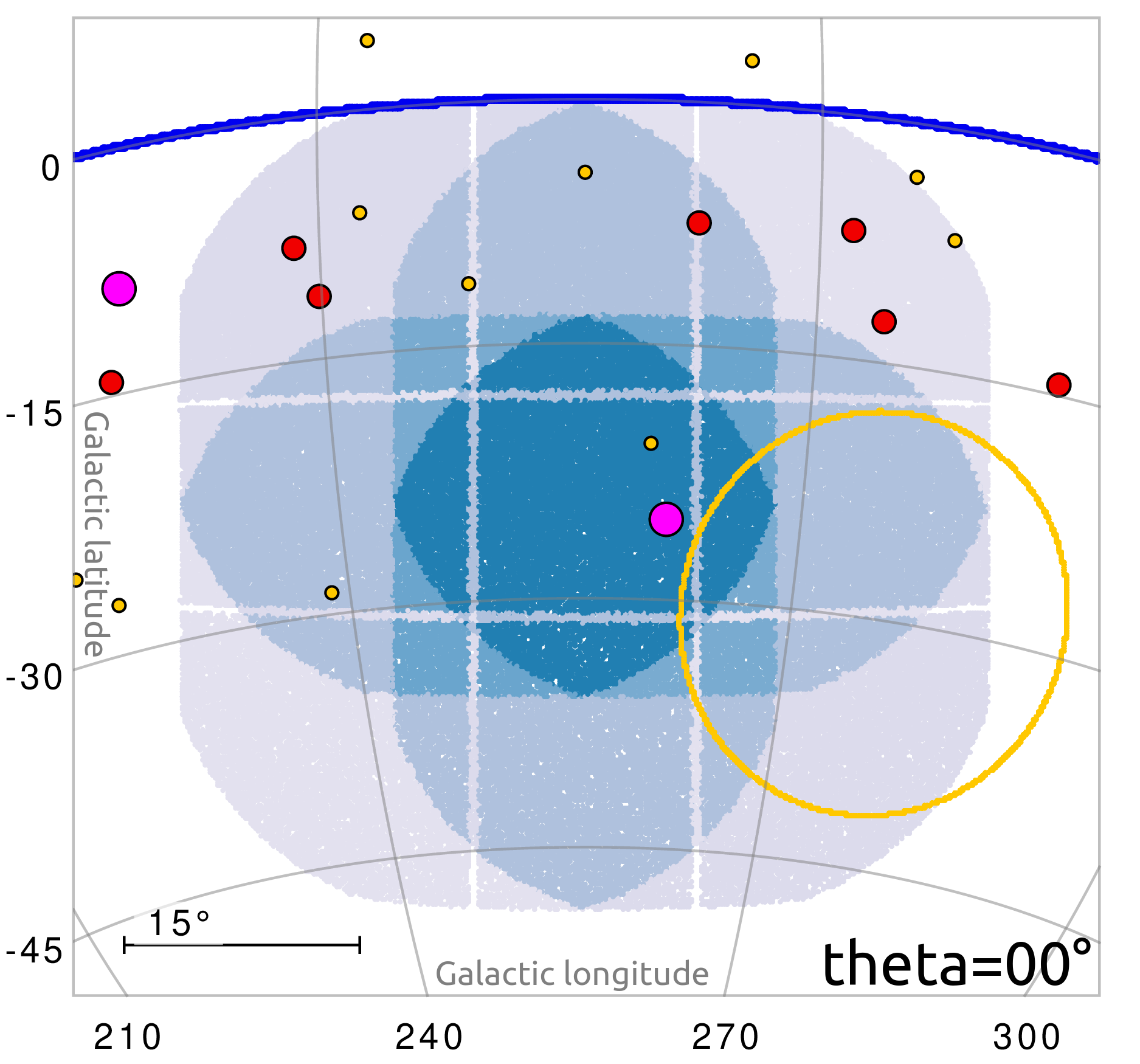}\hspace{5mm}
    \includegraphics[height=0.54\columnwidth]{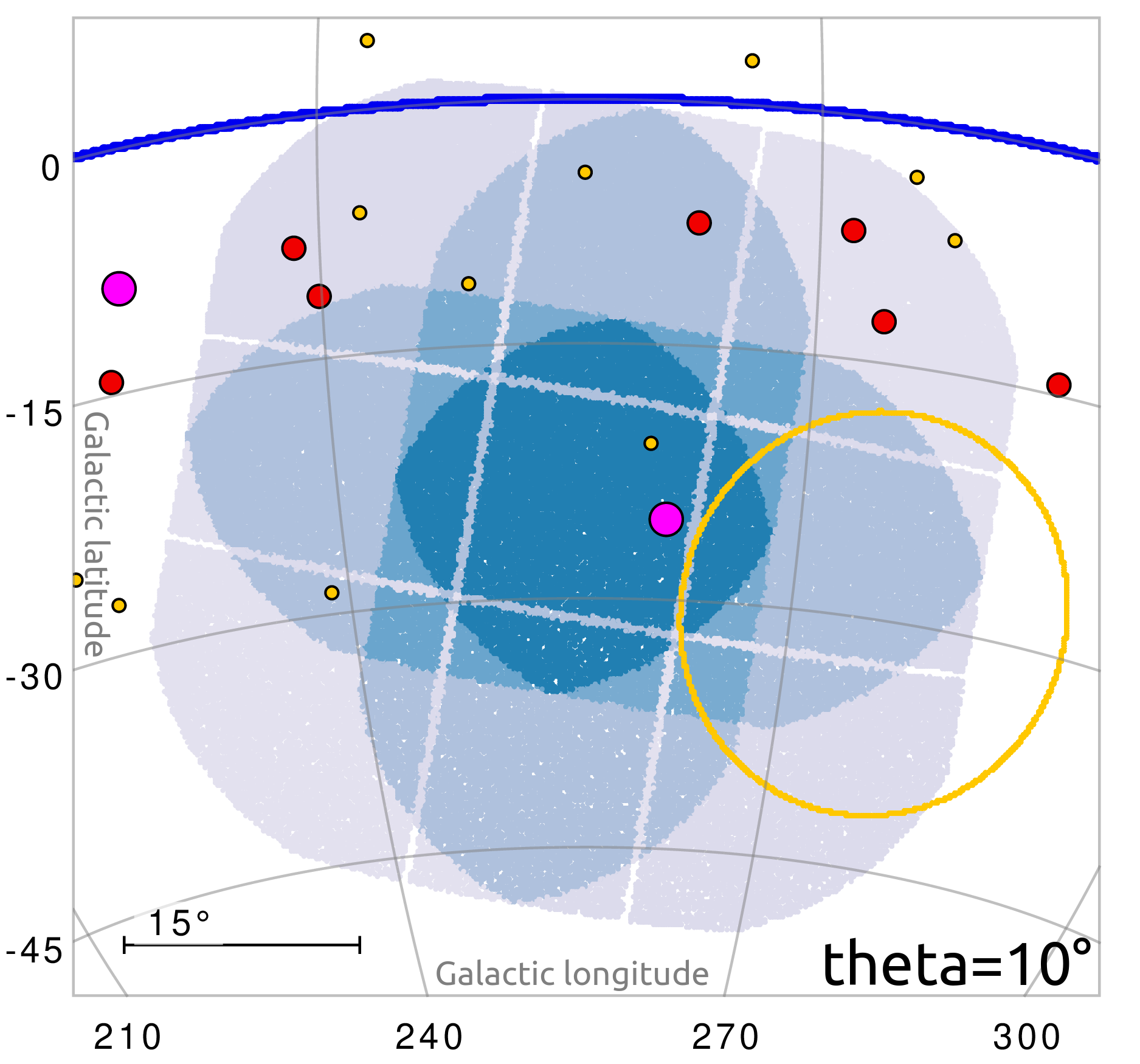}\hspace{5mm}
    \includegraphics[height=0.54\columnwidth]{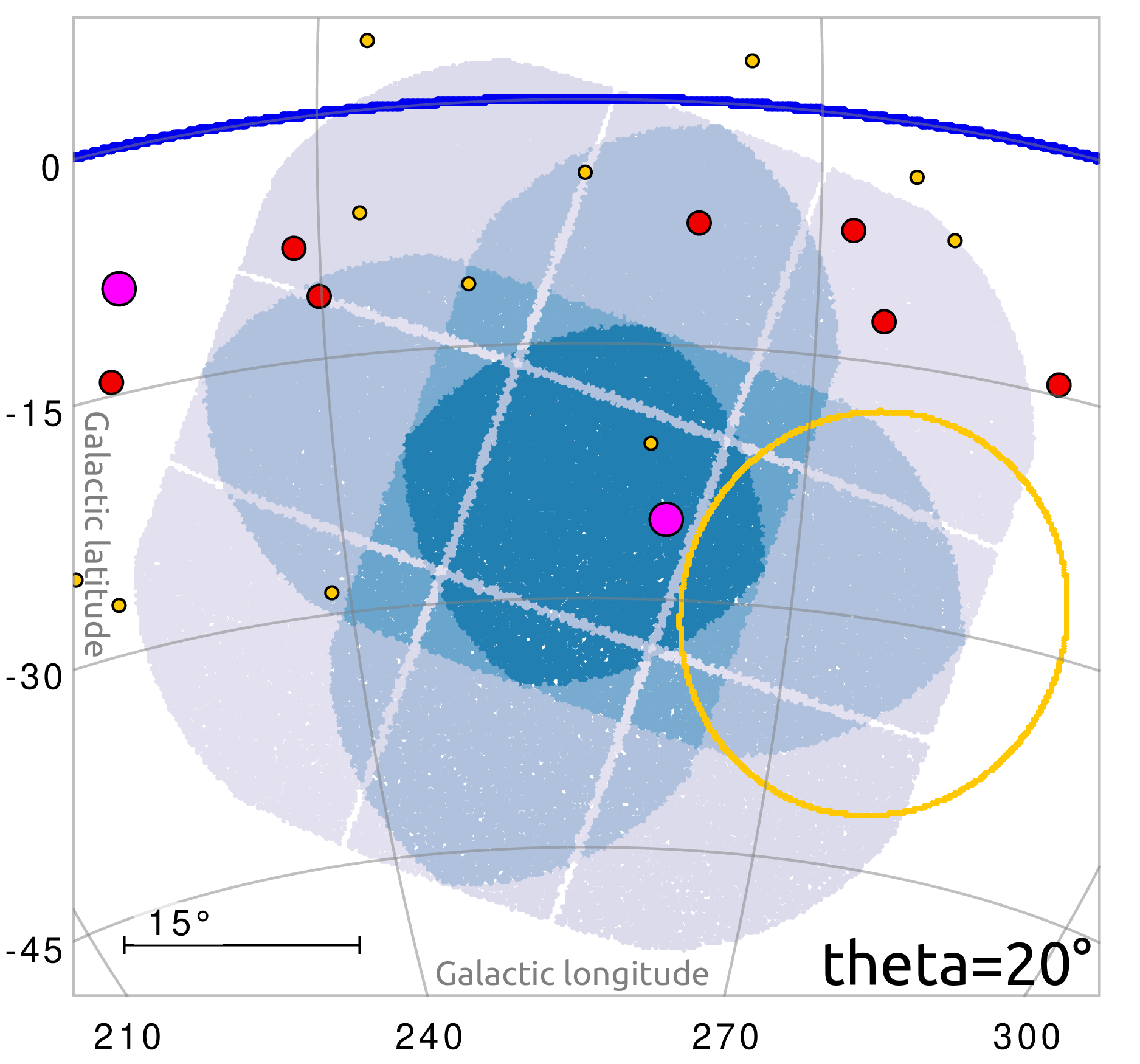}\hspace{5mm}\\ \vspace{3mm}
    \includegraphics[height=0.54\columnwidth]{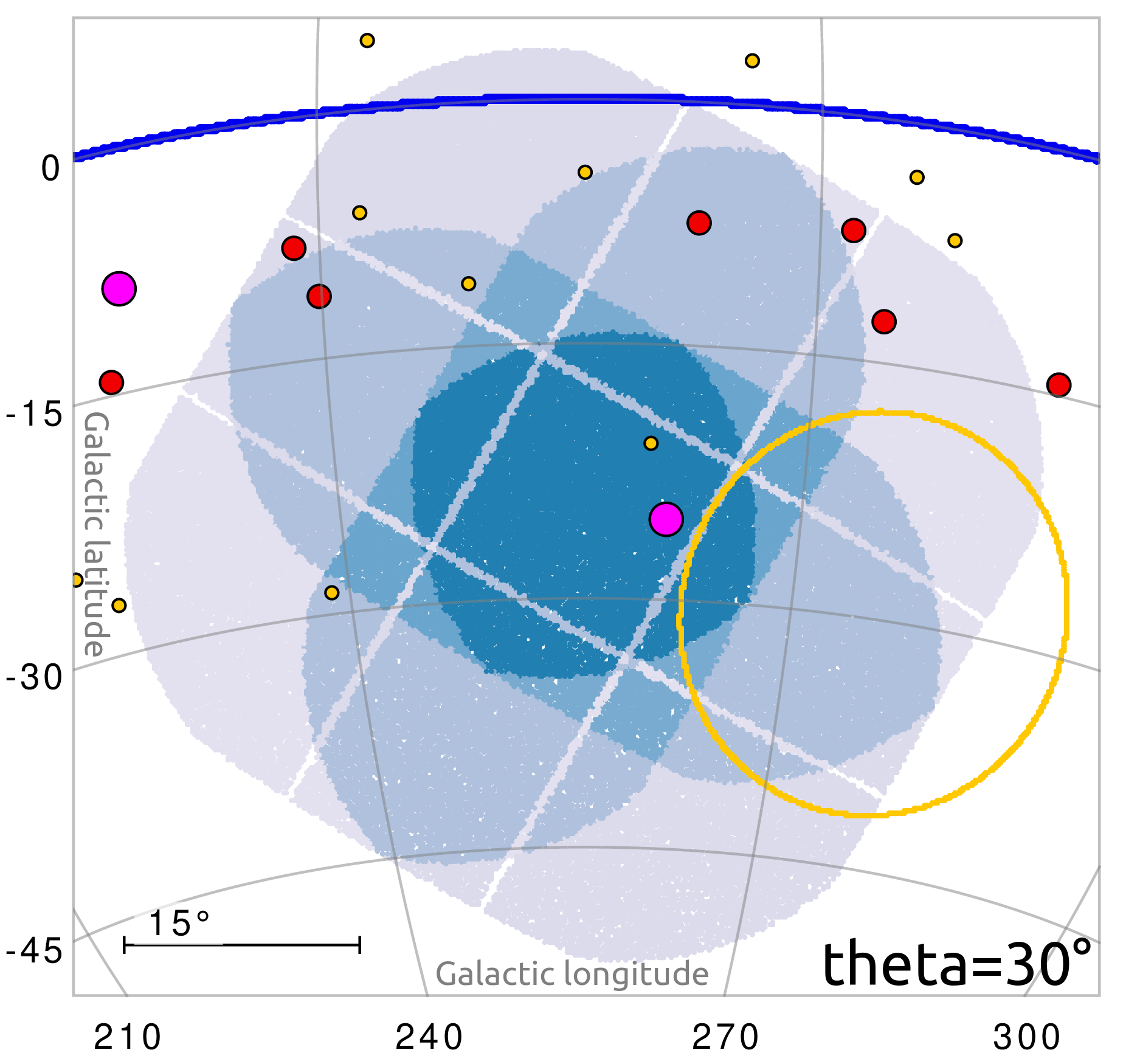}\hspace{5mm}
    \includegraphics[height=0.54\columnwidth]{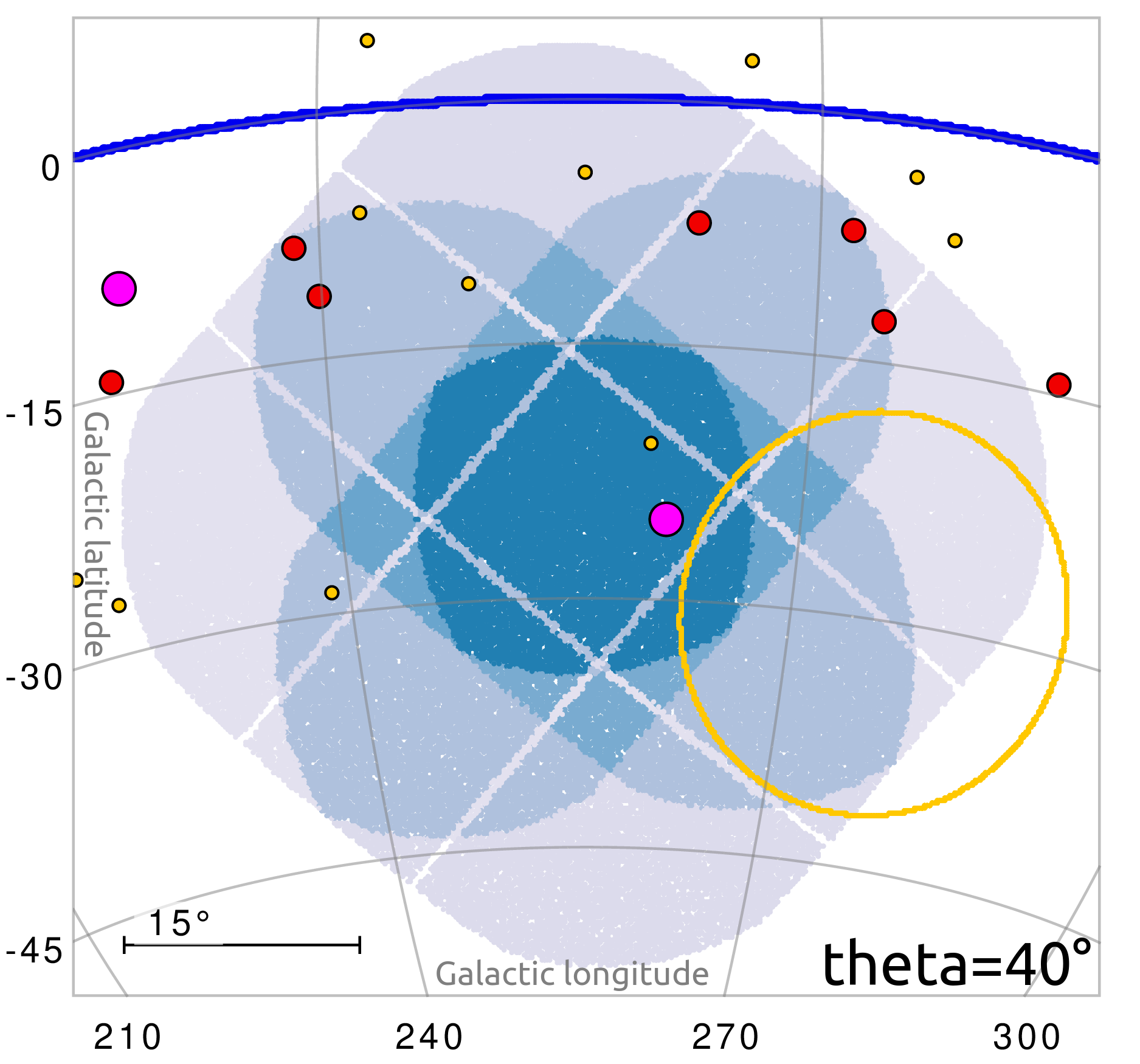}\hspace{5mm}
    \includegraphics[height=0.54\columnwidth]{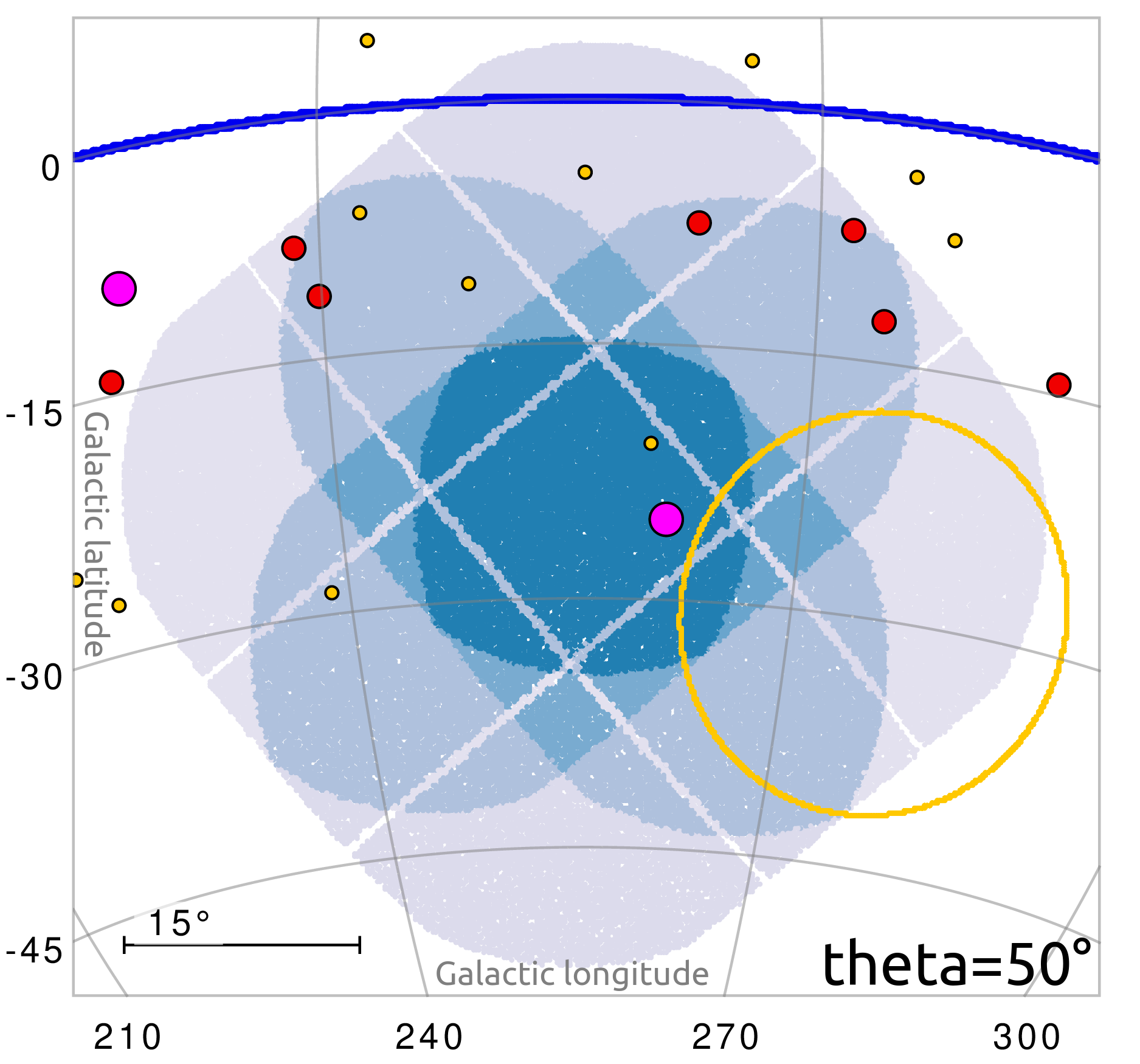}\hspace{5mm}\\ \vspace{3mm}
    \includegraphics[height=0.54\columnwidth]{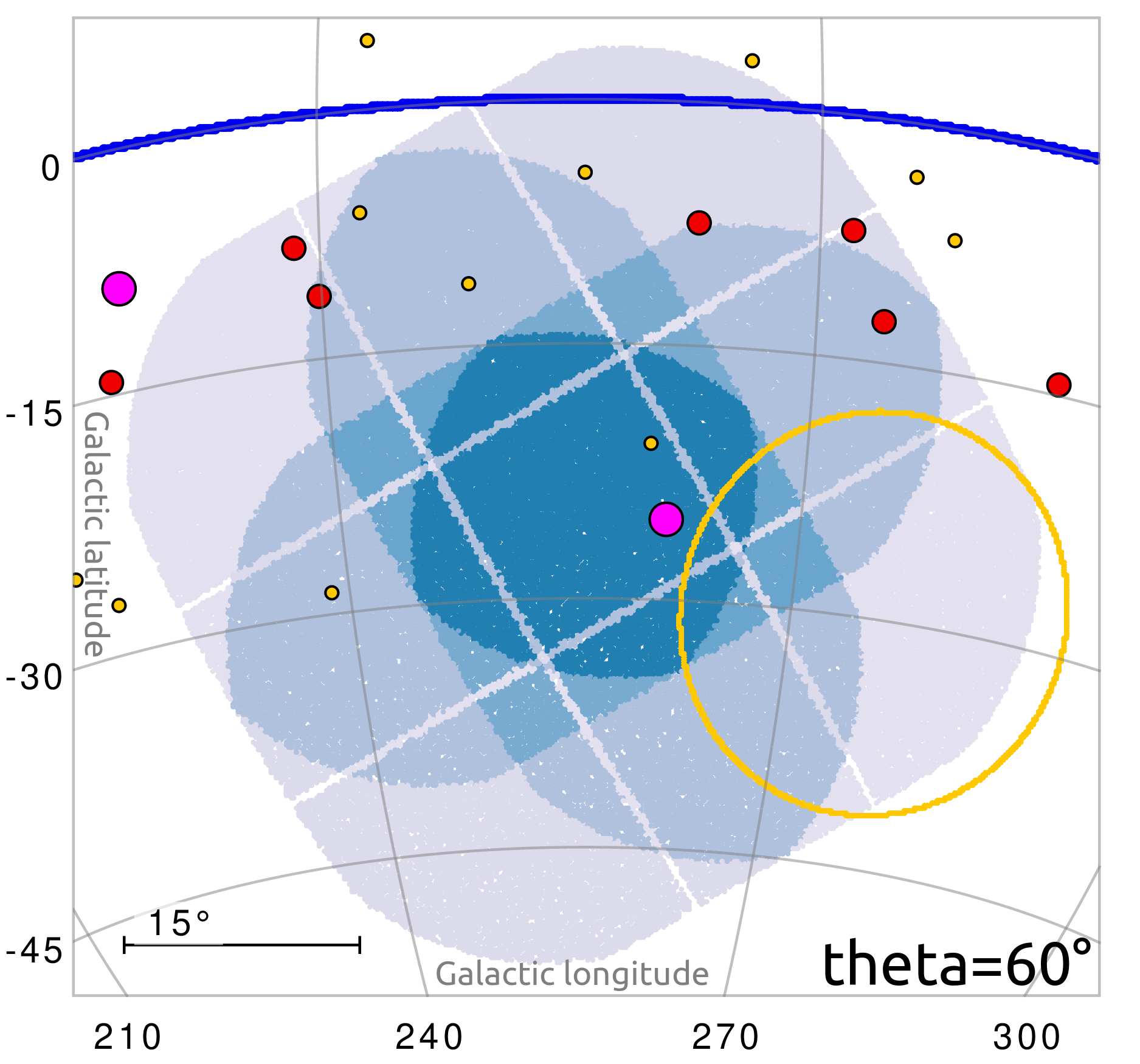}\hspace{5mm}
    \includegraphics[height=0.54\columnwidth]{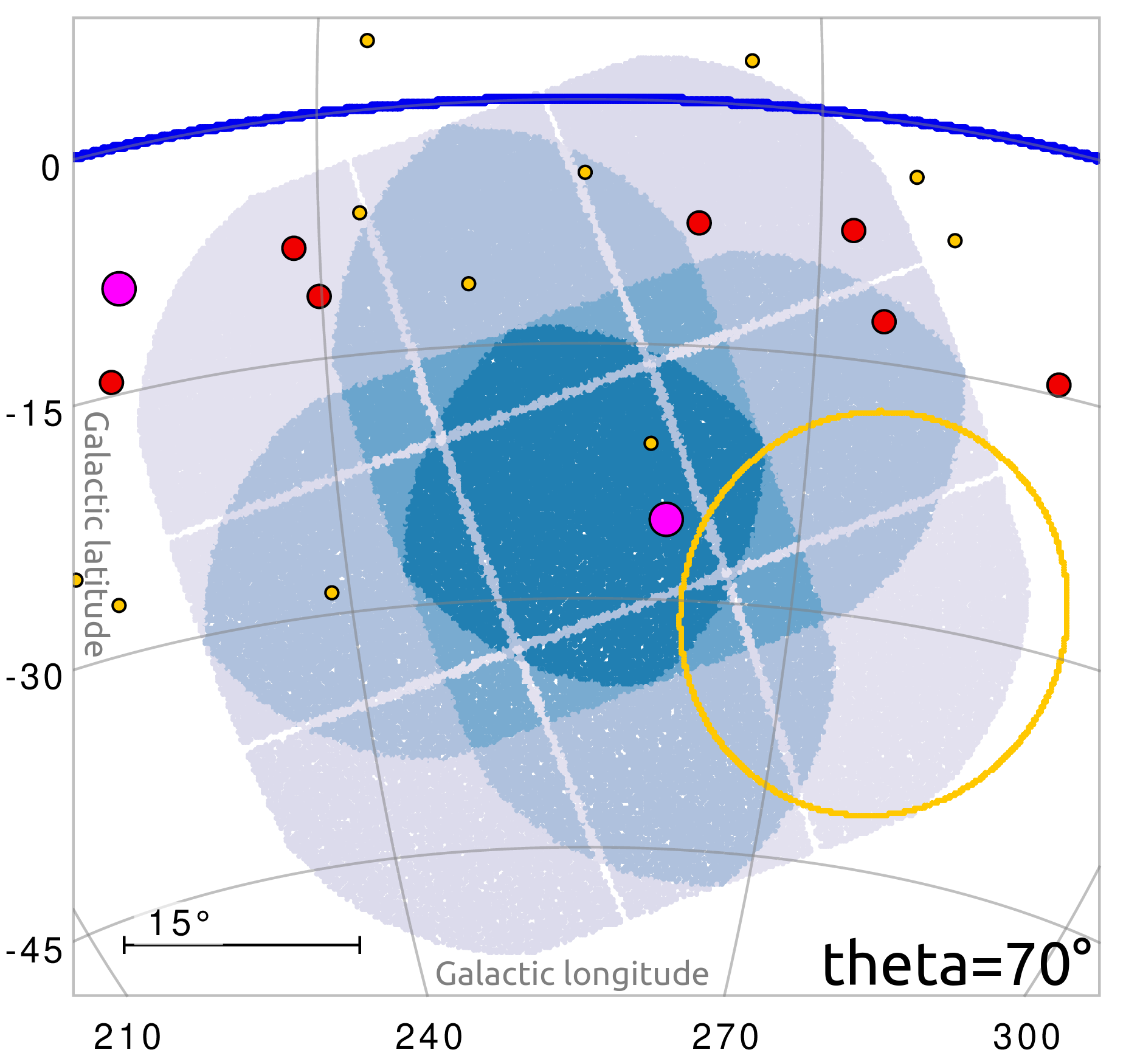}\hspace{5mm}
    \includegraphics[height=0.54\columnwidth]{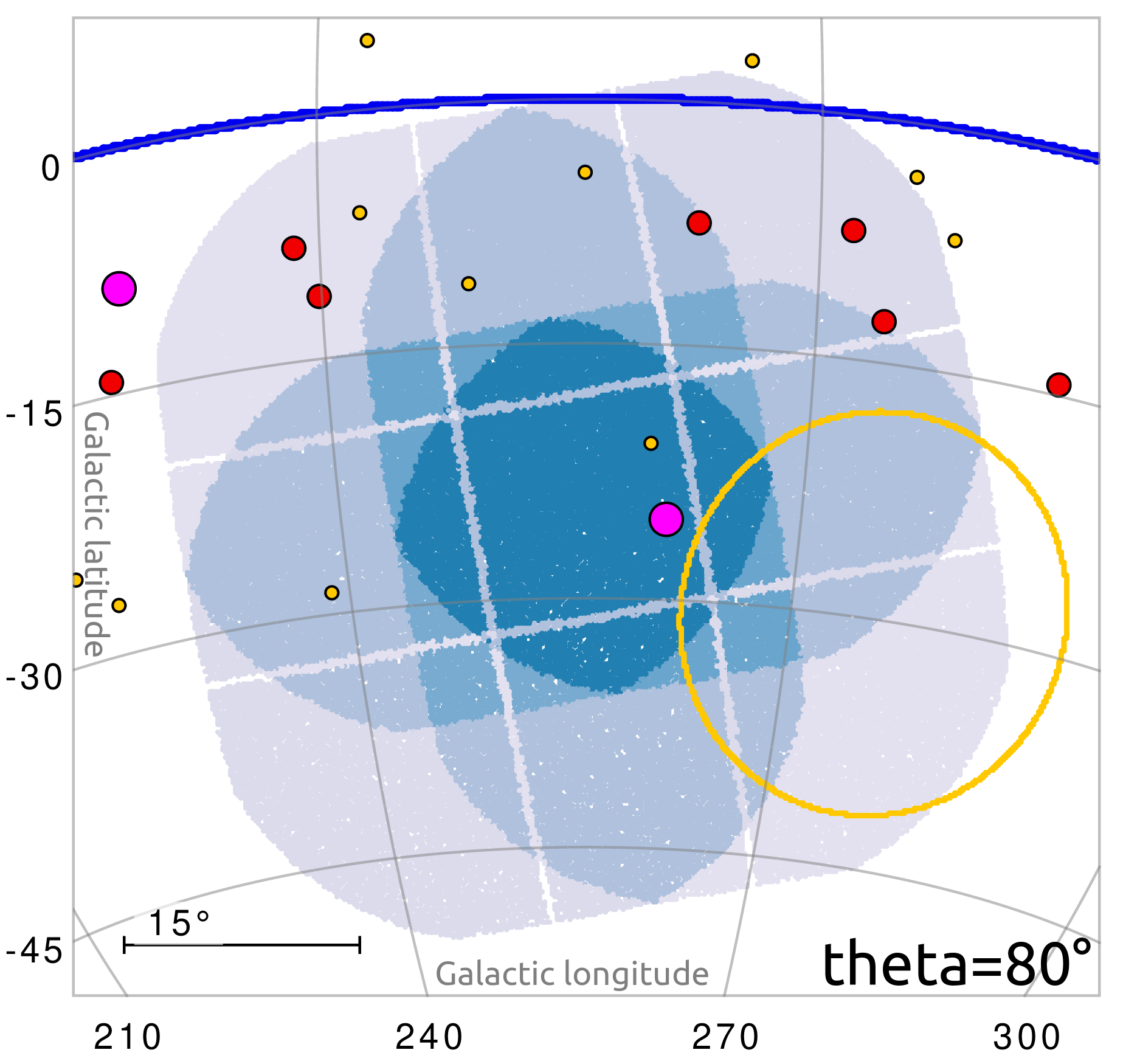}\hspace{5mm}\\
    \caption{Optimization of the PLATO field rotation angle. The PLATO test fields are centered at the LOPS2 coordinates (Table~\ref{tab:lops2}) but rotated by $0^\circ$, $10^\circ$, $20^\circ$, $30^\circ$, $40^\circ$, $50^\circ$, $60^\circ$, $70^\circ$, $80^\circ$ (in reading order); projection is galactic. The $\varphi = 0^\circ$ rotation angle corresponds to the nominal LOPS2 geometry. As in Fig.~\ref{fig:bright}, the brightest stars from the Yale Bright Star Catalog are plotted as circles of different colors: magenta ($V<1$), red ($1<V<2$), yellow ($2<V<3$). The TESS southern CVZ is marked with a yellow circle (see Section~\ref{sec:rotation} for more details).}
    \label{fig:rotation_fields}
\end{figure*}

\begin{figure*}[p]
    \centering
    \includegraphics[height=0.75\columnwidth]{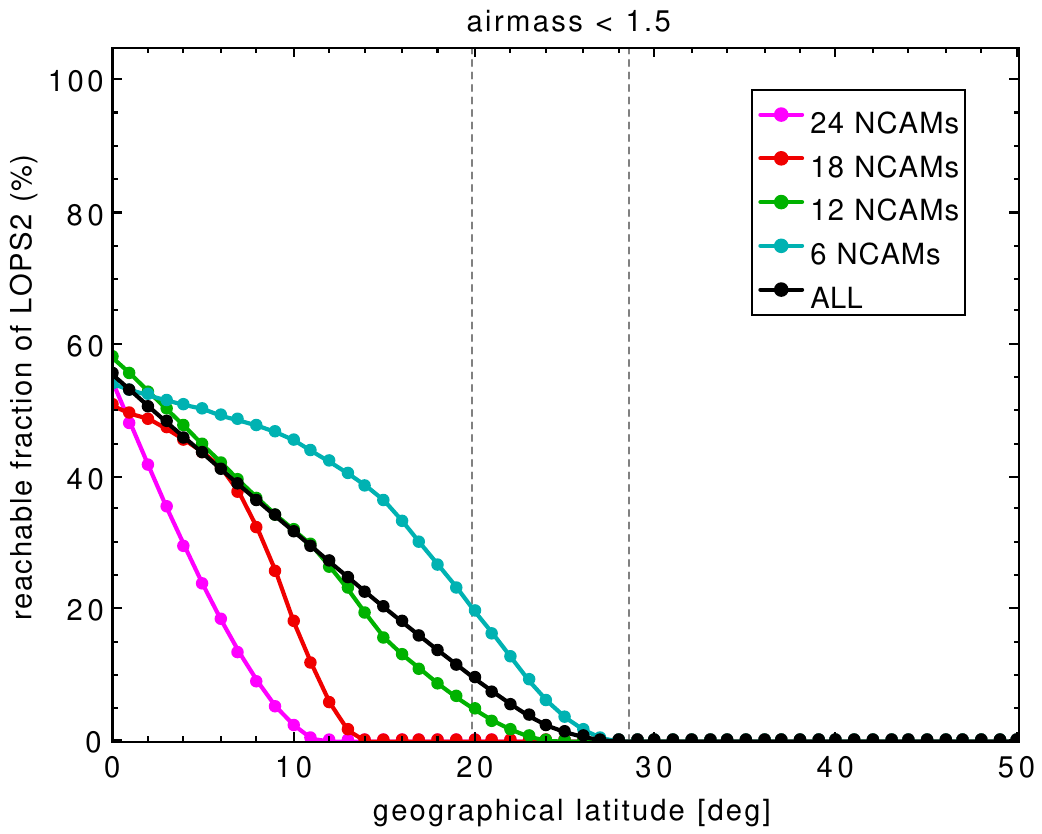}\hspace{10mm}
    \includegraphics[height=0.75\columnwidth]{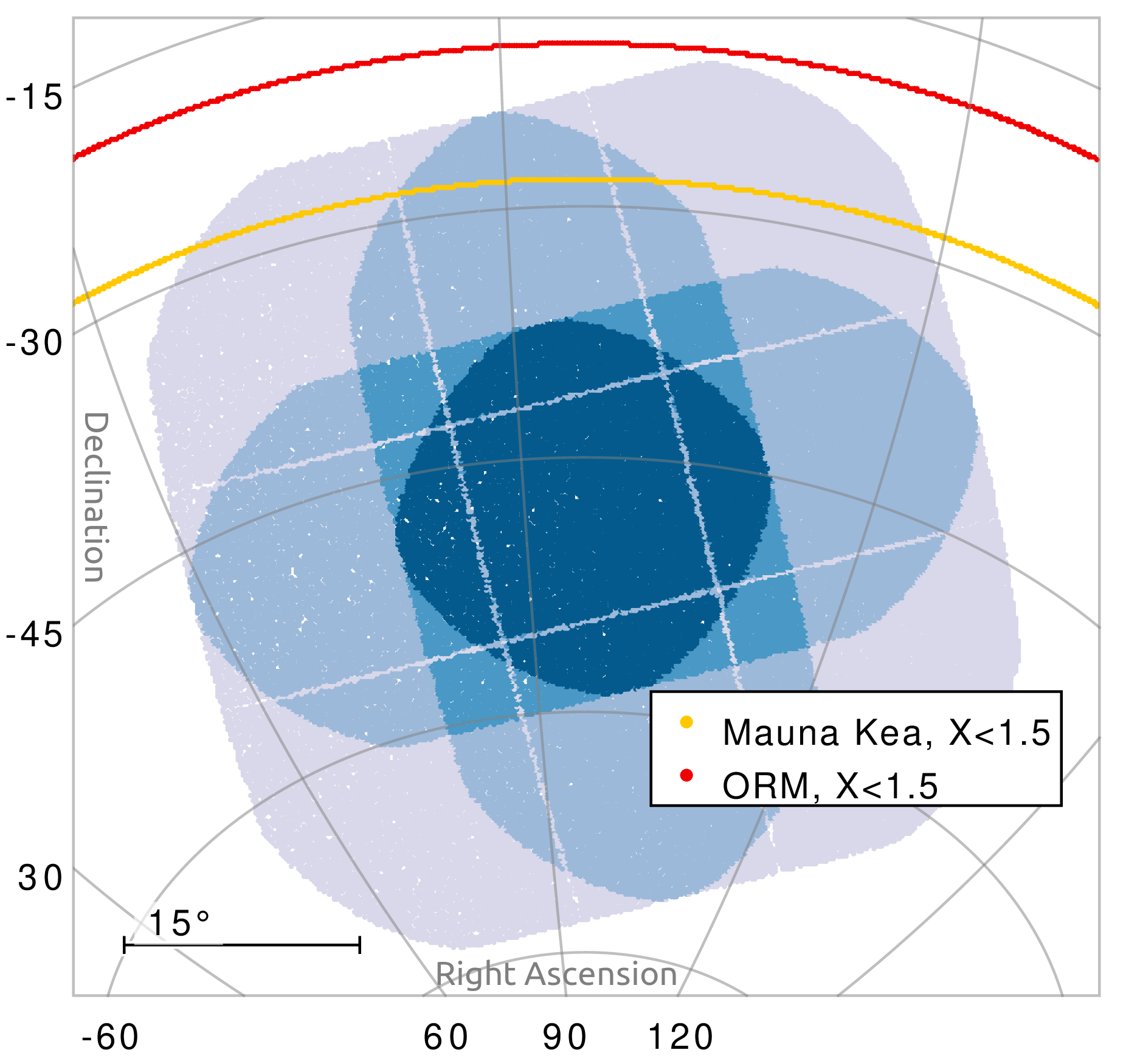}\\ \vspace{5mm}
    \includegraphics[height=0.75\columnwidth]{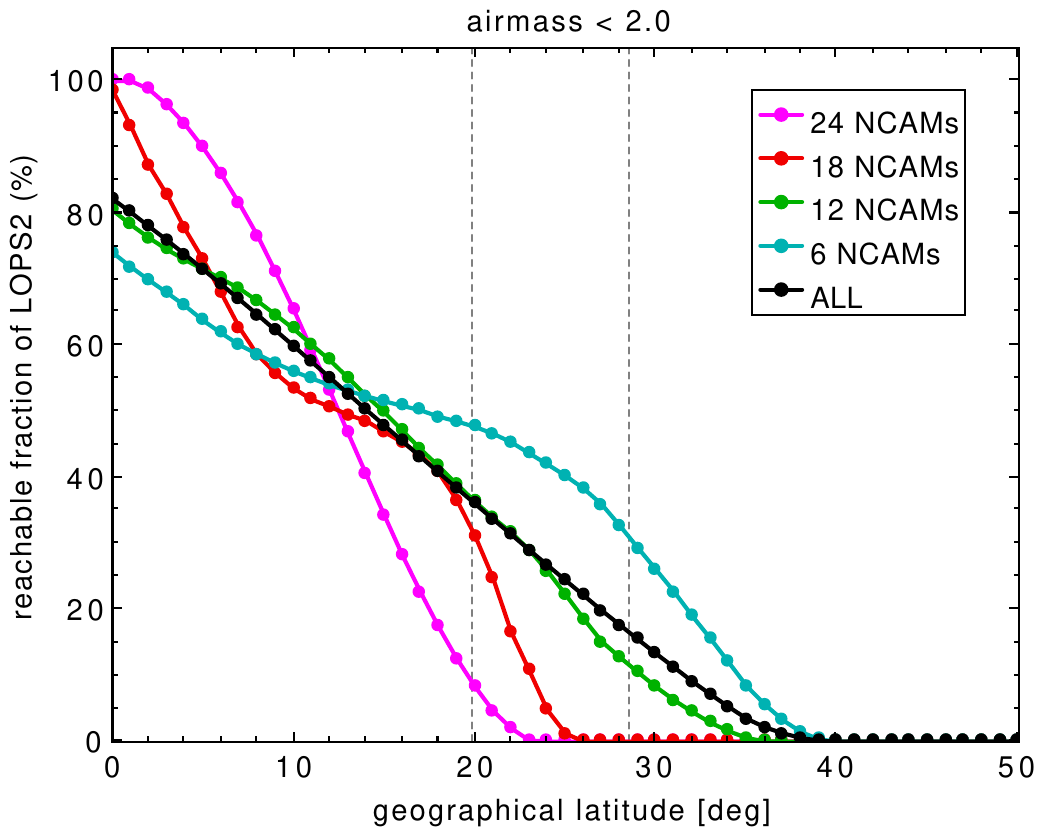}\hspace{10mm}
    \includegraphics[height=0.75\columnwidth]{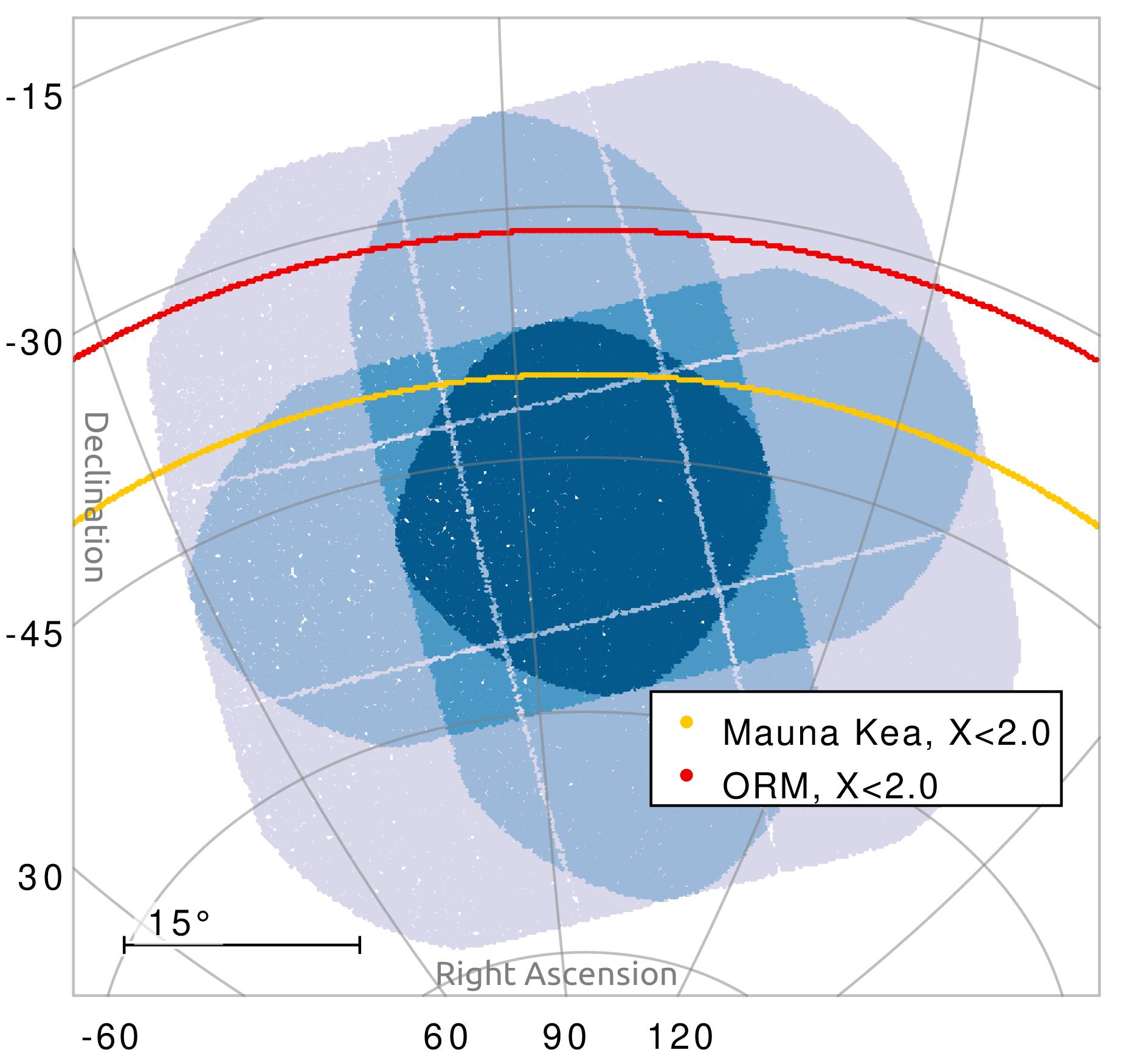}\\ \vspace{5mm}
    \includegraphics[height=0.75\columnwidth]{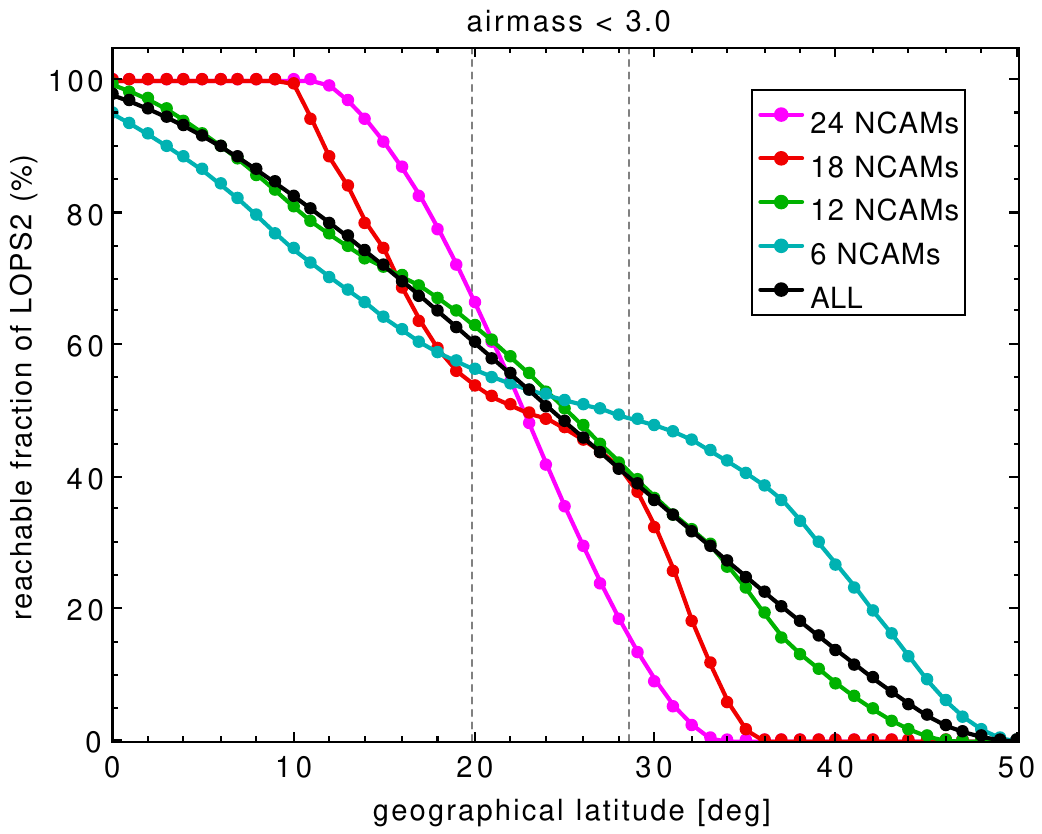}\hspace{10mm}
    \includegraphics[height=0.75\columnwidth]{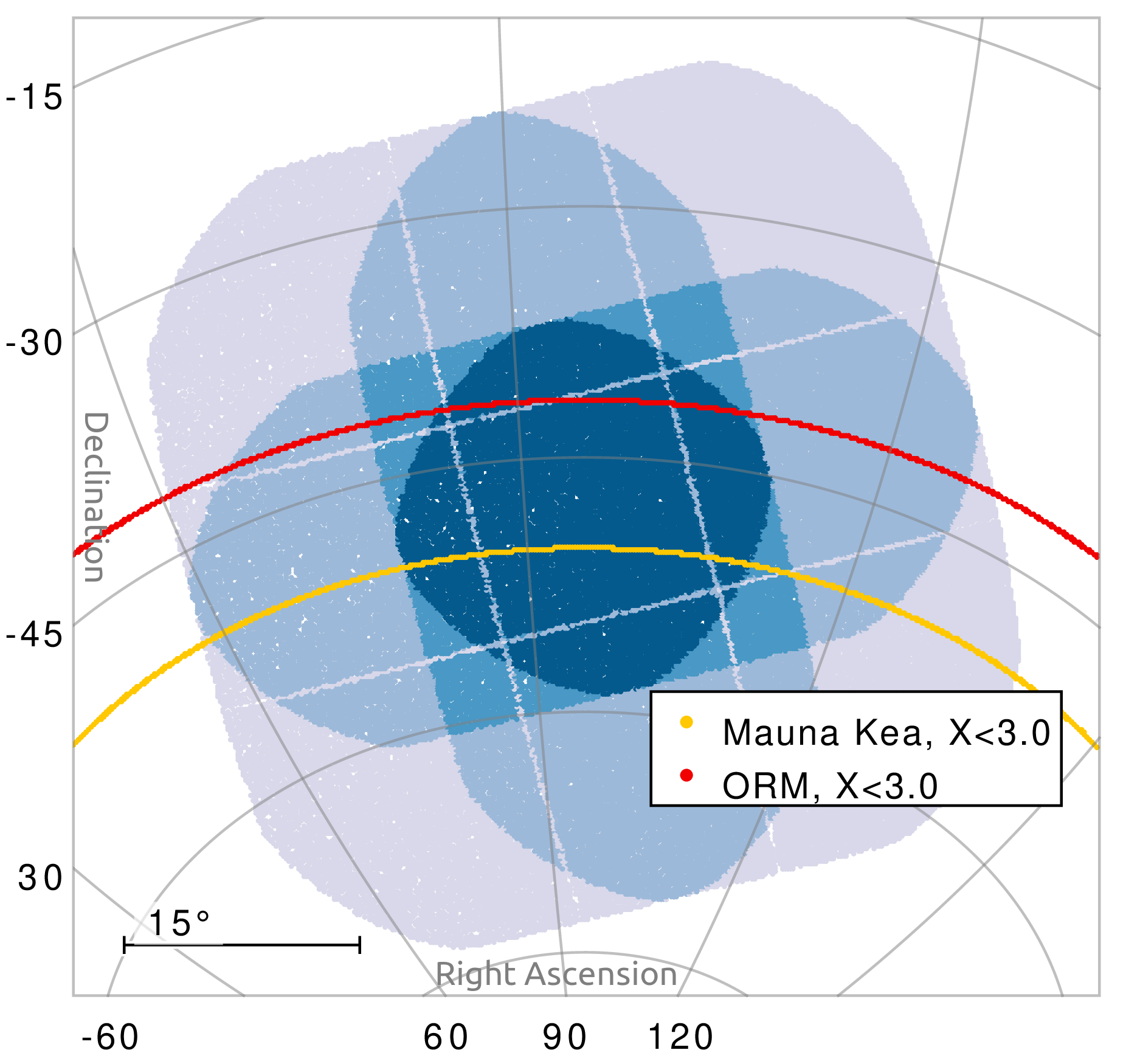}\\ \vspace{5mm}
    \caption{Visibility of LOPS2 from the northern hemisphere. \emph{Left panels}: Reachable fraction of LOPS2 area (both for the whole field, and for its sub-regions observed by $N$ NCAMs) as a function of the geographical latitude of the observatory, at limiting airmass of 1.5 (upper plot), 2.0 (middle plot), 3.0 (lower plot). The latitude of Mauna Kea ($\psi\simeq 19^\circ.8$) and ORM ($\psi\simeq 28^\circ.7$) observatories is marked with two vertical dashed lines. \emph{Right panels:} Sky chart of LOPS2 (ortographic projection in equatorial coordinates) showing the minimum declination reachable from Mauna Kea (yellow circle) or ORM (red circle) at limiting airmass of 1.5 (upper plot), 2.0 (middle plot), 3.0 (lower plot). See also Section~\ref{sec:synergies} for a discussion.}
    \label{fig:reachable}
\end{figure*}

\begin{figure*}
    \centering\vspace{0.3cm}
    \includegraphics[width=1.9\columnwidth]{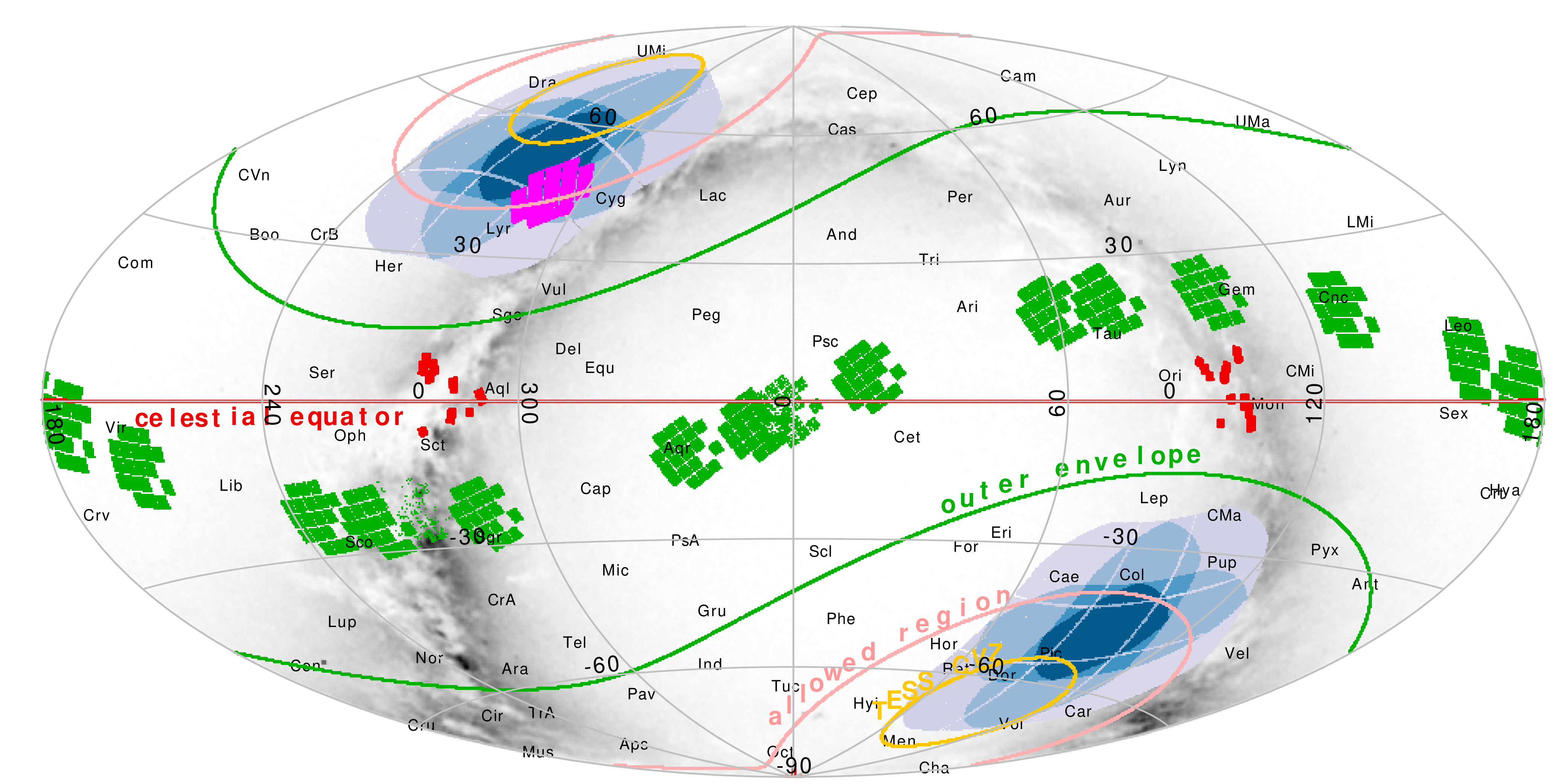}\vspace{0.3cm}
    \caption{All-sky Aitoff projection in Equatorial coordinates of LOPS2 and LOPN1, showing the formal constraints for the selection of the PLATO LOP fields and the synergies with other missions. The two pink circles represent the $|\beta|>63^\circ$ technical requirement for the center of the LOP fields (``allowed region''), implying that the overall envelopes of every allowed field choice are two ecliptic caps at $|\beta|\gtrsim38^\circ$ (green circles). LOPS2 (lower left) and LOPN1 (upper right) are plotted with blue shades according to the number of co-pointing cameras, as in Fig.~\ref{fov}. The footprints of CoRoT (red), \emph{Kepler} (magenta), and K2 (green) are over-plotted together with the TESS continuous viewing zone at $|\beta|\gtrsim78^\circ$ (yellow circle). The background gray layer is color coded according to the areal density of $G<13.5$ stars from \emph{Gaia} DR3. The celestial equator and poles are marked with a red line and crosses, respectively.}
    \label{fig:full_sky_equ}
\end{figure*}

\begin{figure*}
    \centering\vspace{0.3cm}
    \includegraphics[width=1.9\columnwidth]{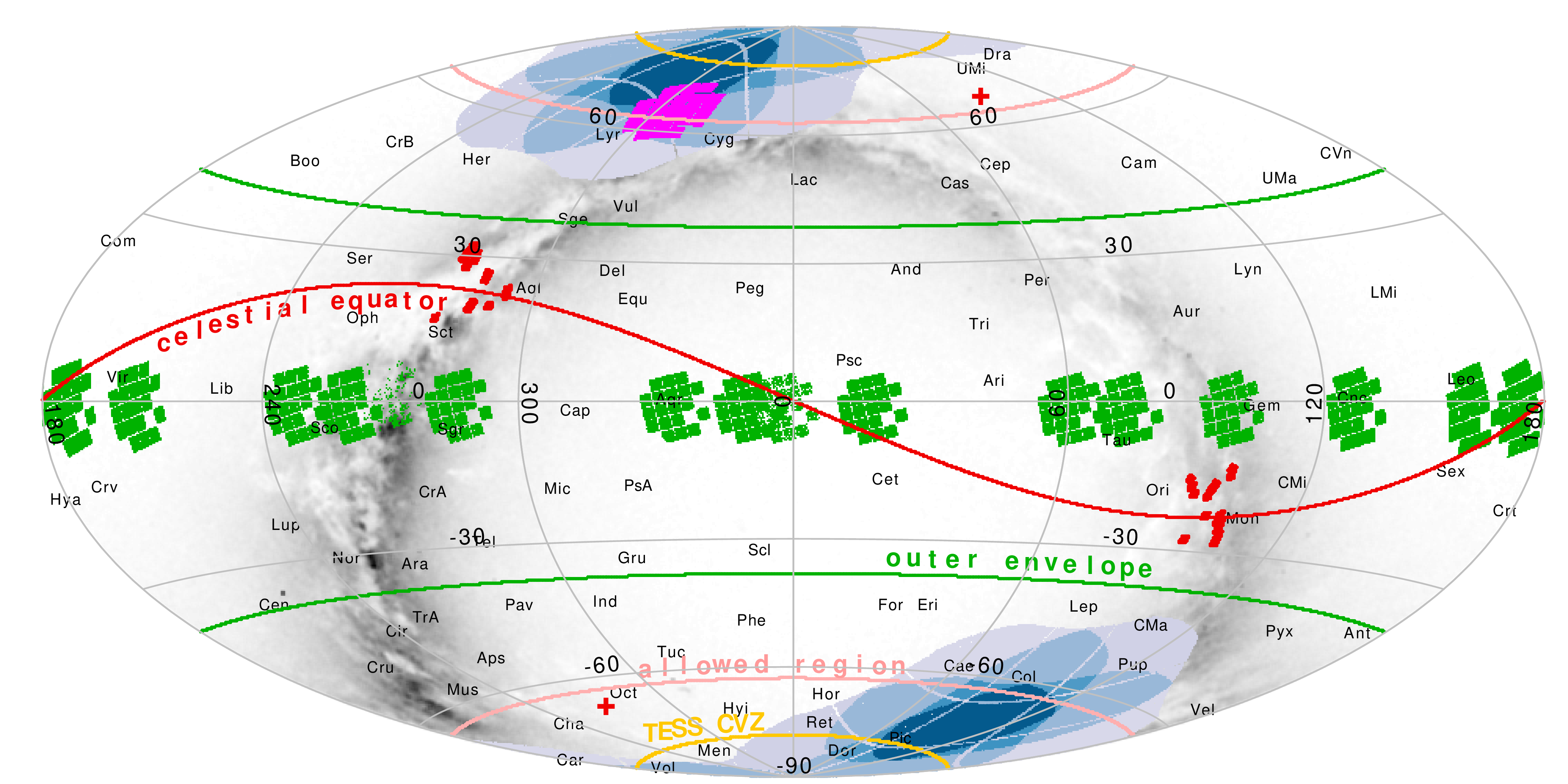}\vspace{0.3cm}
    \caption{All-sky Aitoff projection in Ecliptic coordinates of LOPS2 and LOPN1, showing the formal constraints for the selection of the PLATO LOP fields and the synergies with other missions. The two pink circles represent the $|\beta|>63^\circ$ technical requirement for the center of the LOP fields (``allowed region''), implying that the overall envelopes of every allowed field choice are two ecliptic caps at $|\beta|\gtrsim38^\circ$ (green circles). LOPS2 (lower left) and LOPN1 (upper right) are plotted with blue shades according to the number of co-pointing cameras, as in Fig.~\ref{fov}. The footprints of CoRoT (red), \emph{Kepler} (magenta), and K2 (green) are over-plotted together with the TESS continuous viewing zone at $|\beta|\gtrsim78^\circ$ (yellow circle). The background gray layer is color coded according to the areal density of $G<13.5$ stars from \emph{Gaia} DR3. The celestial equator and poles are marked with a red line and crosses, respectively.}
    \label{fig:full_sky_ecl}
\end{figure*}

\begin{figure*}[p]
    \centering
    \includegraphics[width=0.97\columnwidth]{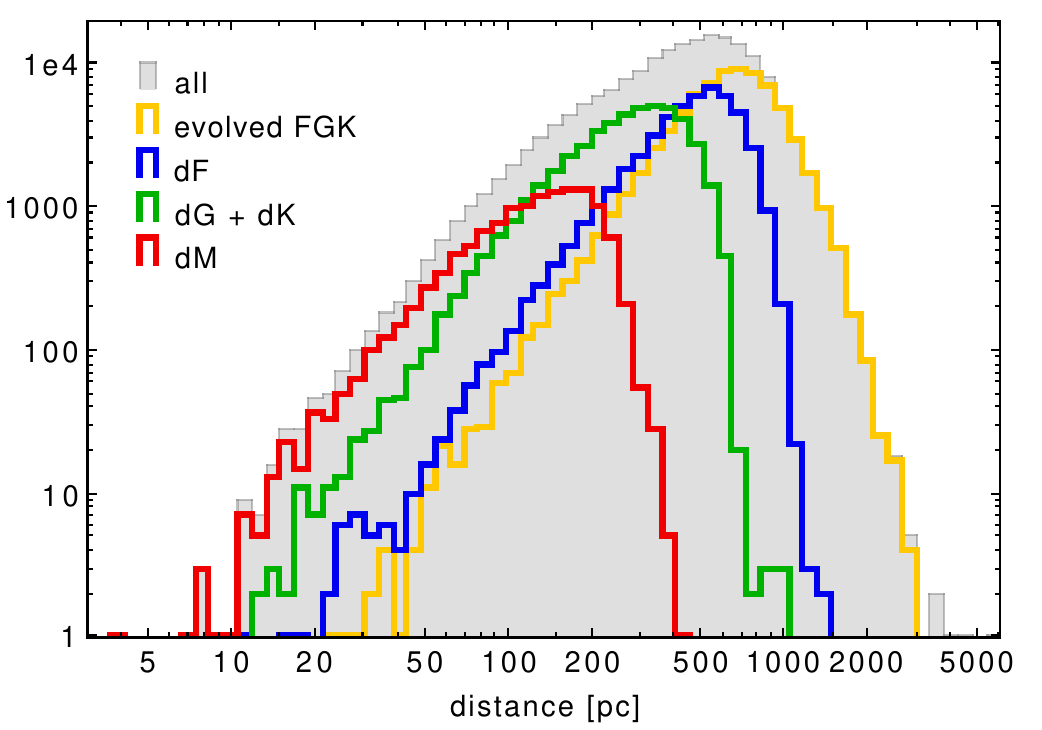}\hspace{5mm}
    \includegraphics[width=0.97\columnwidth]{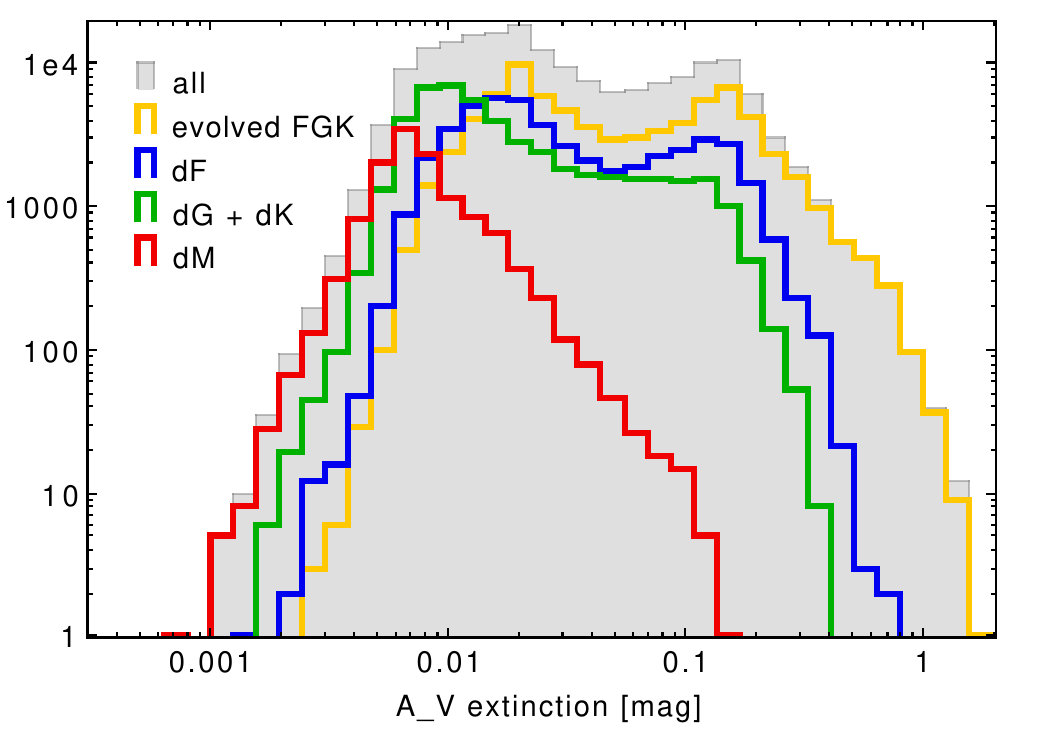}\\ 
    \caption{Distribution in distance and interstellar extinction of LOPS2 targets, from the PIC. Both the distribution of the whole sample (in gray) and for a few subsamples of interest is shown: M dwarfs, i.~e., P4 (in red), main-sequence F stars (blue), main-sequence G and K stars (green) and FGK subgiants (orange). The definition of these sub-samples in terms of $T_\mathrm{eff}$ and $R_\star$ can be found in \citetalias{Nascimbeni2022}, Section 6.3. \emph{Left panel:} Logarithmic histogram of distance. \emph{Right panel:} Logarithmic histogram of the interstellar extinction in the Bessel $V$ band.}
    \label{fig:distance_ext}
\end{figure*}

\begin{figure*}[p]
    \centering
    \includegraphics[width=0.97\columnwidth]{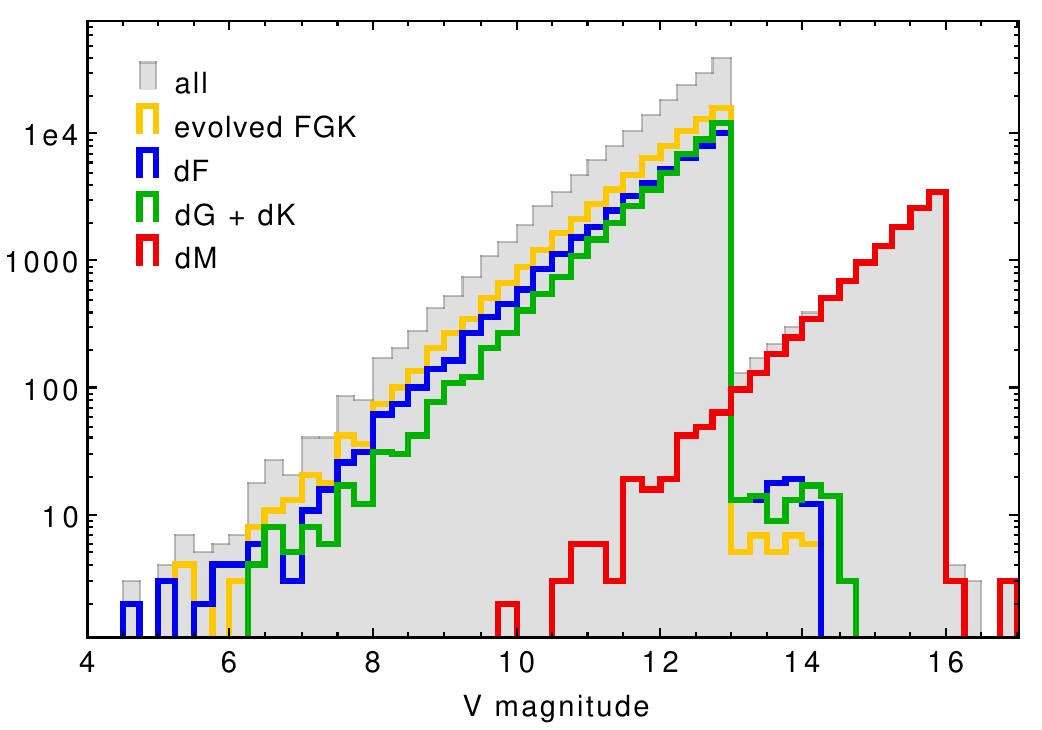}\hspace{5mm}
    \includegraphics[width=0.97\columnwidth]{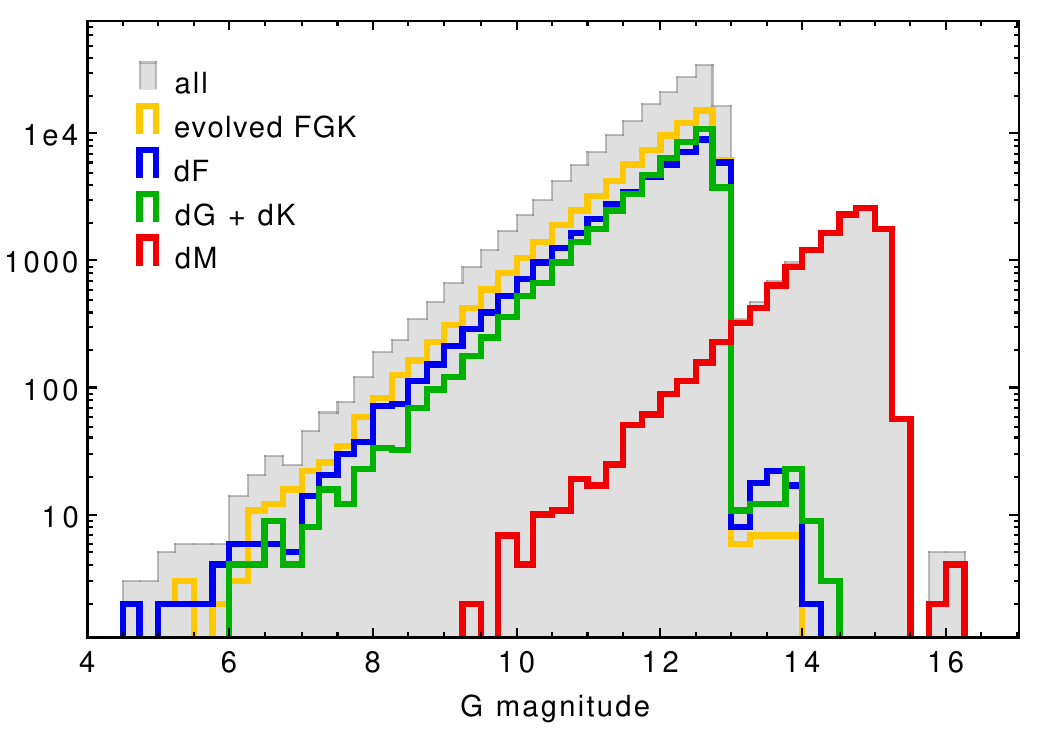}\\ 
    \includegraphics[width=0.97\columnwidth]{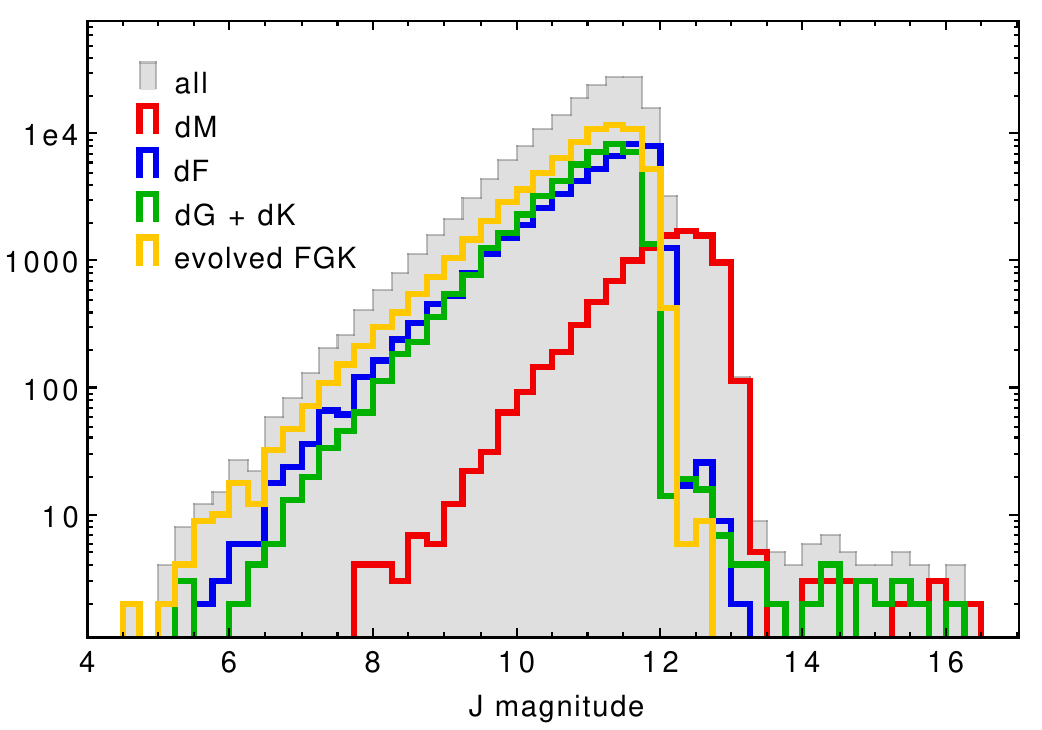}\hspace{5mm}
    \includegraphics[width=0.97\columnwidth]{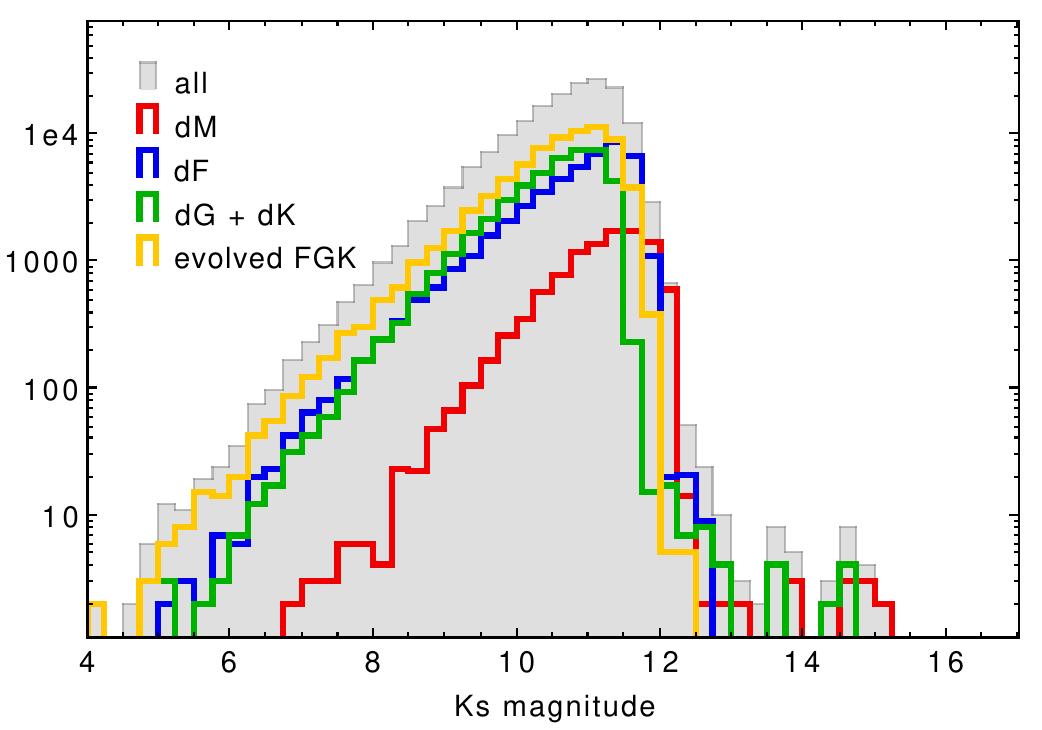}\\ 
    \caption{Distribution in $V$, Gaia $G$, 2MASS $J$, 2MASS $K_s$ magnitude (panels from upper left to lower right) of LOPS2 targets, from the PIC. Both the distribution of the whole sample (in gray) and for a few subsamples of interest is shown: M dwarfs, i.~e., P4 (in red), main-sequence F stars (blue), main-sequence G and K stars (green) and FGK subgiants (orange). }
    \label{fig:magnitudes}
\end{figure*}

\begin{figure*}[p]
    \centering
    \includegraphics[width=0.97\columnwidth]{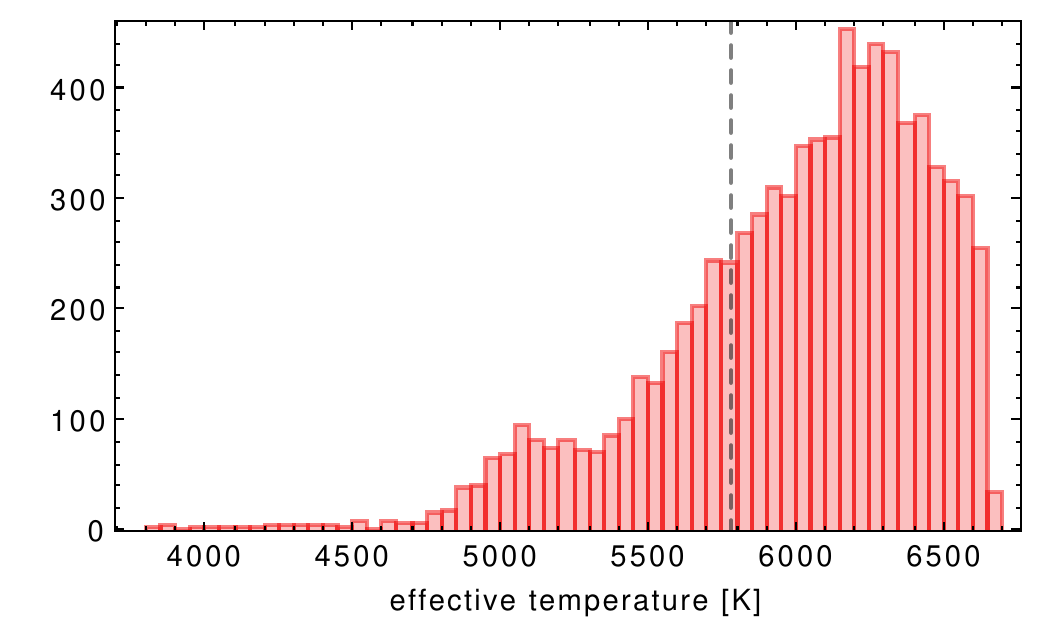}\hspace{1mm}
    \includegraphics[width=0.97\columnwidth]{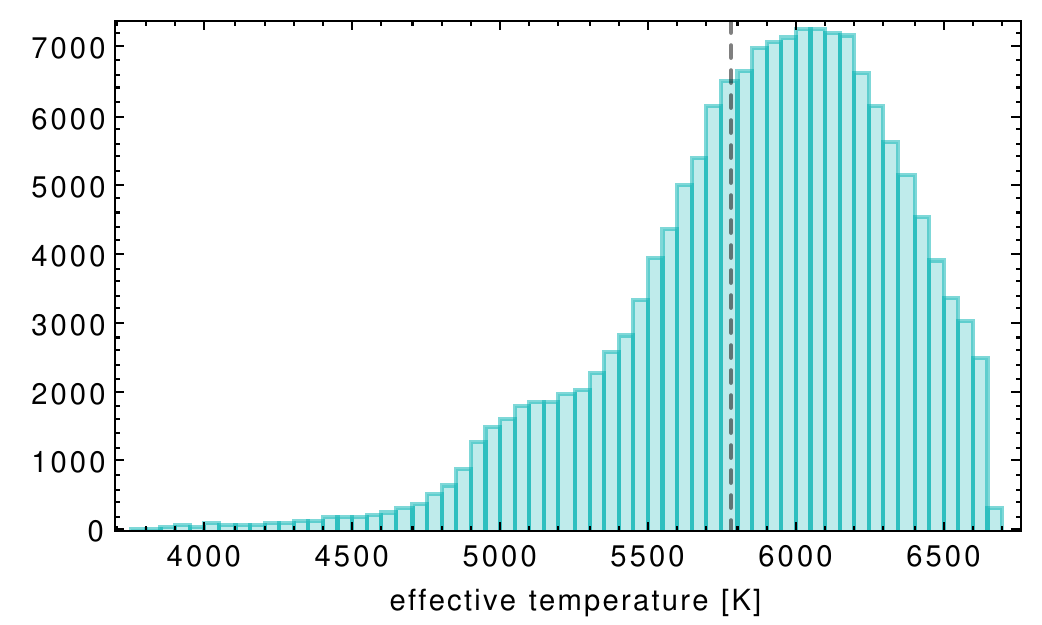}\\ \vspace{5mm}
    \includegraphics[width=0.97\columnwidth]{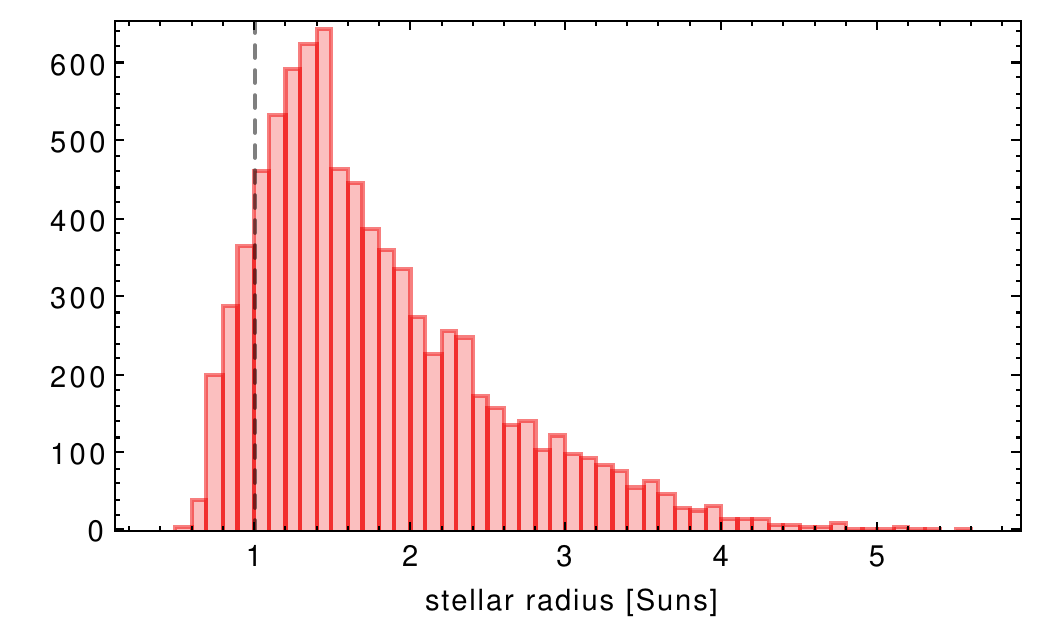}\hspace{1mm}
    \includegraphics[width=0.97\columnwidth]{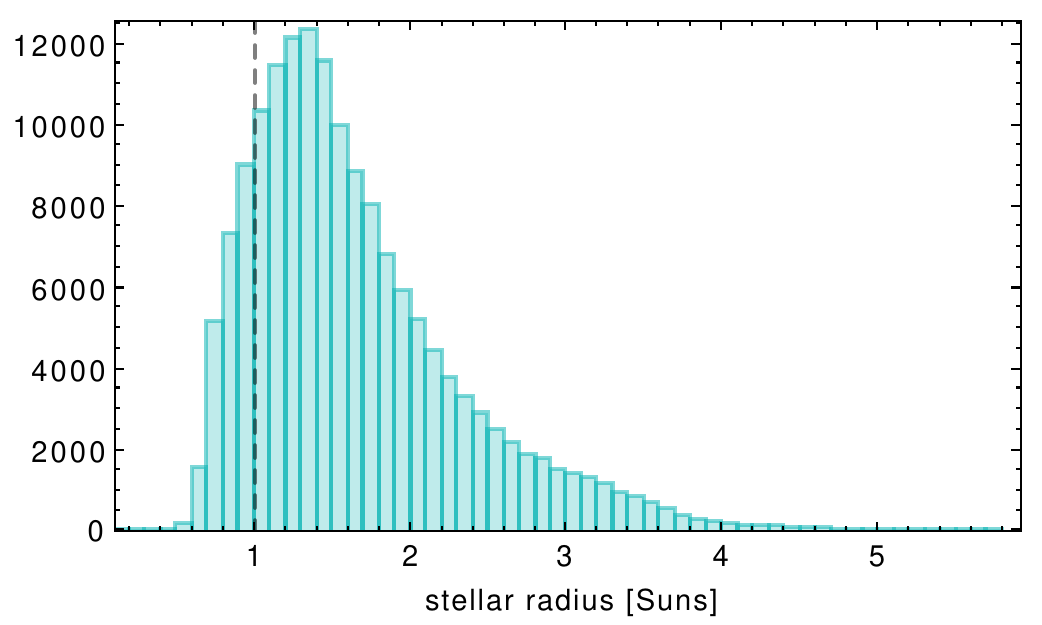}\\ \vspace{5mm}
    \includegraphics[width=0.97\columnwidth]{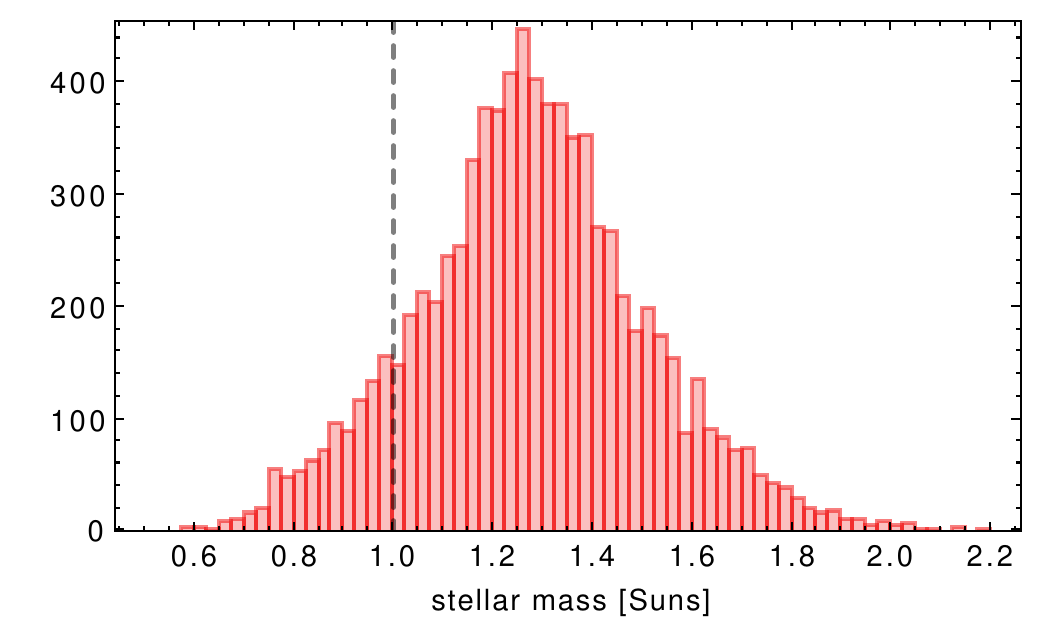}\hspace{1mm}
    \includegraphics[width=0.97\columnwidth]{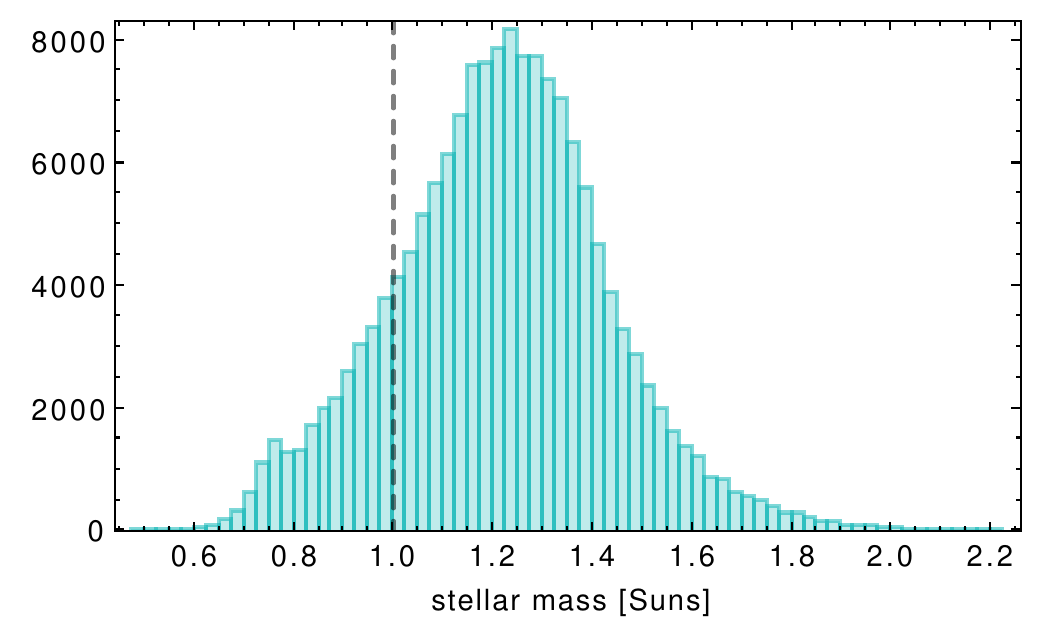}\\ \vspace{5mm}
    \includegraphics[width=0.97\columnwidth]{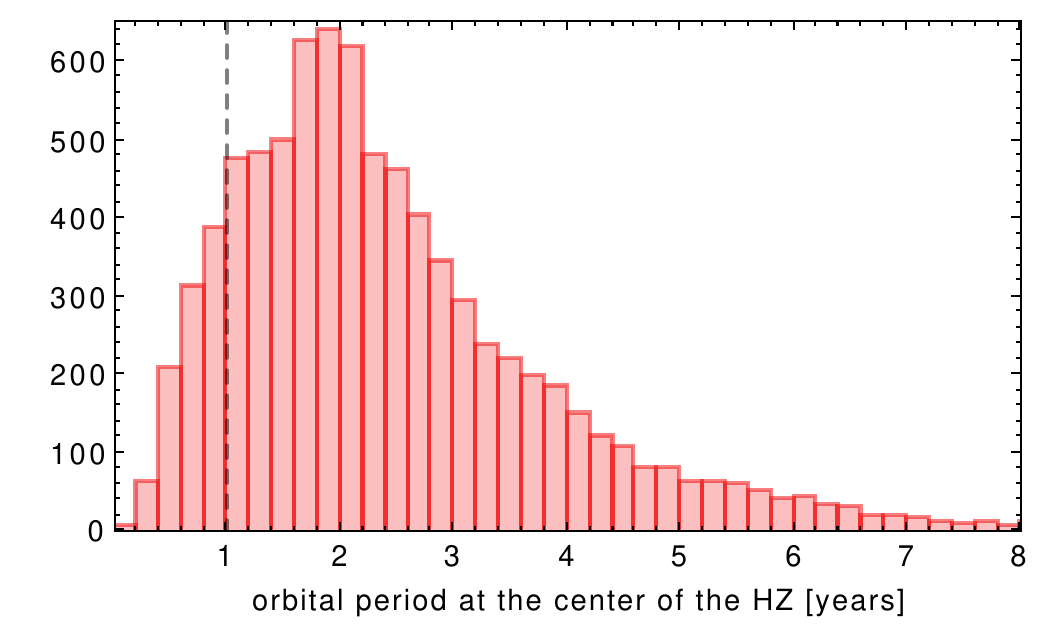}\hspace{1mm}
    \includegraphics[width=0.97\columnwidth]{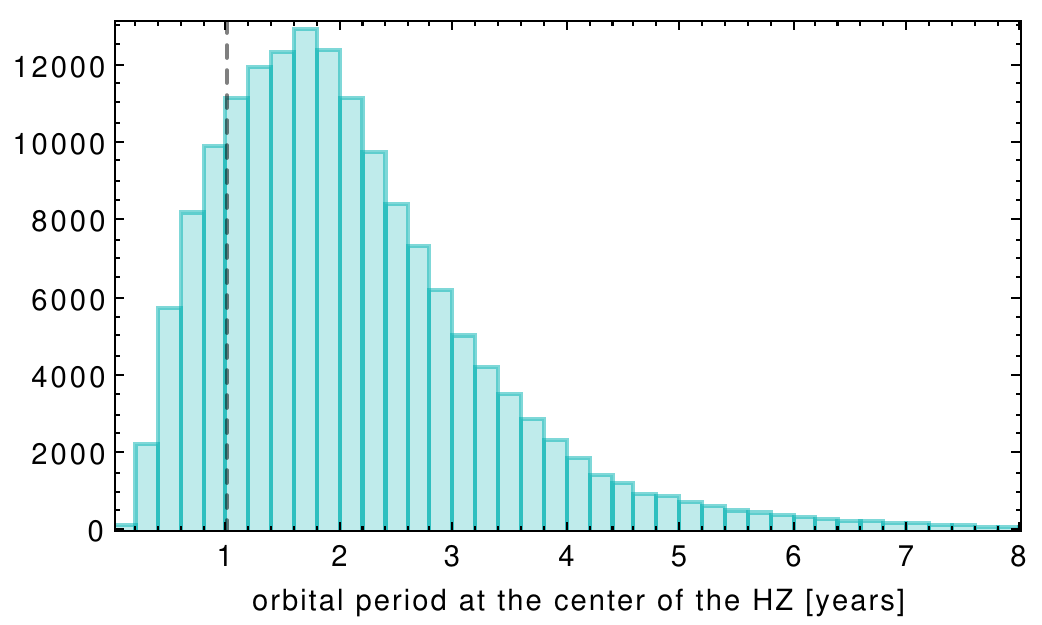}\\ \vspace{3mm}
    \caption{Histograms of some astrophysical quantities of the P1+P2 (red histograms, left side) and P5 (blue histograms, right side) samples within the LOPS2 field (see Section~\ref{sec:main_targets} for details). From top to bottom: effective temperature $T_\mathrm{eff}$ in K, stellar radius $R_\star$ and mass $M_\star$ in solar units, and orbital period $P_\mathrm{HZ}$ corresponding to the center of the habitable zone according to the \citet{Kasting1993} definition ($a_\mathrm{hz}\textrm{ [au]}=\sqrt{L_\star/L_\odot}$) and the Kepler's laws. The Sun/Earth values are marked with dashed vertical line as reference.}
    \label{fig:histograms}
\end{figure*}

\begin{figure*}[p]
    \centering
    \includegraphics[width=0.97\columnwidth]{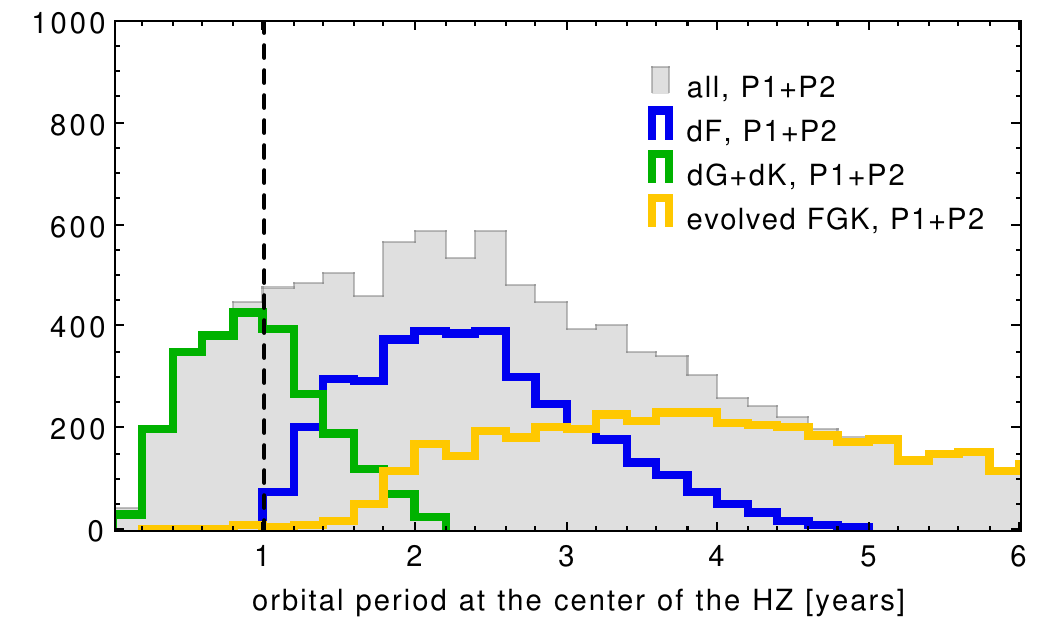}\hspace{1mm}
    \includegraphics[width=0.97\columnwidth]{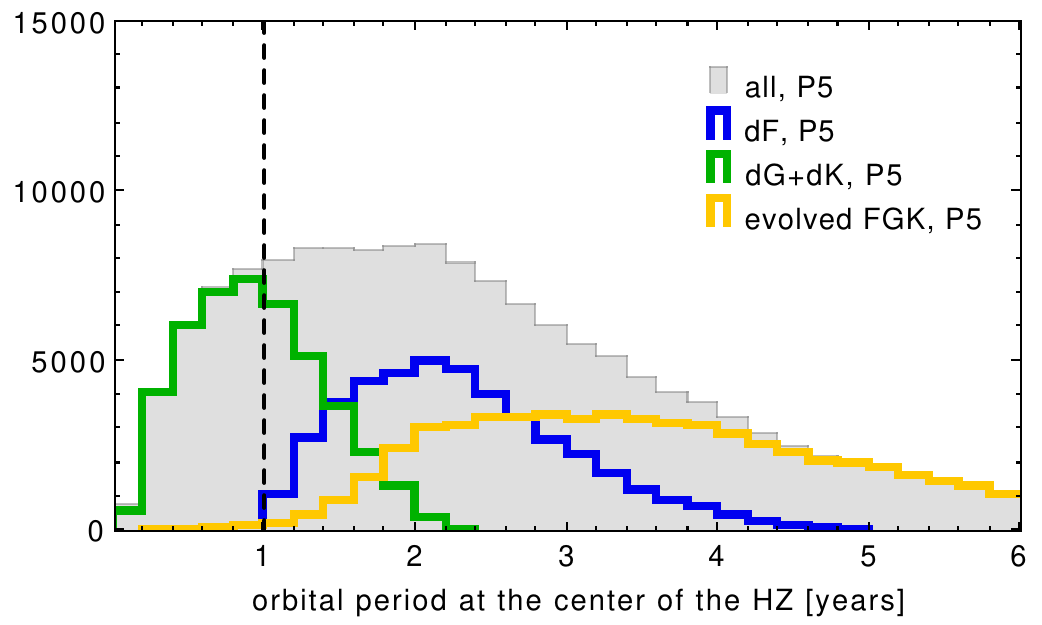}\\ \vspace{5mm}
    \includegraphics[width=0.97\columnwidth]{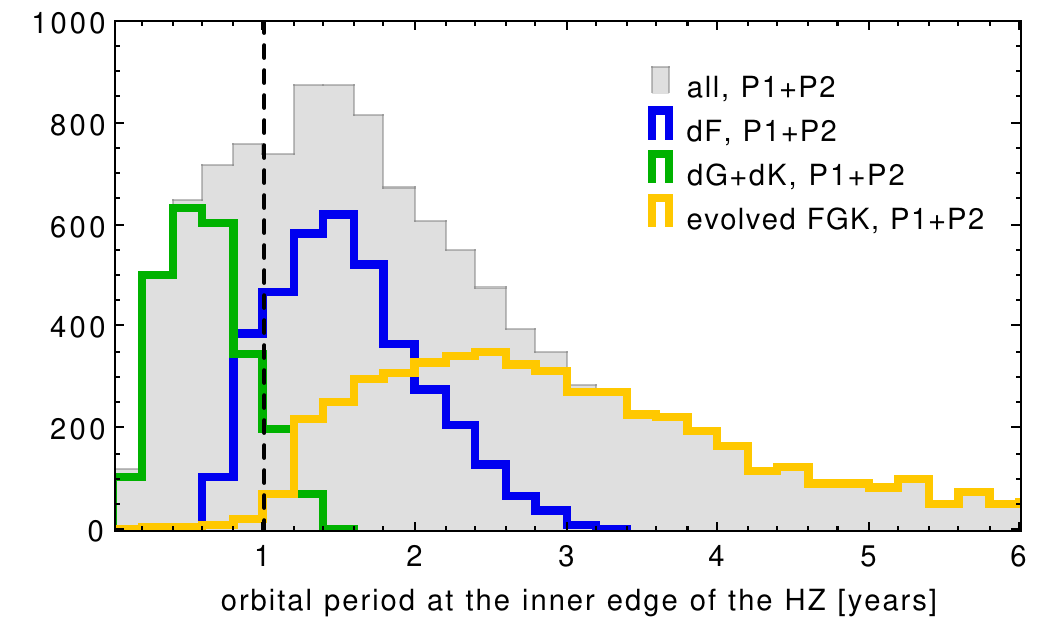}\hspace{1mm}
    \includegraphics[width=0.97\columnwidth]{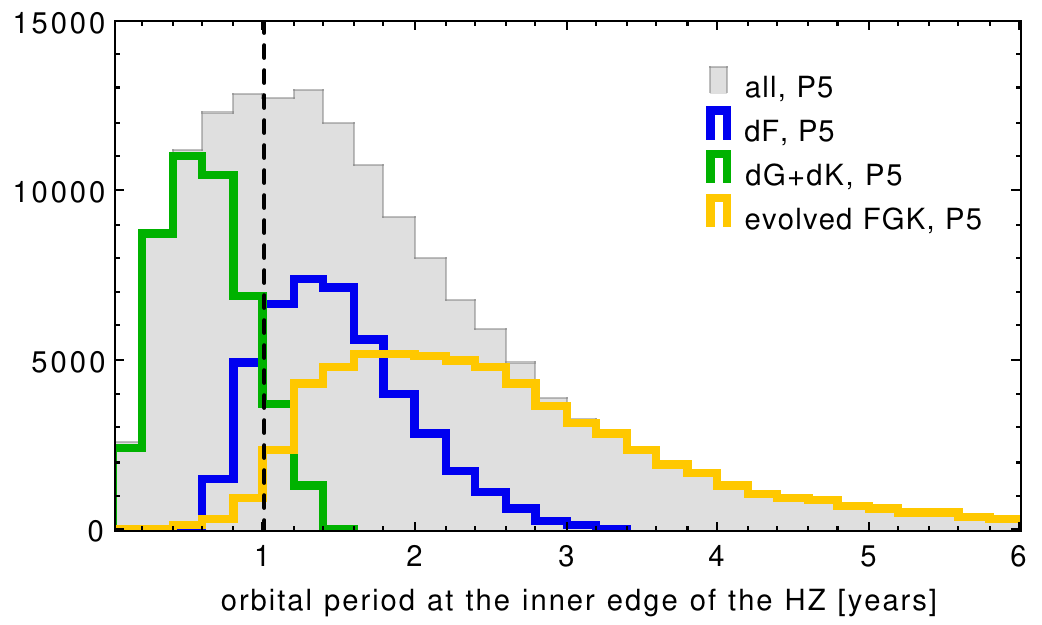}\\ \vspace{5mm}
    \caption{Orbital periods corresponding to the habitable zone around PLATO targets. \emph{Upper panels:} Histograms of the expected orbital period of a planet at the center of the habitable zone according to the the \citet{Kasting1993} definition ($a_\mathrm{hz}\textrm{ [au]}=\sqrt{L_\star/L_\odot}$) and Kepler's laws, for the P1+P2 (left side) and P5 (right side) samples within the LOPS2 field (see Section~\ref{sec:main_targets} for details). The Sun/Earth value (1~yr) is marked with dashed vertical line as reference. Both the distribution of the whole sample (in gray) and for a few subsamples of interest is shown: main-sequence F stars (blue), main-sequence G and K stars (green) and FGK subgiants (orange). The definition of these sub-samples in terms of $T_\mathrm{eff}$ and $R_\star$ can be found in \citetalias{Nascimbeni2022}, Section 6.3. \emph{Lower panels:} Same but for a planet located at the inner edge of the habitable zone according to the the \citet{Kasting1993} definition ($a_\mathrm{hz}\textrm{ [au]}=0.77\times\sqrt{L_\star/L_\odot}$).}
    \label{fig:histograms_hz}
\end{figure*}


\begin{figure*}[p]
    \centering
    \includegraphics[width=0.97\columnwidth]{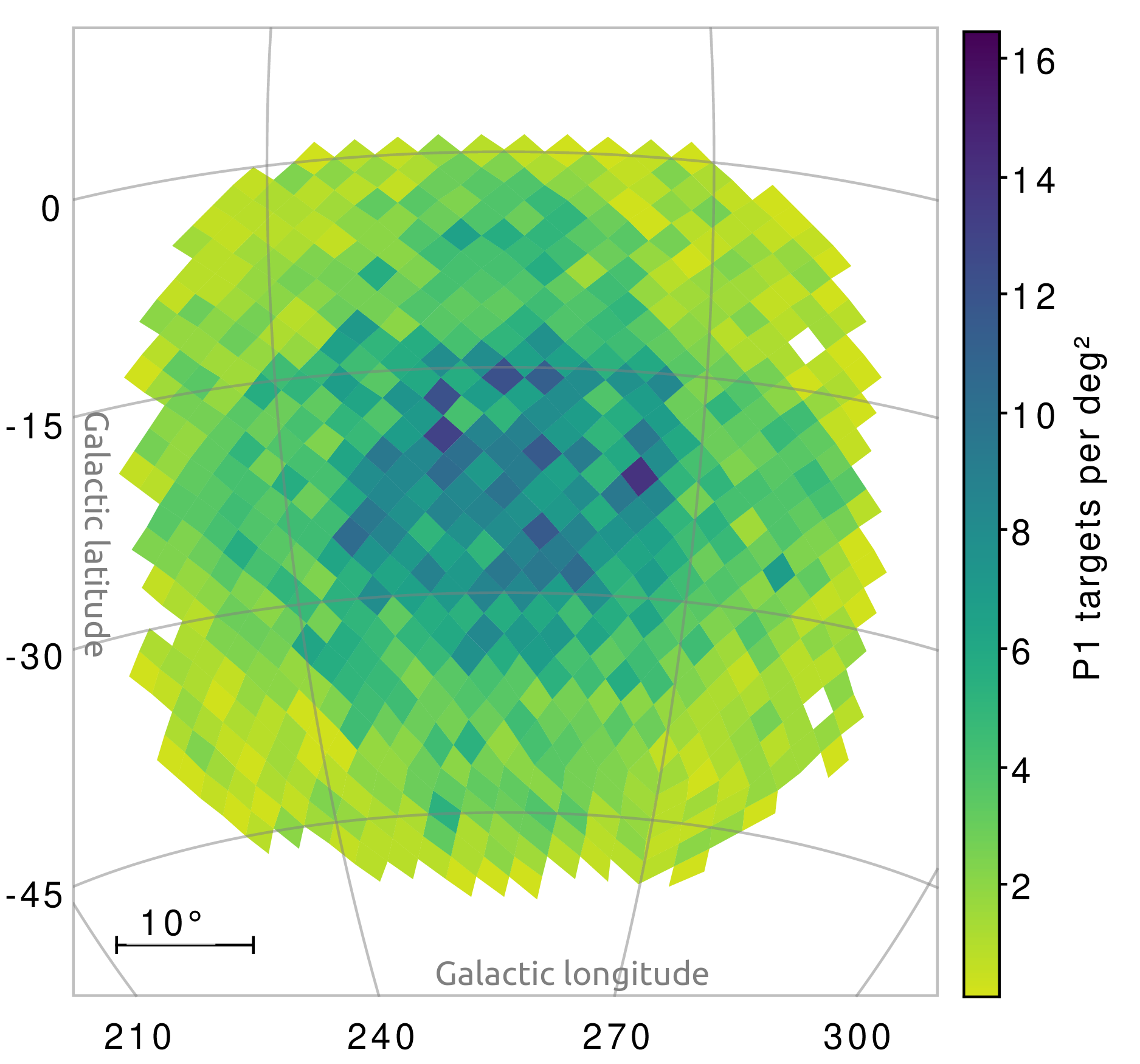}\hspace{1mm}
    \includegraphics[width=0.97\columnwidth]{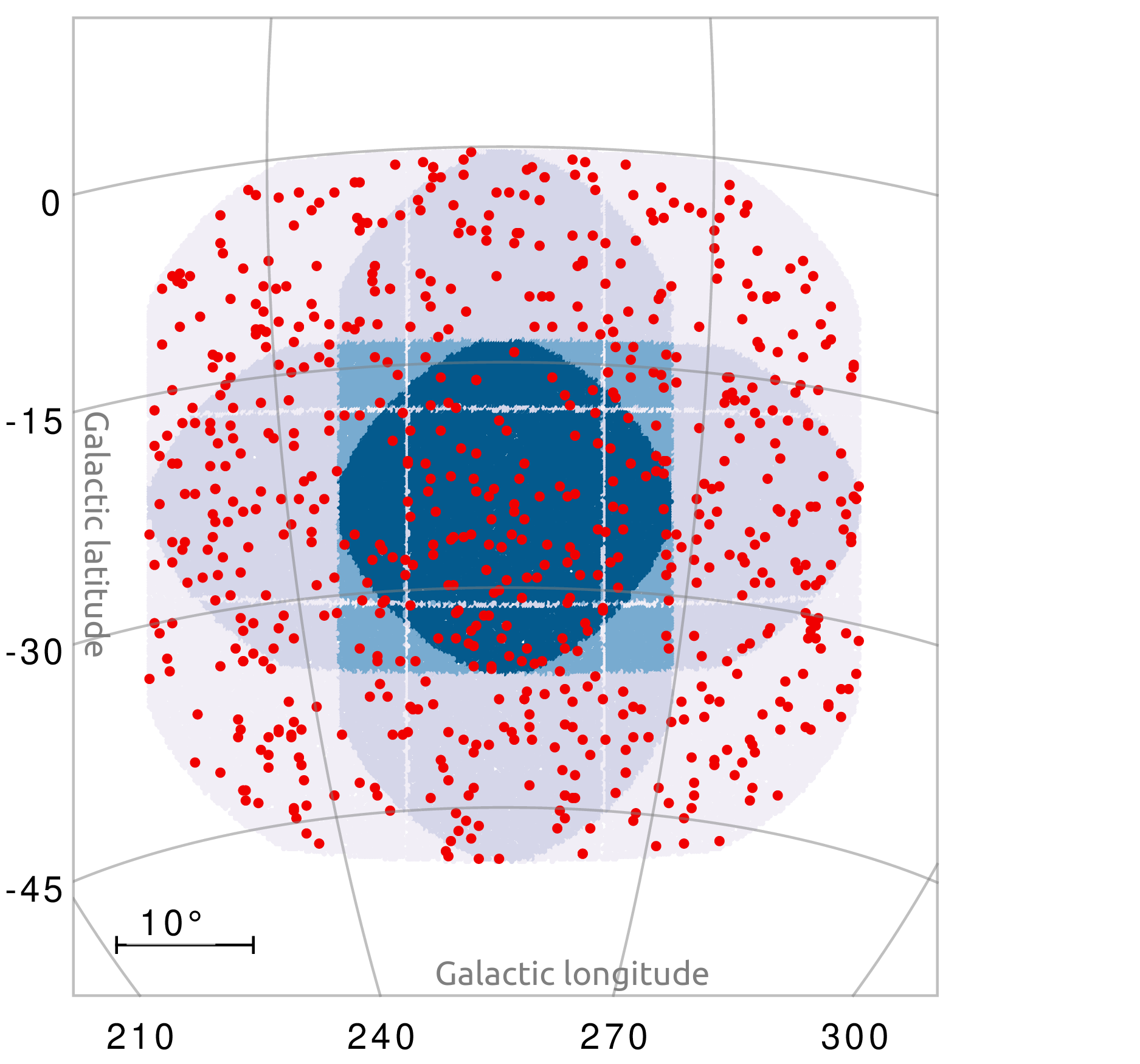}\\ \vspace{5mm}
    \includegraphics[width=0.97\columnwidth]{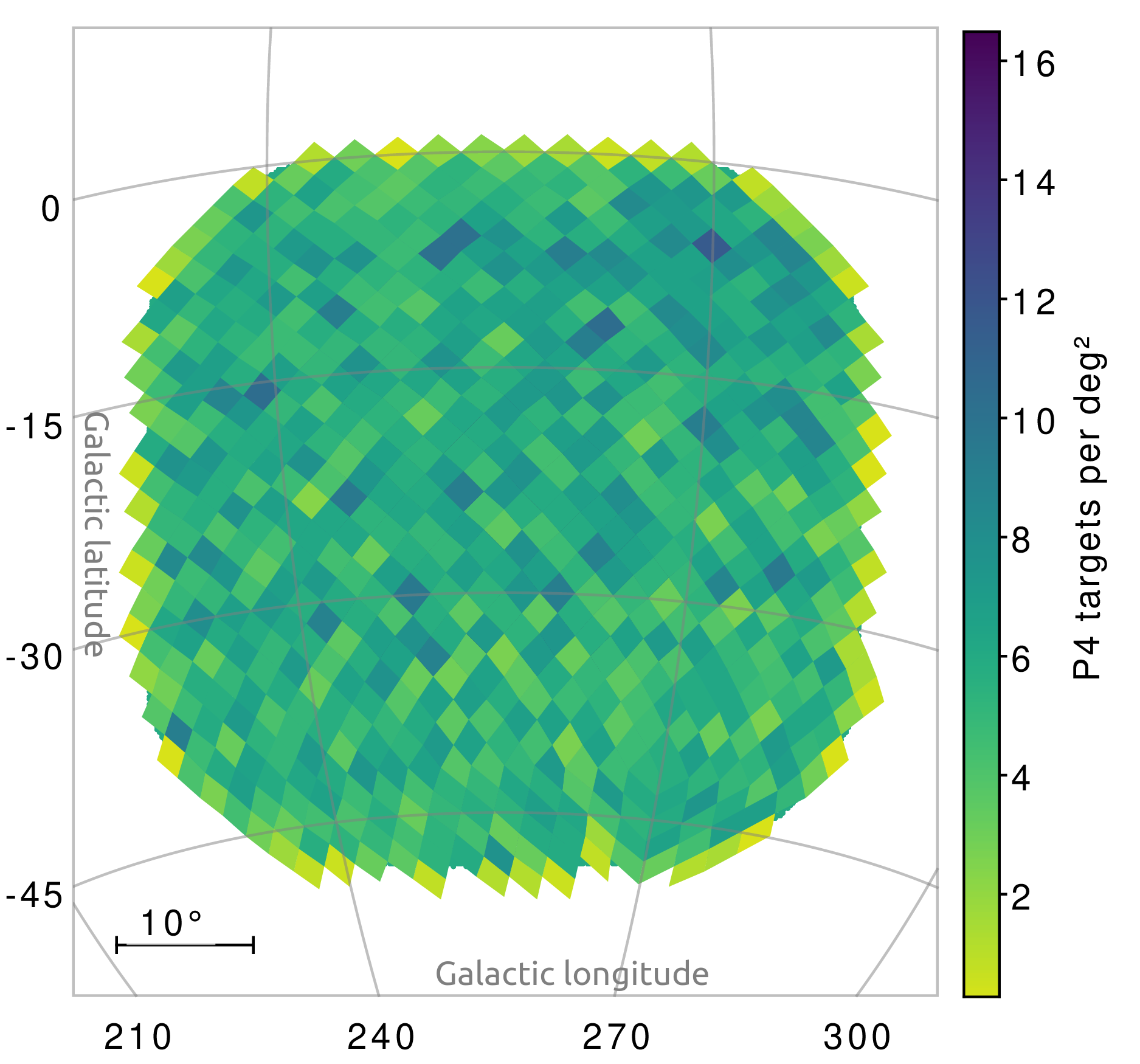} \hspace{1mm}
    \includegraphics[width=0.97\columnwidth]{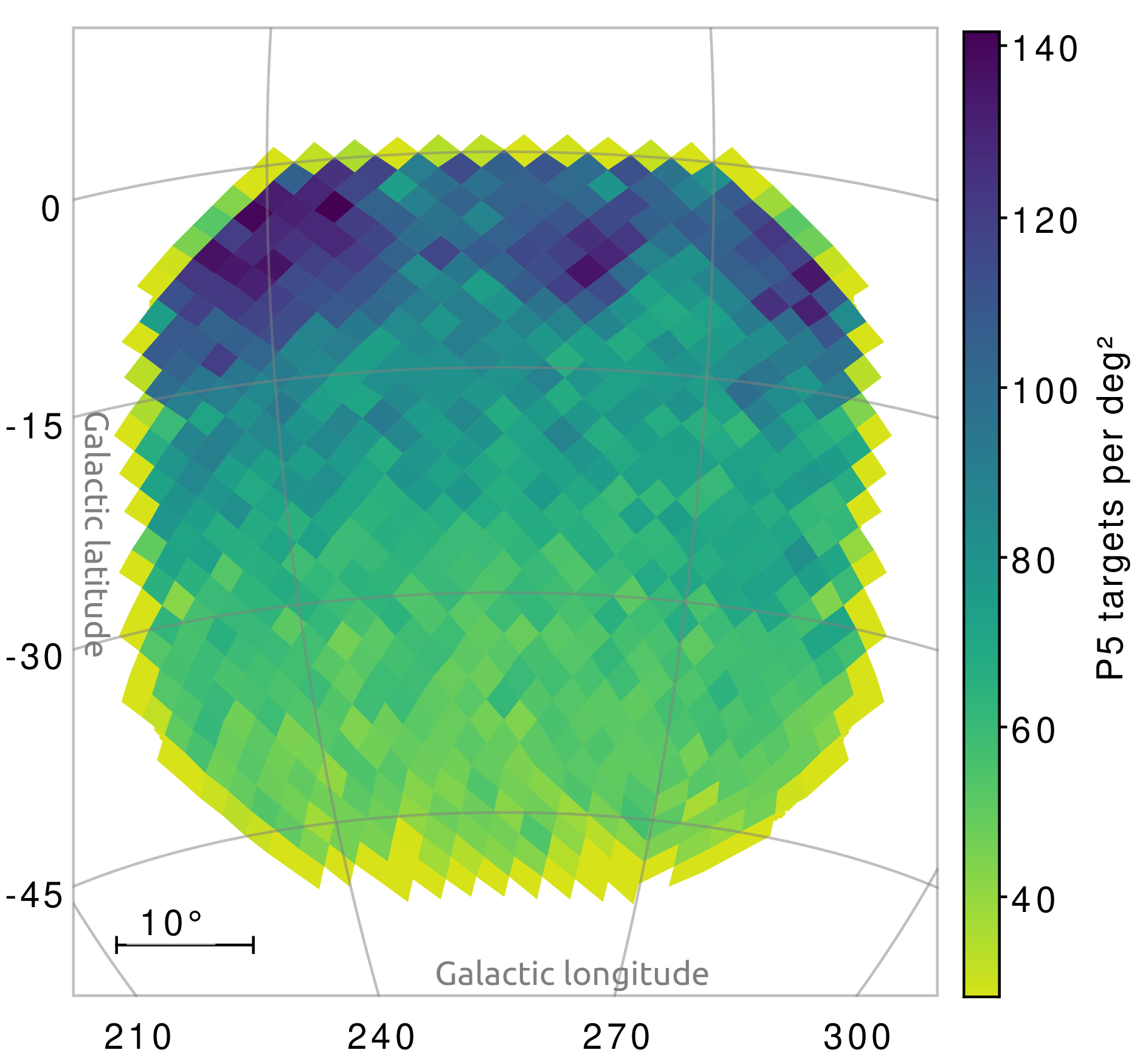}\\ \vspace{5mm}
    \caption{Spatial distribution of the P1-P2-P4-P5 targets (in reading order) over the LOPS2 field. P1, P4 and P5 targets are represented as color coded density maps based on an HEALPix level-5, while the sparse and almost perfectly isotropic distribution of the P2 sample is plotted with individual red points. See also Section~\ref{sec:main_targets} for a discussion of the sample properties.}
    \label{fig:densitymaps}
\end{figure*}


\longtab[1]{\renewcommand{\arraystretch}{1.08}
\begin{longtable}{lcrrrrrcccl}
        \caption{List of confirmed transiting exoplanets in LOPS2 (see Section~\ref{sec:transiting_planets} for details).} \\
    \hline\hline
    planet name & $N_p$ & $P$~[days] & $R_\mathrm{p}/R_\oplus$ & $M_\mathrm{p}/M_\oplus$ & $T_\mathrm{eq}$~[K] & $V$~[mag] & edge? & $N_\mathrm{cam}$ & FCAM? & comment \\
    \hline
    \endfirsthead
    \caption{Continued.} \\
    \hline\hline
    planet name & $N_p$ & $P$~[days] & $R_\mathrm{p}/R_\oplus$ & $M_\mathrm{p}/M_\oplus$ & $T_\mathrm{eq}$~[K] & $V$~[mag] & edge? & $N_\mathrm{cam}$ & F-CAM? & comment \\
    \hline
    \endhead
    \hline
    \endfoot
    \hline
    \endlastfoot
  GJ 238 b & 1 & 1.74 & 0.6 & --- & 750 & 11.57 & no & 24 & yes & \\
  GJ 341 b & 1 & 7.58 & 0.9 & --- & 532 & 9.46 & yes & 6 & no & \\
  HATS-39 b & 1 & 4.58 & 17.6 & 200.2 & 1635 & 12.74 & no & 6 & no & \\
  HATS-40 b & 1 & 3.26 & 17.7 & 505.3 & 2085 & 13.48 & no & 12 & no & \\
  HATS-41 b & 1 & 4.19 & 14.9 & 3083.0 & 1683 & 12.68 & no & 6 & no & \\
  HATS-42 b & 1 & 2.29 & 15.7 & 597.5 & 1842 & 13.68 & no & 12 & no & \\
  HATS-43 b & 1 & 4.39 & 13.2 & 83.0 & 993 & 13.56 & no & 12 & no & \\
  HATS-44 b & 1 & 2.74 & 12.0 & 178.0 & 1164 & 14.4 & no & 12 & no & \\
\end{longtable}
    \tablefoot{The columns give: the name of the planet, the multiplicity of the system, the orbital period $P$ in days, the planetary radius $R_\mathrm{p}$ and mass $M_\mathrm{p}$ in Earth units (when known), the planetary equilibrium temperature in K, the $V$ magnitude of the host star, a flag if the target is close to the LOPS2 outer edge, the number of NCAMs, a flag if the target is also observed with the FCAMs, and a comment field.}
    }


\longtab[2]{\renewcommand{\arraystretch}{1.08}\small
    \begin{longtable}{lcrrrrrp{8cm}}
        \caption{List of TESS candidates in LOPS2, from the TOI/ExoFOP database (see Section~\ref{sec:tois} for details).} \\
    \hline\hline
    TOI ID & $N_p$ & $T$~[mag] & $P$~[days] & $R_\mathrm{p}/R_\oplus$ & $\delta$~[ppm] & $T_\mathrm{eq}$~[K] & comment \\
    \hline
    \endfirsthead
    \caption{Continued.} \\
    \hline\hline
    TOI ID & $N_p$ & $T$~[mag] & $P$~[days] & $R_\mathrm{p}/R_\oplus$ & depth [ppm] & $T_\mathrm{eq}$~[K] & comment \\
    \hline
    \endhead
    \hline
    \endfoot
    \hline
    \endlastfoot
    119.01 & 2 & 9.23 & 5.541 & 2.04 & 633 & 1117 & Potential multi; slight depth aperture correlation with no offset\\
  119.02 & 2 & 9.23 & 10.692 & 1.8 & 535 & 710 & outer candidate in potential multi; variable host\\
  124.01 & 1 & 11.04 & 1.843 & 21.27 & 1950 & 1785 & TFOP SB1/APC; v-shaped\\
  153.01 & 1 & 11.69 & 7.632 & 12.44 & 4580 & 0 & \\
  163.01 & 1 & 10.87 & 4.231 & 14.63 & 7261 & 1584 & TOI-163 b\\
  167.01 & 1 & 11.57 & 4.453 & 0.0 & 5910 & 1027 & no stellar radius; Gaia DR2 Teff + plx indicate sun-like star\\
  171.01 & 1 & 11.54 & 1.239 & 13.58 & 13300 & 1461 & \\
  174.01 & 5 & 8.7 & 17.667 & 1.87 & 854 & 462 & HD 23472 b / TOI 174.01\\
\end{longtable}
    \tablefoot{The columns give: the TOI identifier of the planet, the multiplicity of the system, the TESS magnitude $T$, the orbital period $P$ in days, the planetary radius $R_\mathrm{p}$ in Earth units, the planetary equilibrium temperature in K, and a comment field. All the columns except the second one were copied from the TOI database, 2024-08-20 release \citep{Guerrero2021}}
    }


\end{document}